\documentclass[11pt,letterpaper,oneside]{article}
\usepackage{graphicx}
\usepackage{amssymb,amsmath,amsfonts}
\usepackage{bm}
\usepackage{subfigure}
\usepackage{graphicx}

\usepackage[a4paper,left=1in,top=1in,right=1in,bottom=1in,ignoreheadfoot]{geometry}
\usepackage{algorithm}
\usepackage{setspace}

\providecommand{\bx}{{\bf x}}
\providecommand{\by}{{\bf y}}

\providecommand{\bbeta}{{\bm \beta}}

\providecommand{\bX}{{\bf X}}

\providecommand{\btau}{{\bm \tau}}

\usepackage{threeparttable}
\usepackage{pgf}
\usepackage{tikz}
\usepackage{subfig}
\usetikzlibrary{arrows,automata}

\begin{document}
\setlength{\parindent}{0pt}
\setlength{\parskip}{\baselineskip}

\author{Anthony Lee \footnote{Oxford-Man Institute and Department of Statistics, University of Oxford, UK}
\and Francois Caron \footnote{INRIA Bordeaux Sud-Ouest and Institut de Math\'ematiques de Bordeaux, University of Bordeaux, France}
\and Arnaud Doucet \footnote{Institute of Statistical Mathematics, Japan and University of British Columbia, Department of Statistics and Department of Computer Science, Canada.}
\and Chris Holmes \footnote{Department of Statistics and Oxford-Man Institute and Wellcome Trust Centre for Human Genetics, University of Oxford, and MRC Harwell, UK}
}
\title{Bayesian Sparsity-Path-Analysis of Genetic Association Signal using Generalized t Priors}

\maketitle

\begin{abstract}
We explore the use of generalized t priors on regression coefficients to help understand the nature of association signal within ``hit regions'' of genome-wide association studies. The particular generalized t distribution we adopt is a Student distribution on the absolute value of its argument. For low degrees of freedom we show that the generalized t exhibits `sparsity-prior' properties with some attractive features over other common forms of sparse priors and includes the well known double-exponential distribution as the degrees of freedom tends to $\infty$. We pay particular attention to graphical representations of posterior statistics obtained from sparsity-path-analysis (SPA) where we sweep over the setting of the scale (shrinkage / precision) parameter in the prior to explore the space of posterior models obtained over a range of complexities, from very sparse models with all coefficient distributions heavily concentrated around zero, to models with diffuse priors and coefficients distributed around their maximum likelihood estimates. The SPA plots are akin to LASSO plots of maximum {\it{a posteriori}} (MAP) estimates but they characterise the complete marginal posterior distributions of the coefficients plotted as a function of the precision of the prior. Generating posterior distributions over a range of prior precisions is computationally challenging but naturally amenable to sequential Monte Carlo (SMC) algorithms indexed on the scale parameter. We show how SMC simulation on graphic-processing-units (GPUs) provides very efficient inference for SPA. We also present a scale-mixture representation of the generalized t prior that leads to an EM algorithm to obtain MAP estimates should only these be required.

\end{abstract}

\section{Introduction}
Genome-wide association studies (GWAS) have presented a number of interesting challenges to statisticians \cite{Balding2006}. Conventionally GWAS use single marker univariate tests of association, testing marker by marker in order to highlight ``hit regions'' of the genome showing evidence of association signal. The motivation for our work concerns methods in Bayesian sparse multiple regression analysis to characterise and decompose the association signal within such regions spanning possibly hundreds of markers. Hit-regions typically cover loci in high linkage disequilibrium (LD) which leads to strong correlation between the markers making multiple regression analysis non-trivial. A priori we would expect only a small number of markers to be responsible for the association and hence priors that induce sparsity  are an important tool to aid understanding of the genotype-phenotype dependence structure; an interesting question being whether the association signal is consistent with a single causal marker or due to multiple effects.
We restrict attention to case-control GWAS, i.e. binary phenotypes, these being by far the most prevalent although generalisations to other exponential family likelihoods or non-linear models is relatively straightforward.

In the non-Bayesian literature sparse regression analysis via penalised likelihood has gained increasing popularity since the seminal paper on the LASSO \cite{lasso,tibshirani2011regression}. The LASSO estimates have a Bayesian interpretation as  maximum {\it a posteriori} (MAP) statistics under a double exponential prior on regression coefficients $\pi(\beta) \propto \exp(- \lambda \sum_j | \beta_j |)$ and it is known that, unlike ridge penalties using a normal prior, the Lasso prior tends to produce `sparse' solutions in that as an increasing function of the regulariation penalty $\lambda$ more and more of the MAP estimates will be zero. From a Bayesian perspective the use of MAP estimates holds little justification, see the discussion of \cite{tibshirani2011regression},  and in \cite{park2008bayesian} they explore full Bayesian posterior analysis with Lasso double-exponential priors using Makov chain Monte Carlo (MCMC) to make inference.  While the Lasso penalty / prior is popular there is increasing awareness that the use of identical `penalization' on each coefficient, e.g. $\lambda\sum^p_{j=1}|\beta_j|$, can lead to unacceptable bias in the resulting estimates \cite{penlike}. In particular coefficients get shrunk towards zero even when there is overwhelming evidence in the likelihood that they are non-zero. This has motivated use of sparsity-inducing non-convex penalties which reduce bias in the estimates of large coefficients at the cost of increased difficulty in computing. Of note we would mention the ``adaptive'' methods \cite{adaptive_lasso,one_step} in the statistics literature and iteratively reweighted methods \cite{rwl1,rwl2} in the signal processing literature; although these papers are only concerned with MAP estimation there being no published articles to date detailing the use of full Bayesian analysis.

In the context of GWAS there have been recent illustrations of the use of adaptive sparsity priors and MAP estimation \cite{Hoggart2008}, reviewed and compared in \cite{Ayers2010}. In  \cite{Hoggart2008} they use the Normal-Exponential-Gamma sparsity-prior of \cite{griffin07} to obtain sparse MAP estimates from  logistic regression. The use of continuous sparse priors can be contrasted with an alternative approach using two component mixture priors such as in \cite{George.McCulloch.93} which adopt a distribution containing a spike at zero and another component with broad scale. Coefficients are then {\it{a posteriori}} classified into either state, relating to whether their corresponding variables (genotypes) are deemed predictively irrelevant or relevant. In the GWAS setting the two-component mixture prior has been explored by \cite{Wilson2010,fridley2009bayesian}. For model exploration the sparsity prior has some benefits in allowing the statisticians to visualize the regression analysis over a range of scales / model complexities.

In this paper we advocate the use of the generalized t prior, first considered in \cite{mcdonald1988partially}, on the regression coefficients and a full Bayesian analysis. The generalized t prior we adopt is a Student distribution on the absolute value of the coefficient. We demonstrate it has a number of attractive features as a `sparsity prior' in particular due to its simple analytic form and interpretable parameters. In applications the setting of the scale parameter in the prior is the key task that affects the sparsity of the posterior solution. We believe much is to be gained in exploring the continuum of posterior models formed by sweeping through the scale of the prior, even over regions of low posterior probability, in an exploratory approach we term sparsity-path-analysis (SPA). This leads towards providing graphical representations of the posterior densities that arise as we move from the most sparse models with all coefficient densities heavily concentrated around the origin through to models with diffuse priors and coefficients distributed around their maximum likelihood estimates. We stress that even though parts of this model space have low probability, they have utility and are interesting in gaining a fuller understanding of the predictor-response (genotype-phenotype) dependence structure. SPA is an exploratory tool that is computationally challenging but highly suited to sequential Monte Carlo (SMC) algorithms indexed on the scale parameter and simulated in parallel on graphics cards using graphical processing units (GPUs).

The generalized t as a prior for Bayesian regression coefficients was first suggested by \cite{mcdonald1988partially} and more recently in the context of sparse MAP estimation for normal (Gaussian) likelihoods in \cite{sip,cevher,armagan1}, and using Markov chain Monte Carlo for a full Bayesian linear regression analysis in \cite{armagan1}. In \cite{cevher, armagan1} the prior is referred to as a double-pareto distribution but we prefer to use the terminology of the generalized t as we feel it is more explicit and easier to interpret as such. Our contribution in this paper is to consider a full Bayesian analysis of logistic regression models with generalized t priors in the context of GWAS applications for which we develop SMC samplers and GPU implementation. The GPU implementation is important in practice, in the GWAS analysis in reduces run-time from over five days to hours. We pay particular attention to developing graphical displays of summary statistics from the joint posterior distribution over a wide range of prior precisions; arguing that parts of the model space with low probability have high utility and are informative to the exploratory understanding of regression signal.

In Section 2 we introduce the generalized t prior and show that it has a scale-mixture representation and exhibits sparsity type properties for regression analysis. Section 3 provides a brief description of the genetic association data and simulated phenotypes used in our methods development. Section 4 concerns computing posterior distributions over a range of prior scale parameters using sequential Monte Carlo algorithms simulated on GPUs. Section 5 deals with graphical display of SPA statistics and plots of posterior summary statistics over the path of scales. We conclude with a discussion in Section 6.

\section{Generalized t prior for sparse regression analysis}

In Bayesian logistic regression analysis given a data set $\{\bX, \by\}$, of $(n \times p)$ predictor matrix $\bX$ and $(n \times 1)$ vector of binary phenotypes (or response variables) $\by$, we adopt a prior $p(\bbeta)$ on the $p$ regression coefficients from which the log-posterior updates as,
$$
\log p(\bbeta | \bX, \by, a, c) \propto \log f(\by | \bX, \bbeta, \theta) + \log p(\bbeta)
$$
where $f(\cdot)$ is the likelihood function,
\begin{eqnarray}
f(\by | \bX, \bbeta) & = & \prod_{i=1}^n (p_i)^{y_i} (1 - p_i)^{1-y_i} \nonumber \\
p_i & = &  logit^{-1}(x_i \bbeta) \nonumber
\end{eqnarray}

In this paper we propose the generalized t  distribution as an interesting prior for $\bbeta$, which has the form of a Student density on the $L_q$-norm of its argument, \cite{mcdonald1988partially},
\begin{align}
\label{eqn:gent_prior}
p(\beta |\mu, a,c,q) = \frac{q}{2ca^{1/q} B(1/q,a)}\left ( 1 + \frac{|\beta - \mu|^q}{a c^q} \right)^{-(a+1/q)}
\end{align}
where we note that the usual Student's density is obtained with $\{q=2, c=\sqrt{2}\}$; the exponential power family of distributions arise as degrees of freedom $a \to \infty$; and the double-exponential or Laplace distribution occurs with $\{q=1, a=\infty\}$. The  use of this distribution as a prior on regression coefficients was first described in \cite{mcdonald1988partially}.

We shall only consider the $L_1$-norm case with central location $\mu=0$, and independent priors across the $p$ coefficients for which we can simplify the distribution to,
\begin{align}
\label{eqn:gent_prior}
p(\beta_j |\mu=0, a,c, q=1) = \frac{1}{2c}\left ( 1 + \frac{|\beta_j|}{a c} \right)^{-(a+1)}
\end{align}
where $b>0$ is a scale parameter and $a >0$ is related to the degrees of freedom; we will denote this centred $L_1$ generalized t as $Gt(a, c)$. Note that \cite{cevher, armagan1} refer to this distribution as a double-pareto but we prefer to think of it as an instantiation of a generalized t following the earlier derivation in the context of priors on regression coefficients in \cite{mcdonald1988partially}. This also allows for easier understanding of the parameters and direct connections with the double-exponential distribution and other standard densities.

It is interesting to note that (\ref{eqn:gent_prior}) can be derived from the marginal of a scale-mixture of double-exponential densities, such that, for multivariate $\bbeta = \{\beta_1, \ldots, \beta_p\}$, where,
\begin{eqnarray}
p(\bbeta) &  = & \prod_j p(\beta_j | \tau_j) \nonumber \\
p(\beta_j | \tau_j) & = & \frac{1}{2\tau_j}\exp \left(-\frac{|\beta_j|}{\tau_j} \right) \nonumber \\
\tau_j & \sim & IG(a, b)
\label{eqn:scalemix}
\end{eqnarray}
where $IG(\cdot, \cdot)$ denotes the Inverse-Gamma density, from this we find,
$$
p(\beta_j) = \int_{\tau_j} p(x_j | \tau_j) p(\tau_j | a, b) d \tau_j = Gt(a,b/a).
$$
where we can set $b = a c$ to generate a marginal $Gt(a,c)$ density. Hence as a prior on the set of regression coefficients, one way to think of (\ref{eqn:gent_prior}) is as defining a hierarchical prior whereby each coefficient has a double-exponential prior with a coefficient specific scale parameter distributed a priori as $IG(a,ac)$.  A plot of the log densities for $Gt(1,1)$, and the double-exponential, $DE(0, 1)$ is shown in Figure \ref{fig:gent_dexp} where we can see the greater Kurtosis and heavier tails provided by the generalized t. As mentioned in the Section 1, the relatively light tails of the $DE(\cdot)$ prior is unattractive as it tends to shrink large values of the coefficients even when there is clear evidence in the likelihood that they are predictively important.

We are interested with inferring and graphing the regression coefficents' posterior distributions over a range of values for $c$ for fixed degrees of freedom $a$, an exploratory procedure we call sparsity-path-analysis. But first it is instructive to examine the form of the MAP estimates obtained under the generalized t prior as this sheds light on their utility as sparsity priors.

\subsection{MAP estimates via the EM algorithm}
\label{section:compute_map}
The optimization problem associated with obtaining the MAP estimates for $\bbeta$ under the generalized t-distribution prior is not concave. However, one can find local modes of the posterior by making use of the scale-mixture representation (\ref{eqn:scalemix}) and using the EM algorithm with the hierarchical scale parameters $\btau = \tau_{1:p}$ as latent variables. Indeed, each iteration of EM takes the form
$$\bbeta^{(t+1)} = \arg \max_{\bbeta} \log f(\by | \bX, \bbeta) + \int \log [p(\bbeta|\btau)] p(\btau | \bbeta^{(t)}, a, b) d\btau$$
where $b=a c$. The conjugacy of the inverse-gamma distribution with respect to the Laplace distribution gives
$$\tau_j | \beta_j^{(t)}, a_j, b_j \sim IG(a_j+1,b_j + |\beta_j|)$$
allowing for different prior parameters for each coefficient, and with $p(\bbeta|\btau) = \prod^p_{j=1} p(\beta_j|\tau_j) = \prod^p_{j=1} 1/(2\tau_j)\exp(-|\beta_j|/\tau_j)$ yields
$$\bbeta^{(t+1)} = \arg \max_{\bbeta} \log f(\by | \bX, \bbeta, \theta) - \sum^p_{j=1} |\beta_j| \int \frac{1}{\tau_j} p(\tau_j | \beta_j^{(t)},a_j, b_j) d\tau_j$$
where the expectation of $1/\tau_j$ given $\tau_j \sim IG(a_j+1,b_j+|\beta^{(t)}_j|)$ is $(a_j+1)/(b_j + |\beta^{(t)}_j|)$.

As such, one can find a local mode of the posterior $p(\bbeta | \by, \bX, \bbeta, \theta)$ by starting at some point $\bbeta^{(0)}$ and then iteratively solving
\begin{equation}
\label{eqn:beta_it}
\bbeta^{(t+1)} = \arg \max_{\bbeta} \log f(\by | \bX, \bbeta, \theta) - \sum^p_{j=1} w_j^{(t)}|\beta_j|
\end{equation}
where
\begin{eqnarray}
\label{eqn:adapt_weights}
w_j^{(t)} & = & \frac{a_j+1}{b_j + |\beta^{(t)}_j|} \nonumber \\
& = & \frac{a_j+1}{a_j c_j + |\beta^{(t)}_j|} \nonumber
\end{eqnarray}
For logistic regression the update (\ref{eqn:beta_it}) is a convex optimization problem for which standard techniques exist. We can see the connection to the Lasso as with $a \to \infty$ we find $w_j^{(t)} \to \frac{1}{c_j}$, where $\lambda = 1 / c$ is the usual Lasso penalty.

\subsection{Oracle properties of the MAP}

In the penalized optimization literature, some estimators are justified at least partially by their possession of the oracle property: that for appropriate parameter choices, the method performs just as well as an oracle procedure in terms of selecting the correct predictors and estimating the nonzero coefficients correctly asymptotically in $n$ when the likelihood is Gaussian. Other related properties include asymptotic unbiasedness, sparsity and continuity in the data \cite{penlike}. Penalization schemes with the oracle property include the smoothly clipped absolute deviation method \cite{penlike}, the adaptive lasso \cite{adaptive_lasso} and the one-step local linear approximation method \cite{one_step}.

For the generalized t-distribution, the MAP estimate obtained using this prior has been examined in \cite{sip,cevher,armagan1}, all of which derive the expectation-maximisation algorithm for finding a local mode of the posterior. It is straightforward to establish that the estimate is asymptotically unbiased, sparse when $c < 2\sqrt{(a+1)}/a$ and continuous in the data when $c = \sqrt{(a+1)}/a$ following the conditions laid out in \cite{penlike}. \cite{armagan1} shows that this scheme satisfies the oracle property if simultaneously $a \rightarrow \infty$, $a/\sqrt(n) \rightarrow 0$ and $ac\sqrt(n) \rightarrow C < \infty$ for some constant $C$.

It is worth remarking that the oracle property requires the prior on $\bbeta$ to depend on the number of observations $n$. This is in conflict with conventional Bayesian analysis in that the prior should represent your beliefs about the data generating mechanism irrespective of the number of observations that you may or may not receive. In addition in the context of applications your $n$ and $p$ may well not be in the realm of asymptotic relevance.

\subsection{Comparison to other priors}
There have been a number of sparsity prior representations advocated recently in the literature, although it is important to note that most of the papers deal solely with MAP estimation and not full Bayesian analysis. To name a few important contributions we highlight the normal-gamma, normal-exponential-gamma (NEG) and the Horseshoe priors, see \cite{Griffin2010,Caron2008,griffin07,carvalho2010horseshoe}. Like all these priors the $Gt(\cdot, \cdot)$ shares the properties of sparsity and adaptive shrinkage, allowing for less shrinkage on large coefficients as shown in Fig.\ref{fig:shrink} for a normal observation, and a scale-mixture representation which facilitates an efficient EM algorithm to obtain MAP estimates.

In Table \ref{tab:priors} we have listed the log densities for the sparsity-priors mentioned above. From Table \ref{tab:priors} we see an immediate benefit of the generalized t density is that it has a simple analytic form compared with other sparsity-priors and is well known and understood by statisticians, which in turn aids explanation to data owners. The generalized t is also easy to compute relative to the NEG, normal-gamma and Horseshoe priors which all involve non-standard computation, making them more challenging to implement and less amenable to fast parallel GPU simulation. Finally, the setting of the parameters of the generalized t is helped as statisticians are used to thinking about degrees of freedom and scale parameters of Student's densities.  We believe this will have an impact on the attractiveness of the prior to non-expert users outside of sparsity researchers.

\begin{table}[htdp]
\begin{center}
\begin{tabular}{|c|c|}
\hline
 & $ \propto \log p(\beta_j)$ \\ \hline
  & \\
 double-exponential & $- \lambda | \beta_j | $ \\ &  \\ \hline
 normal-gamma & $(\frac{1}{2} - \lambda) \log | \beta_j| - \log K_{\lambda-1/2} \left( \frac{|\beta_j | }{\lambda} \right) $ \\ & \\
 $NEG$ & $ -\frac{\beta^2}{4 \lambda^2} - \log D_{-2(\lambda+\frac{1}{2})} \left( \frac{|\beta|}{\lambda} \right)$ \\  & \\
 Horseshoe & $\log\left[ \int_{0}^{\infty} (2 \tau^2 \lambda^2)^{-1/2} \exp\left(- \frac{\beta_j^2}{2\tau^2 \lambda^2}\right) (1+ \lambda)^{-2} d \lambda \right]$ \\ & \\
 $Gt(a,b)$ & $-(a+1) \log \left(1 + \frac{|\beta_j|}{ab}\right) $ \\
 \hline
\end{tabular}
\end{center}
\caption{Log densities for the double-exponential and various sparsity-priors. $K(\cdot)$ denotes a Bessel function of the third kind and $D_{v}(z)$ is the parabolic cylinder function. The Horseshoe prior does not have an analytic form.}
\label{tab:priors}
\end{table}%

\section{Genotype Data}

\begin{figure}[pth]
\center
\includegraphics[scale=0.55]{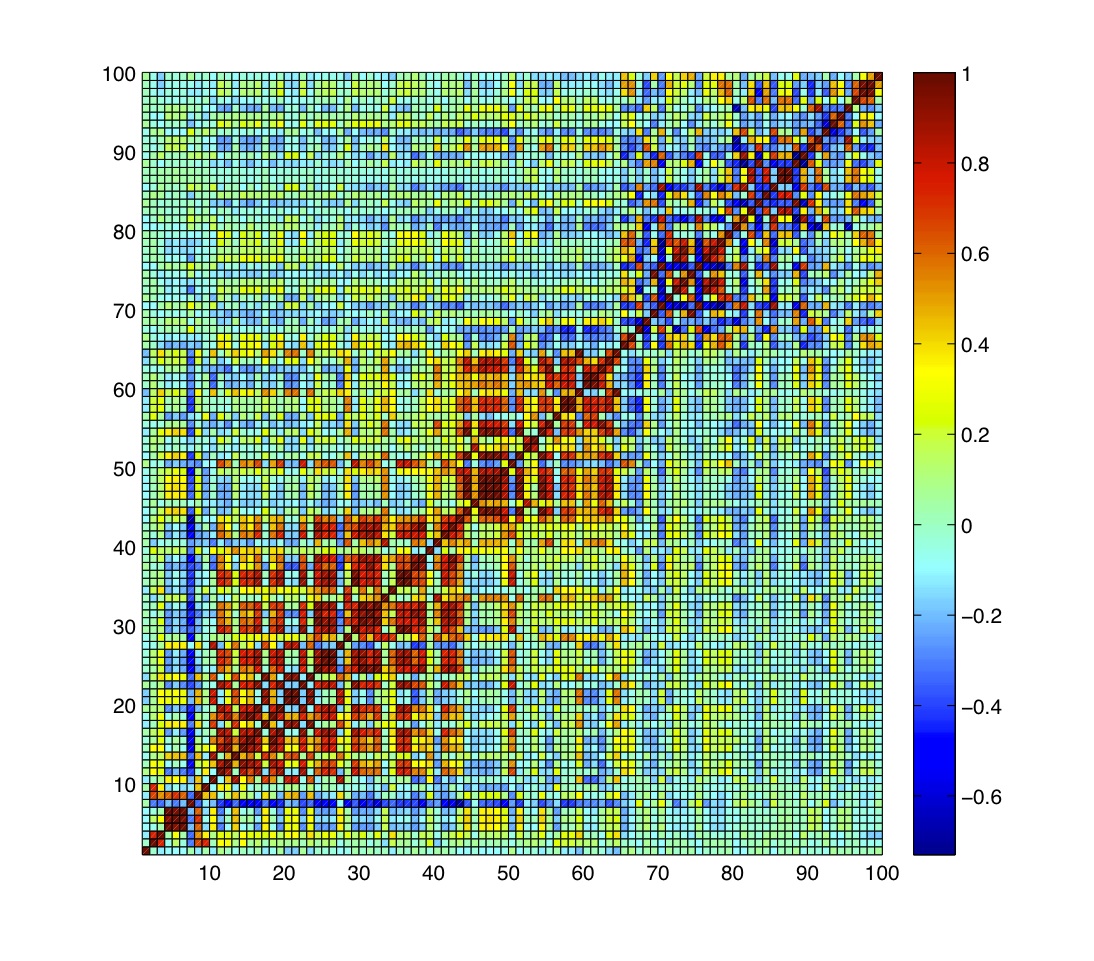}
\caption{Plot of correlation structure across the first 100 adjacent genotype markers from chromosome 18q.}
\label{fig:18qcorr}
\end{figure}

In the development and testing of the methods in this paper we made use of anonymous genotype data obtained from the Wellcome Trust Centre for Human Genetics, Oxford. The data consists of a subset of single-neucleotide polymorphisms (SNPs) from a genome-wide data set originally gathered in a case-control study to identify colorectal cancer risk alleles. The data set we consider consists of genotype data for 1859 subjects on 184 closely spaced SNPs from a hit-region on chromosome 18q. We standardise the columns of $\bX$ to be mean zero and unit standard deviation. A plot of the correlation structure across the first 100 markers in the region is shown in Fig \ref{fig:18qcorr}. We can see the markers are in high LD with block like strong correlation structure that makes multiple regression analysis challenging.

We generated pseudo-phenotype data, $\by$, by selecting at random five loci and generating coefficients, $\beta \sim N(0,0.2)$, judged to be realistic signal sizes seen in GWAS \cite{stephens2009bayesian}. All other coefficients were set to zero. We then generated phenotype, case-control, binary $\by$ data by sampling Bernoulli trials with probability given by the logit of the log-odds, $\bx_i \bbeta$, for an individuals genotype $\bx_i$.  We considered two scenarios. The first when we construct a genotype matrix of 500 individuals using the five relevant markers and an additional 45 ``null'' markers whose corresponding coefficients were zero. The second data set uses all available markers and all subjects to create a $(1859 \times 184)$ design matrix.

\section{Sequential Monte Carlo and graphic card simulation}

We propose to explore the posterior distributions of $\bbeta | \bX, \by, a, c$ with varying $c = b/a$ via Sequential Monte Carlo (SMC) sampler methodology \cite{smcs}. In particular, we let the target distribution at time $t$ be $\pi_t(\bbeta) = p(\bbeta | \bX, \by, a, b_t / a)$ with $t = \{1,\ldots,T\}$ and some specified sequence $\{b_t\}^T_{t=1}$ where $b_i > b_{i+1}$ and $b_T$ is small enough for the prior to dominate the likelihood with all coefficients posterior distributions heavily concentrated around zero, and $b_1$ large enough for the prior to be diffuse and have limited impact on the likelihood. In order to compute posterior distributions over the range of $\{b_j\}_{j=1}^T$ we make use of sequential Monte Carlo samplers indexed on this sequence of scales.

It is important to note that we cannot evaluate
$$\pi_t(\bbeta) = \frac{p(\by|\bX,\bbeta)p(\bbeta|a,b_t)}{\int p(\by|\bX,\bbeta)p(\bbeta|a,b_t) d\bbeta}$$
since we cannot evaluate the integral in the denominator, the marginal likelihood of $\by$. However, we can evaluate $\gamma_t(\bbeta) = Z_t \pi_t(\bbeta)$ where $Z_t = \int p(\by|\bX,\bbeta)p(\bbeta|a,b_t) d\bbeta$.

At time $t$ we have a collection of $N$ particles $\{\bbeta^{(i)}_t\}^N_{i=1}$ and accompanying weights $\{W^{(i)}_t\}^N_{i=1}$ such that the empirical distribution
$$\tilde{\pi}_t^N(d\bbeta) = \sum^N_{i=1}W_t^{(i)}\delta_{\bbeta^{(i)}_t}(d\bbeta)$$
is an approximation of $\pi_t$. The weights are given by
\begin{equation}
\label{eqn:normalize_weights}
W_t^{(i)} = \frac{W^{(i)}_{t-1} w_t(\bbeta^{(i)}_{t-1},\bbeta^{(i)}_{t})}{\sum^N_{j=1} W^{(j)}_{t-1} w_t(\bbeta^{(j)}_{t-1},\bbeta^{(j)}_{t})}
\end{equation}
where
\begin{equation}
\label{eqn:weights}
w_t(\bbeta^{(i)}_{t-1},\bbeta^{(i)}_{t}) = \frac{\gamma_t(\bbeta^{(i)}_t)L_{t-1}(\bbeta^{(i)}_t,\bbeta^{(i)}_{t-1})}{\gamma_{t-1}(\bbeta^{(i)}_{t-1})K_t(\bbeta^{(i)}_{t-1},\bbeta^{(i)}_t)}
\end{equation}
We use an Markov chain Monte Carlo (MCMC) kernel $K_t$ with invariant distribution $\pi_t$ to sample each particle $\bbeta^{(i)}_{t}$, ie. $\bbeta^{(i)}_{t} \sim  K_t(\bbeta^{(i)}_{t-1},\cdot)$. The backwards kernel $L_{t-1}$ used to weight the particles is essentially arbitrary but we use the reversal of $K_t$
$$L_{t-1}(\bbeta_t,\bbeta_{t-1}) = \frac{\pi_t(\bbeta_{t-1})K_t(\bbeta_{t-1},\bbeta_t)}{\pi_t(\bbeta_t)}$$
so that the weights in (\ref{eqn:weights}) are given by
$$w_t(\bbeta^{(i)}_{t-1},\bbeta^{(i)}_{t}) = \frac{\gamma_t(\bbeta^{(i)}_{t-1})}{\gamma_{t-1}(\bbeta^{(i)}_{t-1})}$$
as suggested in \cite{crooks,ais,smcs}.

Since the weights do not depend on the value of $\bbeta_t$, it is better to resample the particles before moving them according to $K_t$, when resampling is necessary. We resample the particles when the effective sample size \cite{liu_chen} is below $3/4N$. Also note that even if we knew the normalizing constants $Z_{t-1}$ and $Z_t$, they would have no impact since the weights are normalized in (\ref{eqn:normalize_weights}). We can however obtain an estimate of $Z_t/Z_{t-1}$ at time $t$ via
\begin{align}
\frac{Z_t}{Z_{t-1}} &= \int \frac{\gamma_t(\bbeta_{t-1})}{\gamma_{t-1}(\bbeta_{t-1})}\gamma_{t-1}(\bbeta_{t-1})d\bbeta_{t-1} \\
                    &\approx \sum^N_{i=1}W_{t-1}^{(i)} w_t(\bbeta^{(i)}_{t-1},\bbeta^{(i)}_{t}) = \widehat{\frac{Z_t}{Z_{t-1}}}
\end{align}
which allows one to estimate the unnormalized marginal density $Z_{t}/Z_1$ for each value of $b_t$ via $\widehat{\frac{Z_t}{Z_{1}}} = \prod^t_{t=2} \widehat{\frac{Z_t}{Z_{t-1}}} $.

If we additionally give $\log b$ a uniform prior, then these unnormalized marginal densities can also be used to weight the collections of particles at each time, allowing us to approximate the distribution of $\bbeta$ with $b$ marginalized out.
\begin{align}
\label{eqn:all_particles}
\tilde{\pi}_t^N(d\bbeta) = \frac{1}{\sum^T_{t=1}\widehat{\frac{Z_t}{Z_{1}}}}\sum^T_{t=1} \widehat{\frac{Z_t}{Z_{1}}} \sum^{N}_{i=1}W_t^{(i)}\delta_{\bbeta^{(i)}_t}(d\bbeta)
\end{align}
Note that this is equivalent to systematic sampling of the values of $b$.

\subsection{Further Details}
For the experiments we used the data with $n=500$ and $p = 50$, we use $N = 8192$ particles and we consider two settings of the degrees of freedom  $a = 1$ and $a = 4$ with $T = 450$ and $T = 350$ respectively, with the lower $T$ for $a=4$ explained by the low marginal likelihood associated with small values of $b$ when $a$ is large. We set $b_t = 2(0.98)^{t-1}$ for $t \in \{1,\ldots,T\}$. Given the nature of the posterior, a simple $p$-dimensional random walk Metroplis-Hastings kernel does not seem to converge quickly. We found that using a cycle of kernels that each update one element of $\bbeta$ using a random walk proposal with variance $0.25$ mixed better than a $p$-dimensional random walk proposal. To ensure that our kernels mix fast enough to get good results we construct $K_t$ by cycling this cycle of kernels $5$ times. This is computationally demanding since each step of the algorithm consists of $5Np$ MCMC steps, each of which has a complexity in $O(n)$, since the calculation of the likelihood requires us to update $\bX\bbeta$ for a change in one element of $\bbeta$.

In fact, computation on a 2.67GHz Intel Xeon requires approximately 20 hours of computation time with these settings and $T = 450$. However, we implemented the algorithm to run on an NVIDIA 8800 GT graphics card using the Compute Unified Device Architecture (CUDA) parallel computing architecture, using the SMC sampler framework in \cite{cuda_mc} to take advantage of the ability to compute the likelihood in parallel for many particles. On this hardware, the time to run the algorithm is approximately 30 minutes for $T = 450$, giving around a $40$ fold speedup.

The initial particles for $t=1$ are obtained by running an MCMC algorithm targeting $\pi_1$ for a long time and picking suitably thinned samples to initialize the particles. The weights of each particle start off equal at $1/N$. While not ideal, trivial importance sampling estimates have a very low effective sample size for reasonable computational power and, in any case, the SMC sampler methodology requires us to have a rapidly mixing MCMC kernel for our resulting estimates to be good.

The particle with the highest posterior density at each time is used as an initial value for the MAP algorithm described in Section \ref{section:compute_map}. The density of the estimated MAP from this initial value is compared with the MAP obtained by using the previous MAP as an initial value and the estimate with the higher posterior density is chosen as the estimated MAP. This has the effect of both removing variables that the original algorithm has not removed and keeping variables in that the original algorithm has not, as it is not possible for the MAP algorithm to explore the multimodal posterior density fully.

While the independence of the weights of the particles makes it possible to adapt the selection of $b_t$ allow aggressive changes in $b$ while maintaining a given effective sample size \cite{adapt1,adapt2}, we found that this scheme had a poor effect on the quality of our samples. This is primarily because we discovered that our cycle length for $K_t$ was not large enough to get the particles to approximate $\pi_t$ after a large change in $b$, despite a small change in effective sample size. Increasing the cycle length has an essentially linear effect on the computational resources required, so this appears to be a poor option for reducing the computational complexity of the method. In addition, having $b_t$ on a logarithmic schedule enables us to use the simple approximation to the posterior of $\bbeta | \bX, \by, a$ given by (\ref{eqn:all_particles}).

For the larger experiment with $n=1859, p = 184$ and $a = 4$, we use the same parameters except that we used the schedule $b_t = (0.98)^{t-1}$ with $T = 250$. This was sufficient to capture the a wide range of model complexities. Nevertheless, the running time for this experiment on our GPU was 7.5 hours. This demonstrates the key practical benefit to using GPUs, as a similar run on a single CPU would take upwards of 5 days.

\subsection{Tuning the SMC sampler}
Following exploratory analysis we found that a sequence of scales as $b_t = b_1(0.98)^{t-1}$ worked well, and that setting $b_1$ equal to one or two produced a diffuse enough prior when sample size was in the 100s or 1000s, as is the situation in GWAS.

In practice, one will have to experiment with the parameters of the SMC sampler in order to ensure satisfactory results. In particular, given an MCMC kernel $K_t$, one must decide how many cycles should be performed per time step to mix adequately in relation to how quickly one wishes to decrease $b$. In an ideal scenario, $K_t$ produces essentially independent samples from $\pi_t$ regardless of the current set of particles. This would allow one to set $\{b_t\}^T_{t=1}$ such that computational power is only spent in areas of $b$ that are of interest. In practice, however, $K_t$ will often only mix well locally and so small steps in $b$ are desirable to ensure that the Monte Carlo error in the scheme does not dominate the results. It is difficult to characaterize the relationship between the number of cycles of $K$ to perform and how aggressively one can decrease $b$ in general. However, in this context, one might not want to decrease $b$ too quickly anyway, so that a smooth picture of the relationship between successive posteriors can be visualized.

The number of particles, $N$, is a crucial parameter that should be set at such a level that some fraction of $N$ can provide a suitable approximation of each intermediate target density. In practice, the number of effective samples is lower than $N$ and the effective sample size is an attempt to ensure that this number is controlled. It is standard practice to choose to resample particles when the effective sample size is in $(2/3, 1)N$. We choose to resample the particles when the effective sample size fell below $3/4N$.

A recommended way to test whether the sampler is performing acceptably is to run an MCMC scheme with fixed $b = b_t$ for some intermediate value of $t$ and compare the samples one obtains after convergence with those produced by the SMC sampler. This is possible in this context because the posterior is not particularly troublesome to sample from for fixed $b$ with a sensible kernel.

Finally it is worth noting that if the aim was to infer a single distribution given a single setting of $c$, one could achieve better results via one long MCMC run. The benefit of the SMC sampler is that it provides a unified method to obtain all of the exploratory statistics and estimates of the marginal likelihood in a single run.

\section{Sparsity-Path-Analysis}

Lasso plots of MAP estimates as functions of the penalty parameter $\lambda$ are a highly informative visual tool for statisticians to gain an understanding of the relative predictive importance of covariates over a range of model complexities \cite{hastie2009elements}. However, MAP statistics are single point estimates that fail to convey the full extent of information contained within the joint posterior distribution. One key aim of this report is to develop full Bayesian exploratory tools that utilize the joint posterior information.

In this section we describe graphical displays of summary statistics contained in SPA results by first considering the $n=500$ $p=50$ data set described above. In particular the indices of the variable with non-zero $\bbeta$'s are ${\cal{I}}=\{10, 14, 24, 31, 37\}$ with corresponding $\bbeta_{\cal{I}} = \{-0.2538, 0.4578, -0.1873, -0.1498, 0.0996\}$. In the $Gt(a,c)$ prior we found the $Gt(a=1,c)$, or $Gt(a=4,c)$, to be good default settings for the degrees of freedom both leading to heavy tails. We ran the SMC-GPU sampler described above and then generated a number of plots using the set of particles and their weights.

\subsection{Path plots}

\begin{figure}[htp]
\center
\subfigure[MAPs]{
	\includegraphics[scale=0.3]{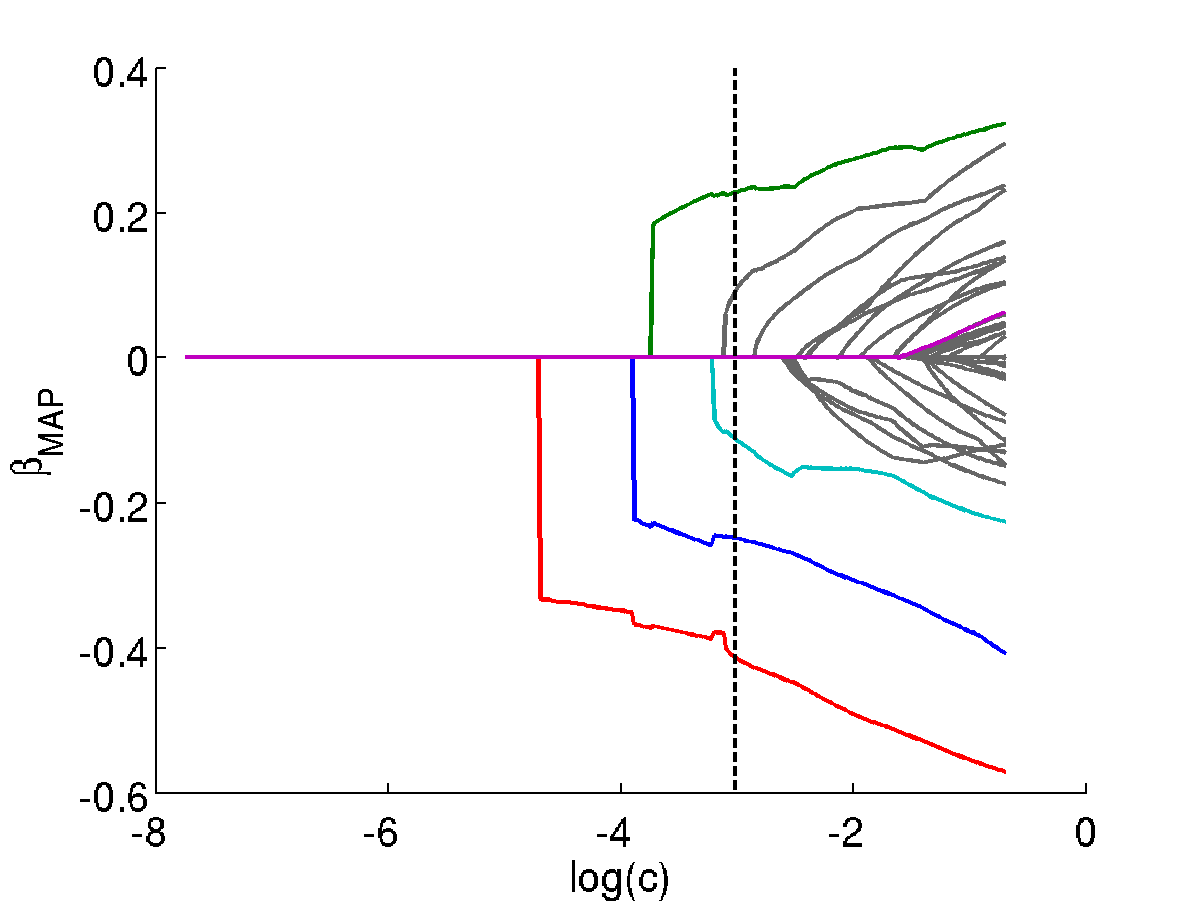}
}
\subfigure[absolute medians]{
	\includegraphics[scale=0.3]{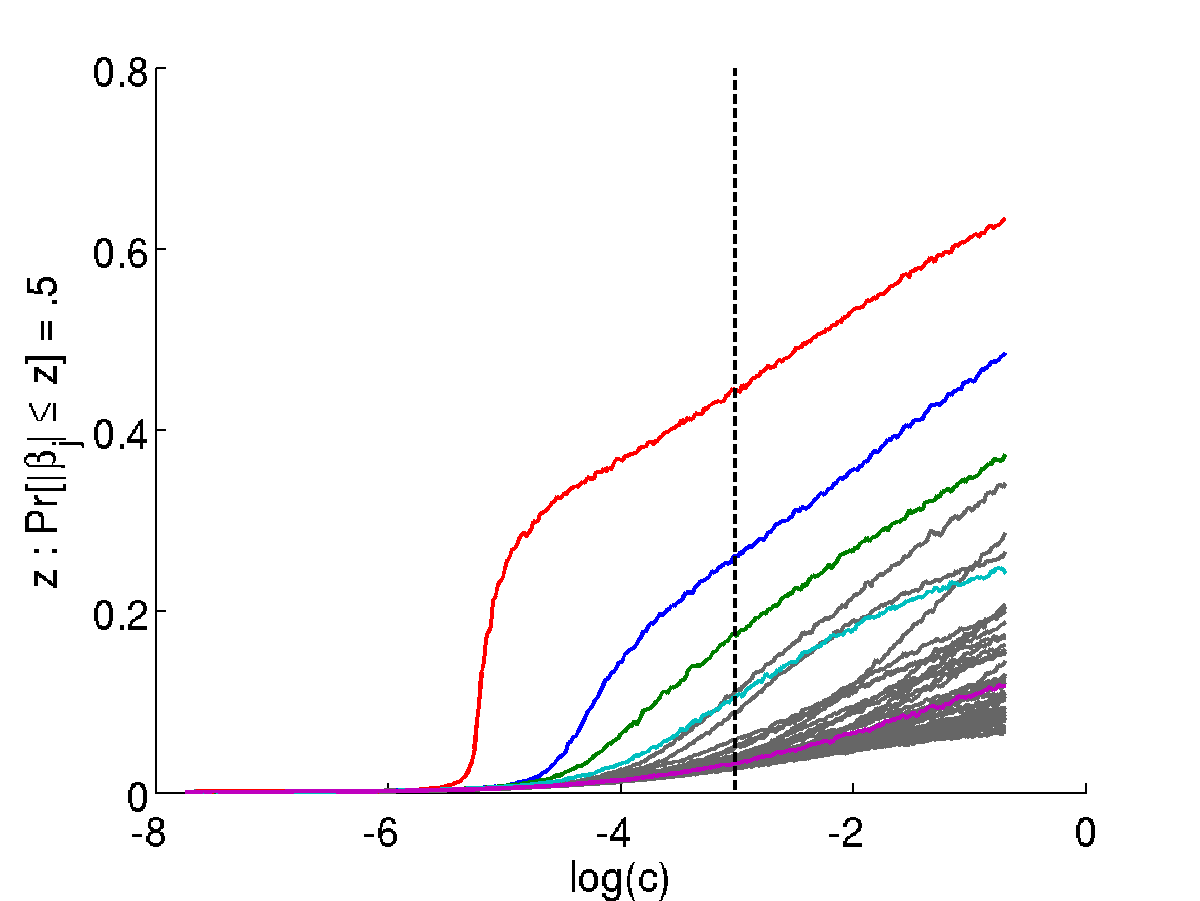}
}
\subfigure[posterior density]{
	\includegraphics[scale=0.3]{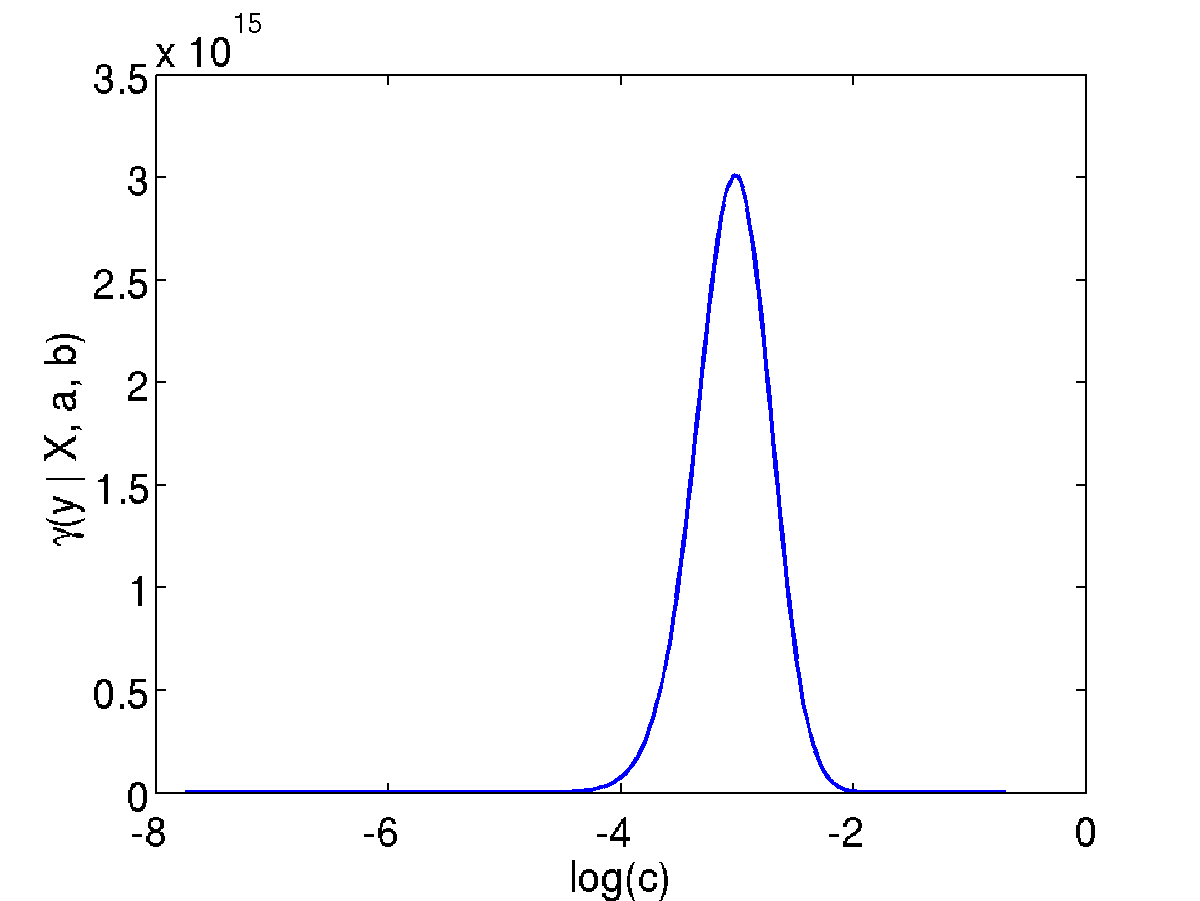}
}
\subfigure[concentrations]{
	\includegraphics[scale=0.3]{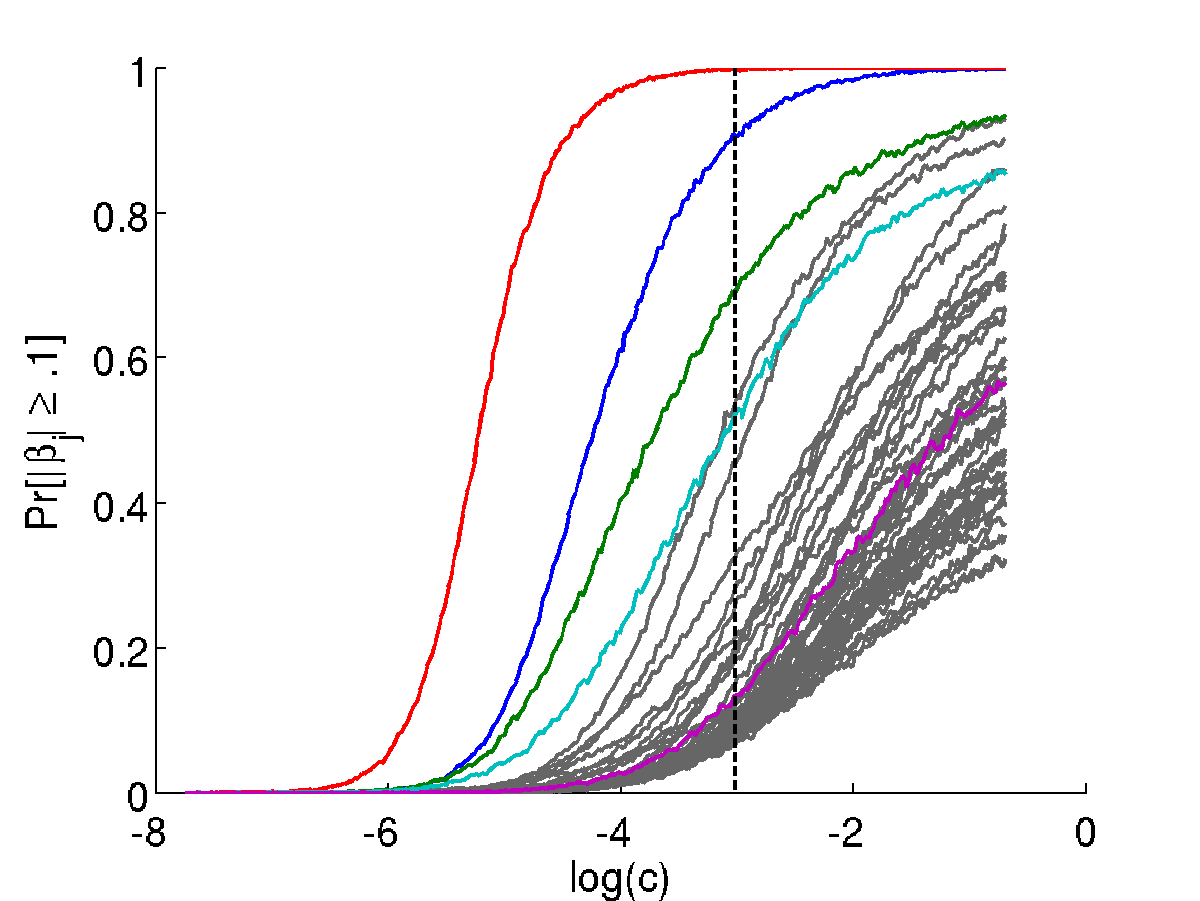}
}
\caption{SPA plots with $a=4$ degrees of freedom; using 50 markers around 18q. The 5 non-zero coefficients are indicated by non-grey lines. The vertical dashed line indicates the mode of the marginal likelihood, $\pi(c |\bX, \by, a)$, for $c$ as shown in plot (c). }
\label{fig:spa_a4}
\end{figure}

\begin{figure}[pth]
\center
\includegraphics[scale=0.6]{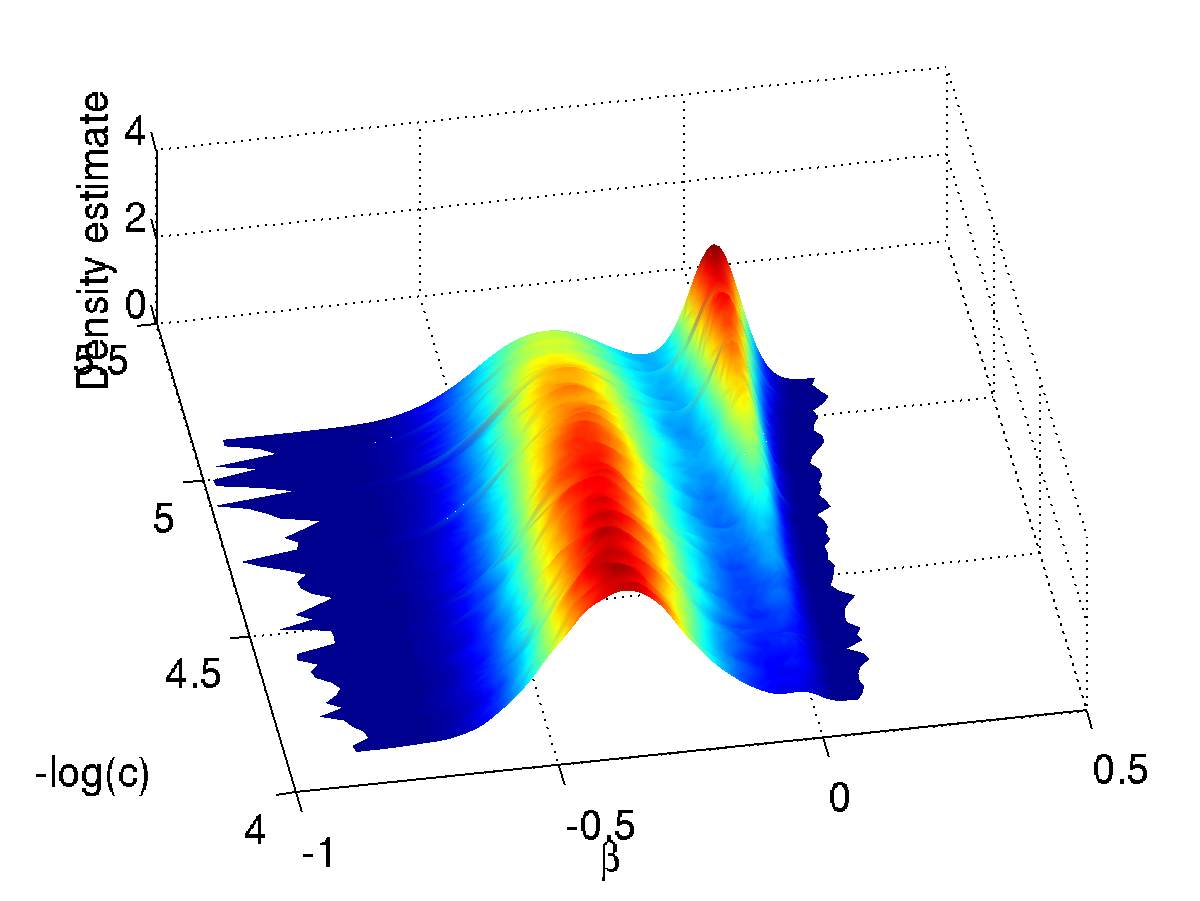}
\caption{Plot of marginal posterior of $\beta_{24}$ as a function of $c$, c.f. the red curve in Fig.~\ref{fig:spa_a4}(a). We can see the bimodality and hence the jumping of the MAP mode in Fig.~\ref{fig:spa_a4}(a) as $\log c$ changes from (-5, -4).}
\label{fig:mode_hop}
\end{figure}

In Fig.~\ref{fig:spa_a4} we plot four graphs of summary statistics obtained from the SMC sampler across a range of scales $\log(c) \in (-8, 0)$.
\begin{itemize}
\item[(a)] MAP plots: In Fig.~\ref{fig:spa_a4}(a) we plot the MAP paths of all 50 coefficients moving from the most sparse models with $c=e^{-8}$ where all $\hat{\bbeta}_{MAP}=0$ through to $c=1$ when all $\hat{\bbeta}_{MAP} \ne 0$. The true non-zero coefficients are plotted in colors other than gray. One can observe that the MAP paths are non-smooth in $\log(c)$ such that the modes suddenly jump away from zero.This is a property of the non-log-concavity of the posterior distributions; such that the marginal posterior densities can be bimodal with one mode at $\beta_j=0$ and another away from zero. As $c$ increases the MAP mode at zero decreases and at some point the global mode switches to the mode at $\beta_j \ne 0$. This can be seen clearly in Fig.~\ref{fig:mode_hop} were we plot the marginal posterior distribution of $\beta_{24}$ over the range $\log c \in (-5, 4)$; c.f. $\beta_{24}$ is shown as as the red curve in Fig.~\ref{fig:spa_a4}(a). We can see in Fig.~\ref{fig:mode_hop} the non-concavity of the posterior density and how the global mode jumps from 0 to $\approx -.04$ as $\log c$ is around $-4.75$. As mentioned above the MAP is found by iterating the EM algorithm beginning from the particle with largest posterior density.
    \item[(b)] Median plots: In Fig.~\ref{fig:spa_a4}(b) we plot the absolute value of the median of the marginal distribution of $\beta_j$'s. This is a plot of $z_j(\log[c])$ vrs $\log c$, for $j=1, \ldots, p$, where,
$$
z_j(c) = | \hat{F}^{-1}_{\beta_j}(0.5 | c) |
$$
where $F^{-1}_{\beta_j}(\cdot)$ is the inverse cumulative posterior distribution of $\beta_j$,
$$
F_{\beta_j}(x | c) = \int_{-\infty}^x \pi(\beta_j | \bX, \by, a, c) d \beta_j
$$
and where $\pi(\beta_j | \bX, \by, a, c)$ is the marginal posterior distribution of $\beta_j$ given $c$;   $\pi(\beta_j | \bX, \by, a, c)= \int_{-j} \pi(\beta_j, \bbeta_{-j} | \bX, \by, a, c) d \bbeta_{-j}$ where index $\{-j\}$ indicates all indices other than the $j$th. The plot of absolute medians gives an indication of how quickly the posterior distributions are moving outward away from the origin as the precision of the prior decreases. By plotting on the absolute scale we are better able to compare the coefficients with one another and we also see that, unlike the MAPs, the medians are smooth functions of $c$.
\item[(c)] Posterior for scale $c$: In Fig.~\ref{fig:spa_a4}(c) we show the marginal posterior distribution $p(c | \bX, \by, a)$,
 $$
 p(c | \bX, \by, a) = \int_{\bbeta} p(c, \bbeta | \bX, \by, a) d \bbeta .
 $$
The posterior on $c$ graphs the relative evidence for particular prior scale. The mode of Fig.~\ref{fig:spa_a4}(c),
 $$
 \tilde{c} = \arg \max_c \pi(c | \bX, \by, a)
 $$
 is indicated on plots (a),(b),(d) by a vertical dashed line.

    \item[(d)] Concentration plots: In Fig.~\ref{fig:spa_a4}(d) we plot the concentration of the marginal posteriors of $\beta_j$'s around the origin. In particular, for user specified tolerance $\Delta$, this is a plot of $V(c)$ vrs $c$ where,
        $$
        V(c) = 1 - \int_{-\Delta}^{\Delta} \pi(\beta_j | \bX, \by, a, c) d \beta_j = 1 - Pr(\beta_j \in (-\Delta, \Delta) | \bX, \by, a, c)
        $$
        This is a direct measure of the predictive relevance of the corresponding covariate (genotype). In Fig.~\ref{fig:spa_a4}(d)  we have set $\Delta = 0.1$ although this is a user set parameter that can be specified according to the appropriate level below which a variable is not deemed important.
\end{itemize}

\begin{figure}[htp]
\center
\subfigure[MAPs]{
	\includegraphics[scale=0.3]{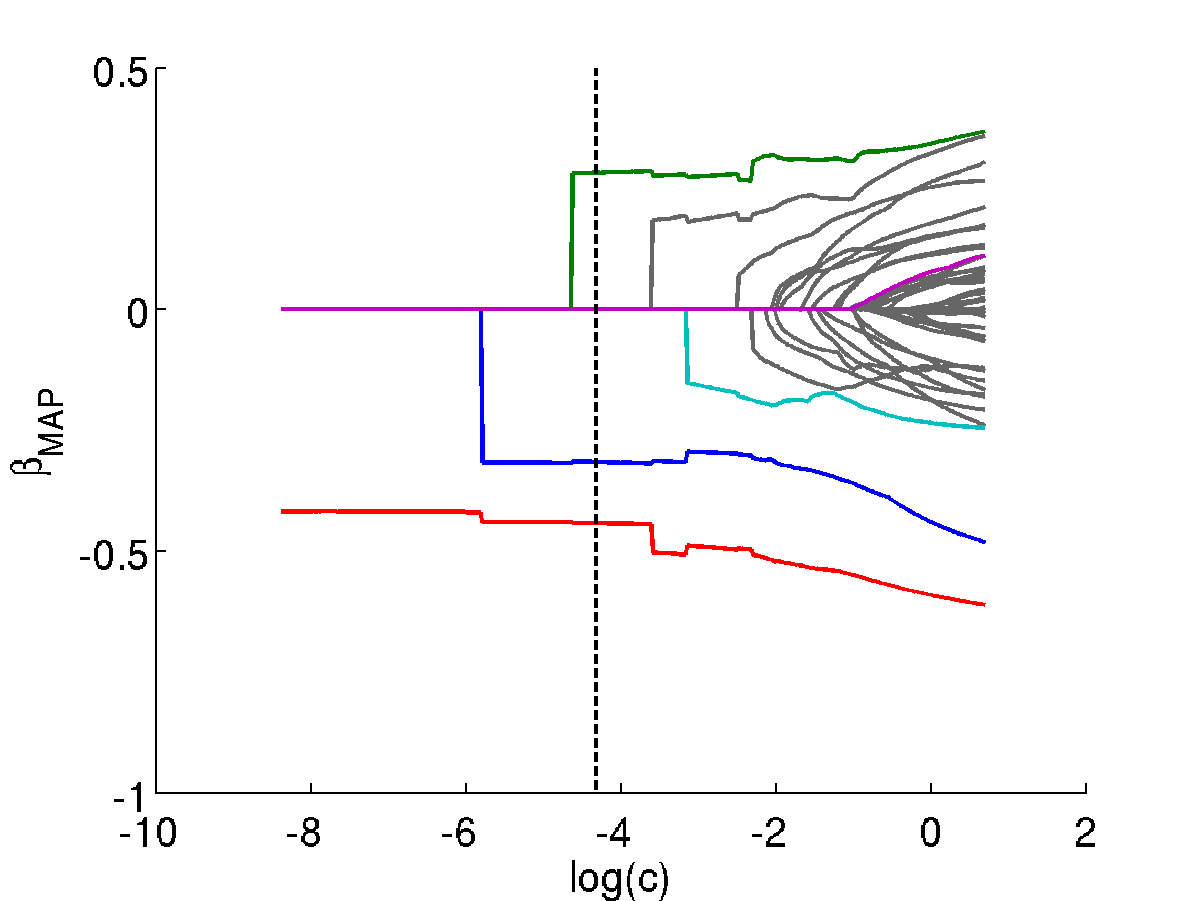}
}
\subfigure[absolute medians]{
	\includegraphics[scale=0.3]{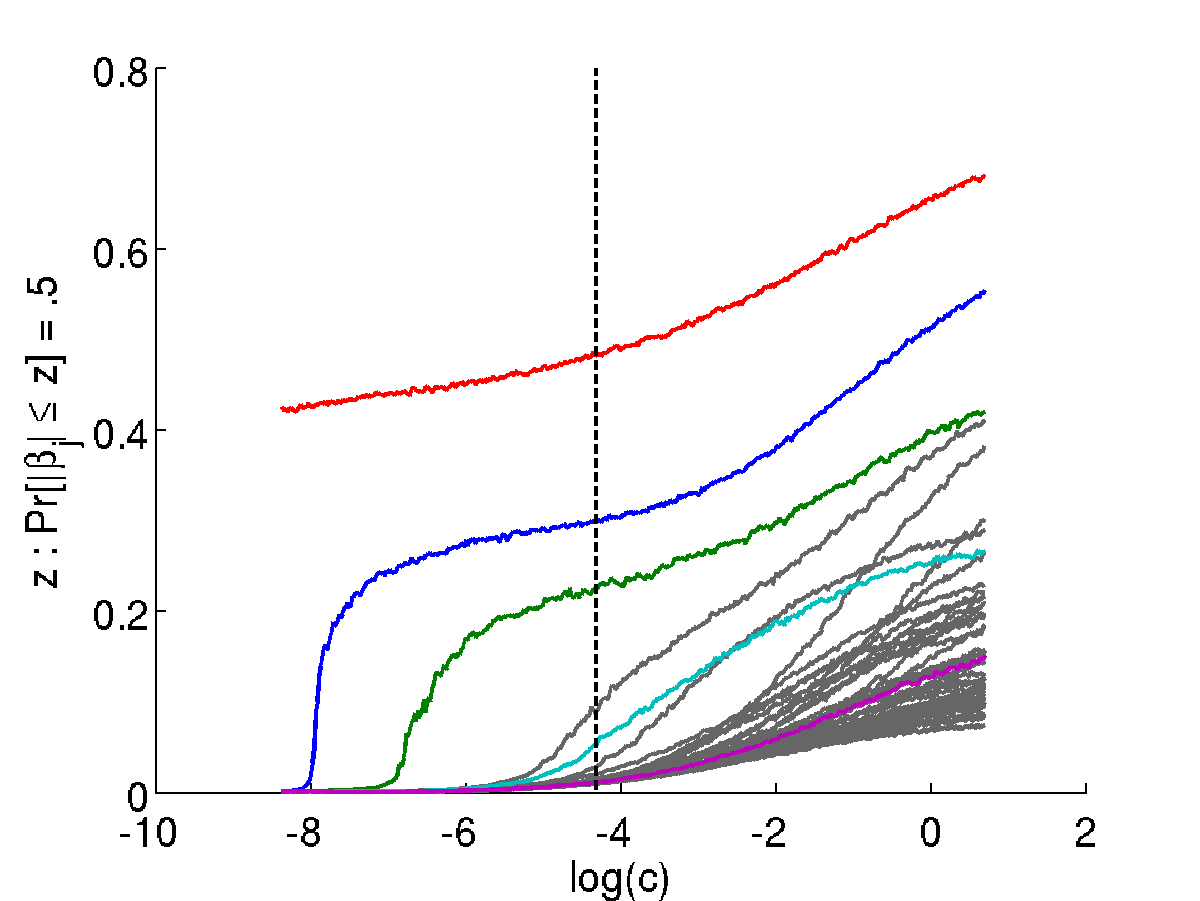}
}
\subfigure[posterior density]{
	\includegraphics[scale=0.3]{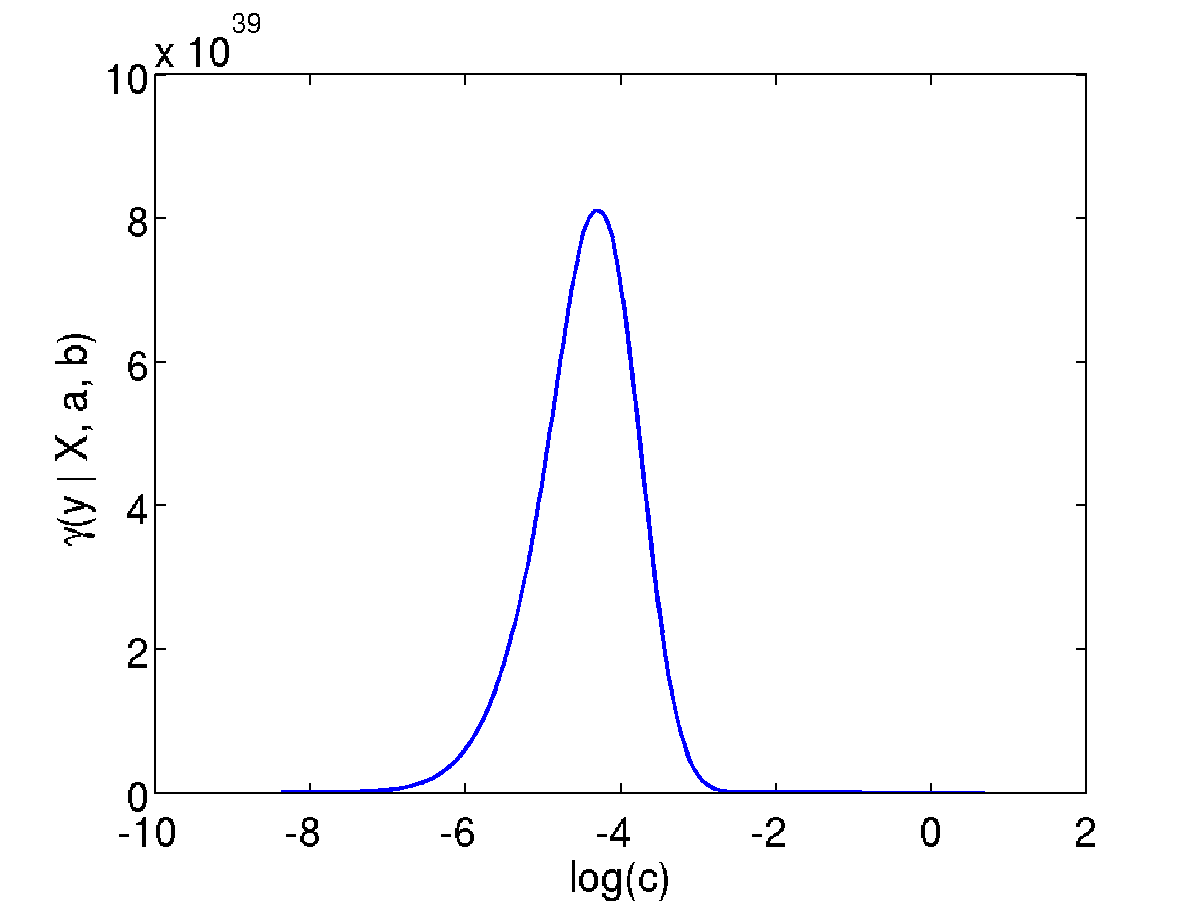}
}
\subfigure[concentrations]{
	\includegraphics[scale=0.3]{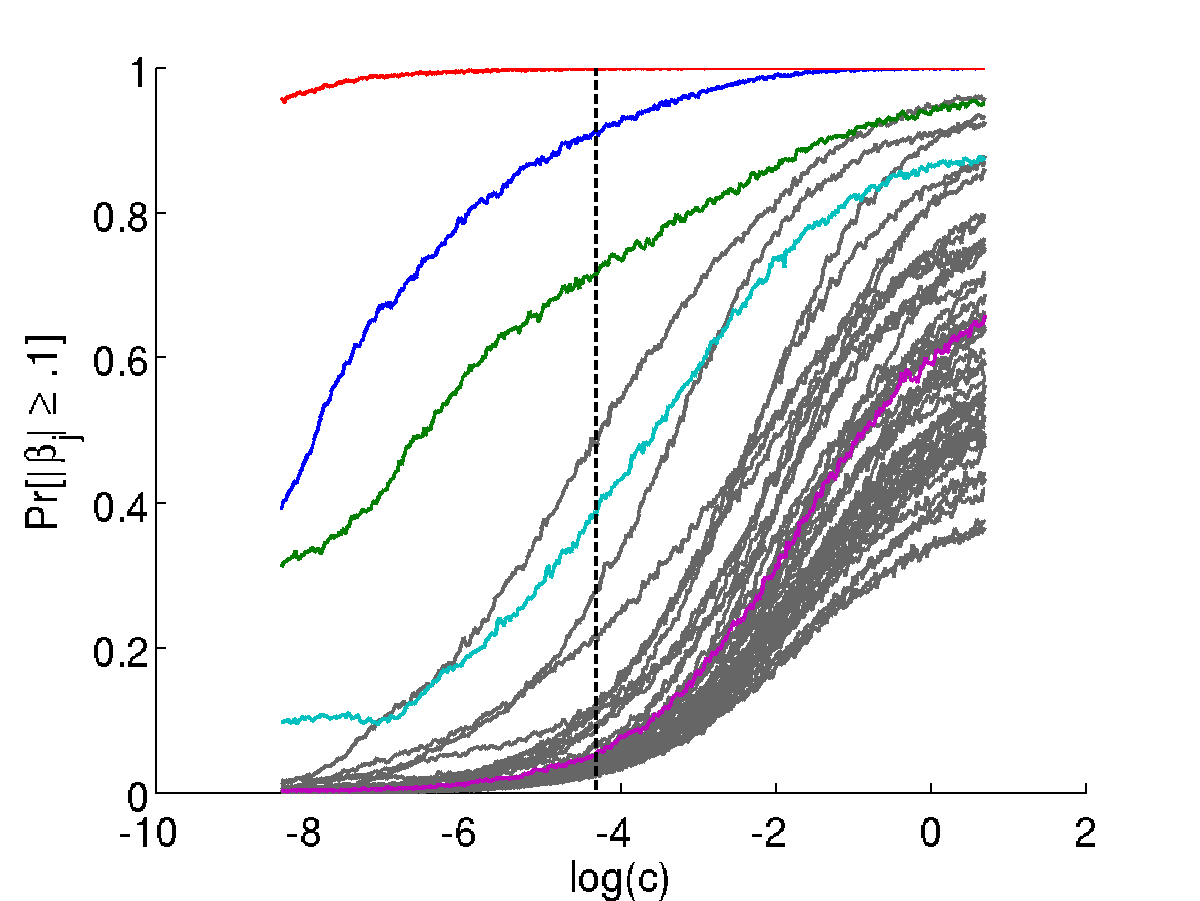}
}
\caption{SPA plots with $a=1$ degrees of freedom; using 50 markers around 18q. The 5 non-zero coefficients are indicated by non-grey lines. The vertical dashed line indicates the mode of the marginal likelihood, $\pi(c |\bX, \by, a)$, for $c$ as shown in plot (c). This Figure should be compared with Fig.~\ref{fig:spa_a4} using $a=4$ degrees of freedom. }
\label{fig:spa_a1}
\end{figure}

Taken together we see the SPA plots highlight the influential genotypes and the relative evidence for their predictive strength. We can also gain an understanding of the support in the data for differing values of $\log(c)$.  Having generated the SPA plots it is interesting to compare the results of changing the degrees of freedom from the $Gt(a=4,c)$ to $Gt(a=1, c)$. In Fig.~\ref{fig:spa_a1} we show plots corresponding to Fig.~\ref{fig:spa_a4} but using $a=1$ degrees of freedom. We can see that, as expected, $Gt(a=1,c)$ produces sparser solutions with only three non-zero MAP estimates at the mode of the posterior for $\tilde{c}$. Moreover, comparing the concentration plots Fig.~\ref{fig:spa_a1}(d) with Fig.~\ref{fig:spa_a4}(d) at the marginal MAP estimate of $c$ we see that for $a=1$ we see much greater dispersion in the concentration plot.

\subsection{Individual coefficient plots}

\begin{figure}[htp]
\center
\subfigure[$a=4$, $j=14$]{
	\includegraphics[scale=0.3]{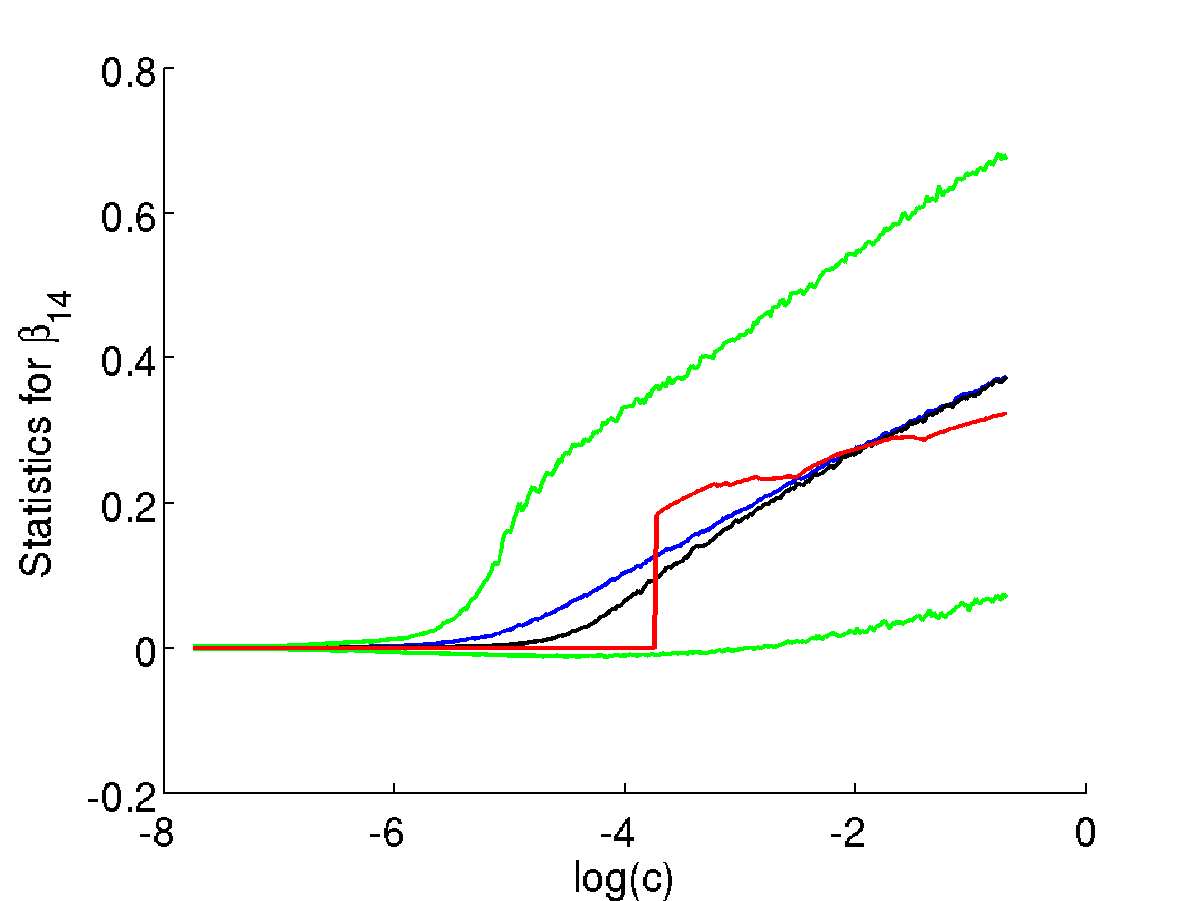}
}
\subfigure[$a=4$, $j=24$]{
	\includegraphics[scale=0.3]{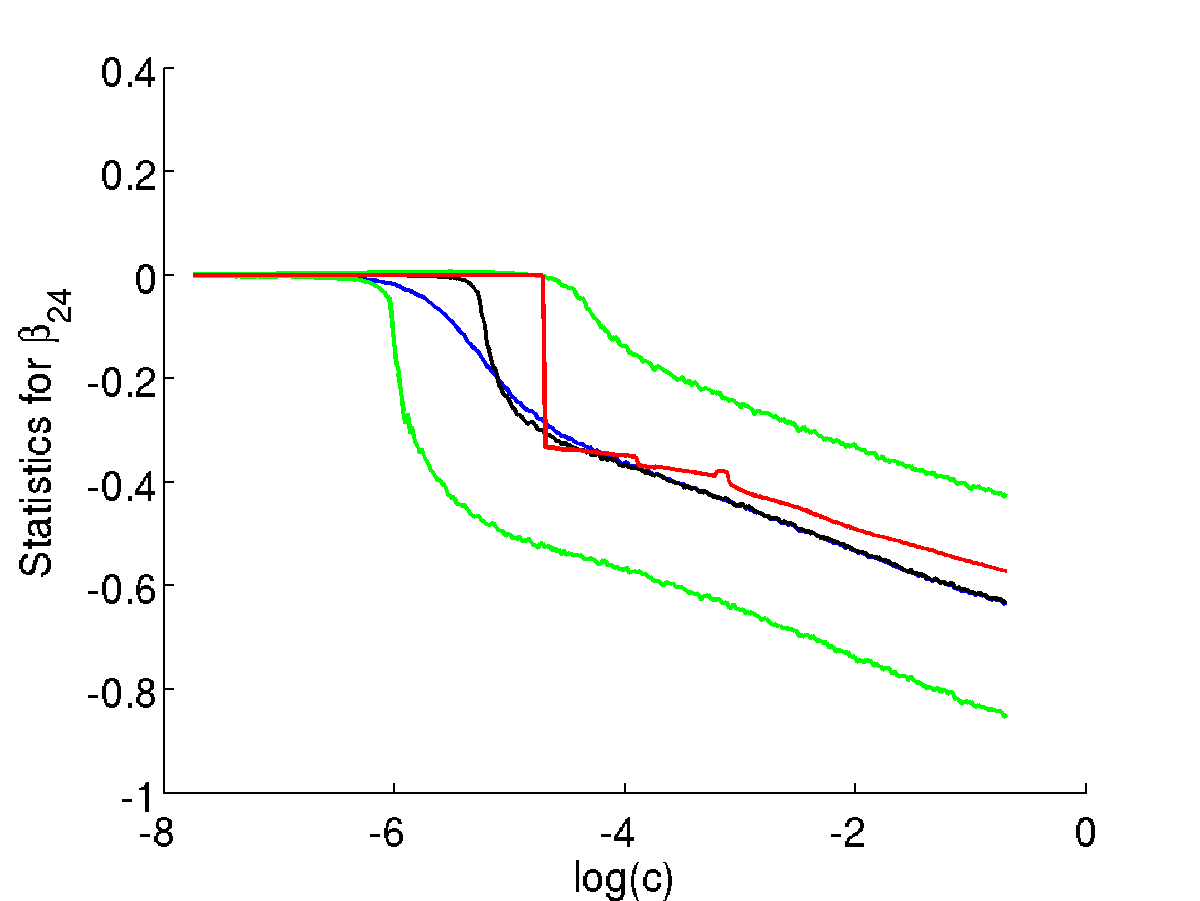}
}
\subfigure[$a=4$, $j=31$]{
	\includegraphics[scale=0.3]{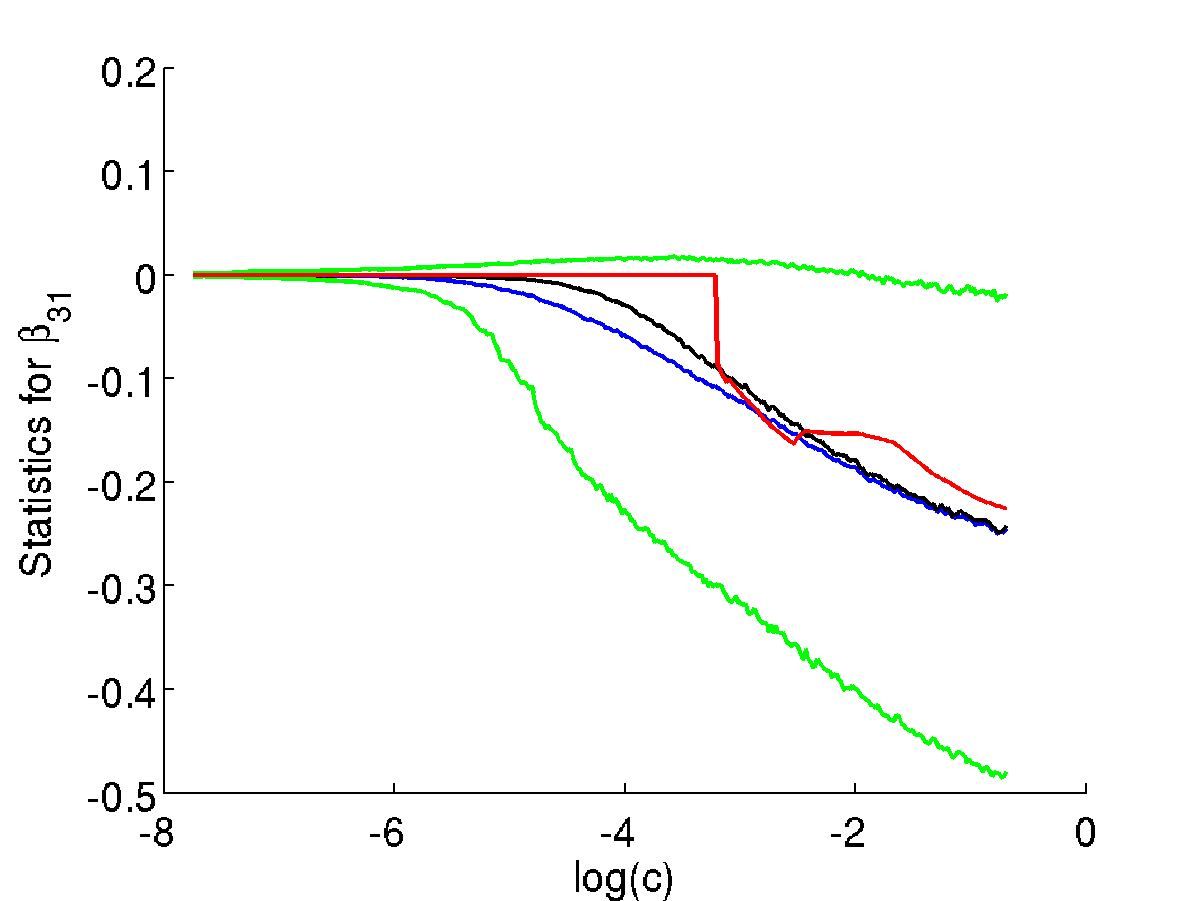}
}
\subfigure[$a=4$, $j=1$]{
	\includegraphics[scale=0.3]{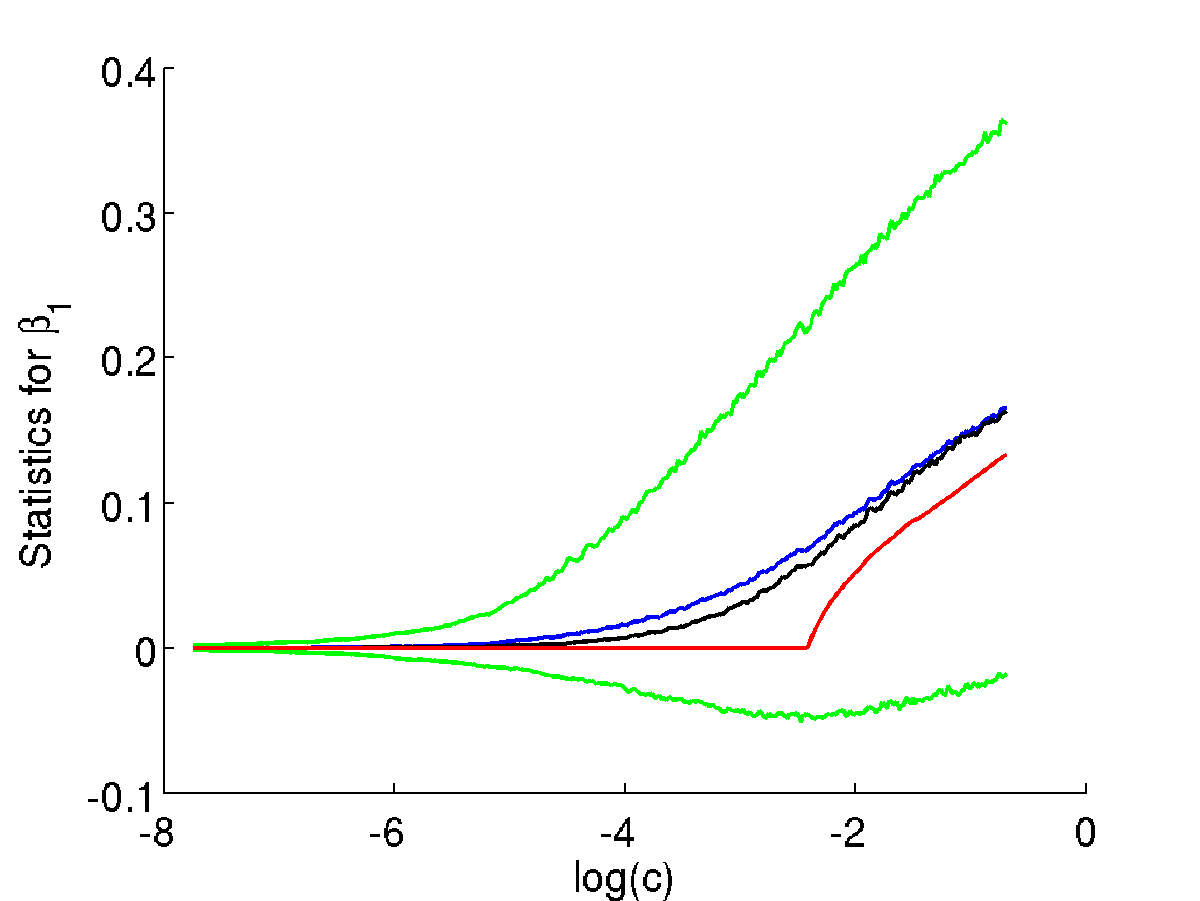}
}
\caption{Stats for individual coefficients from Fig. \ref{fig:spa_a4} with $a=4$. For each coefficient we plot $90\%$ credible intervals (green), median (black), mean (blue) and MAP (red). }
\label{fig:a4_index_1}
\end{figure}

\begin{figure}[htp]
\center
\subfigure[$a=4$, $j=14$]{
	\includegraphics[scale=0.3]{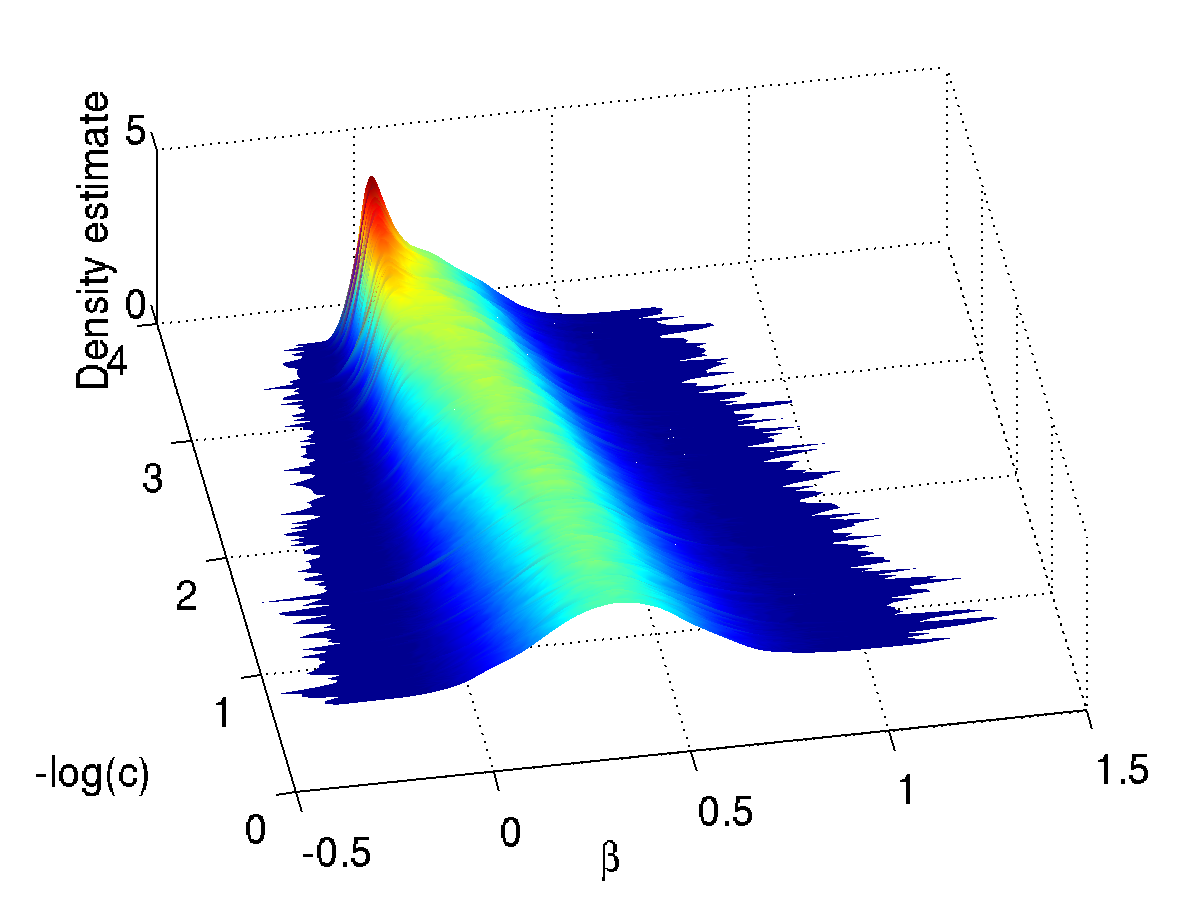}
}
\subfigure[$a=4$, $j=24$]{
	\includegraphics[scale=0.3]{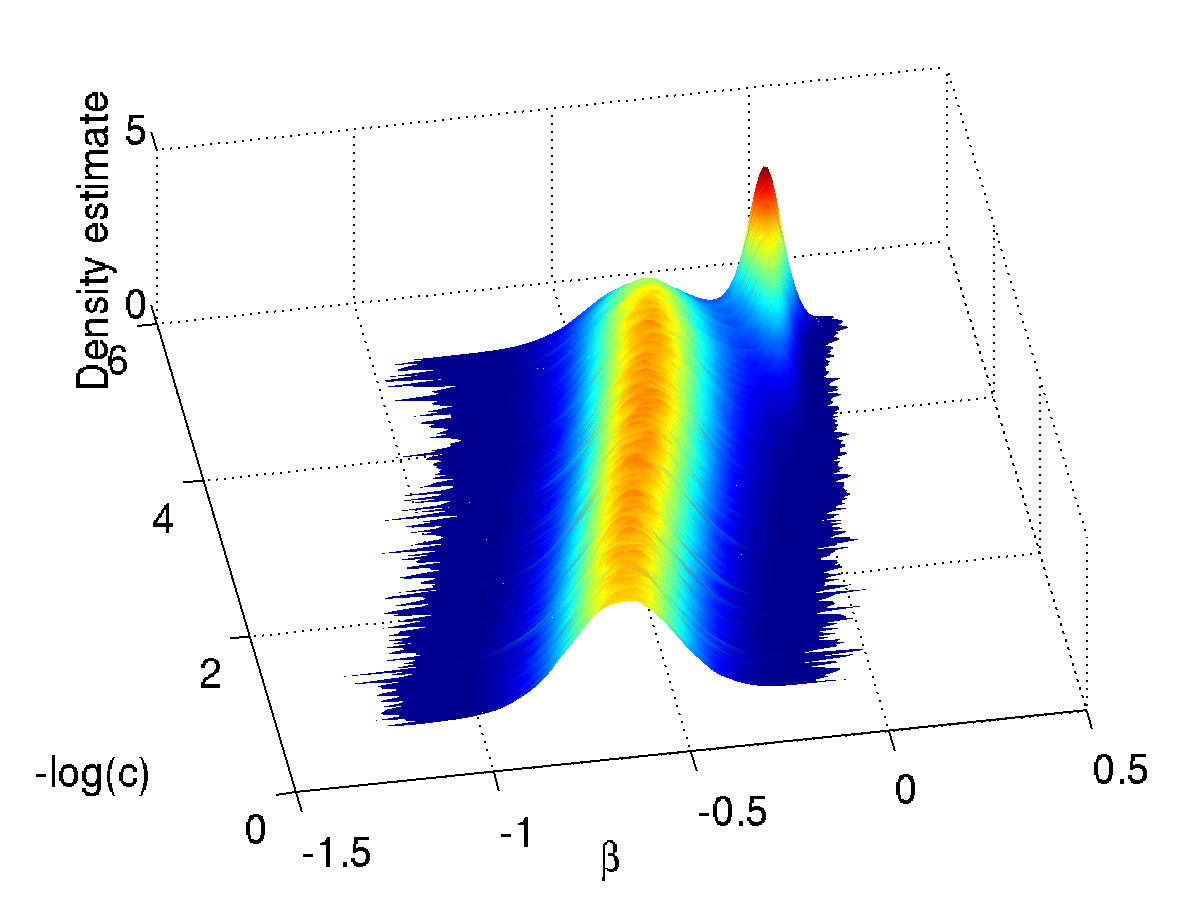}
}
\subfigure[$a=4$, $j=31$]{
	\includegraphics[scale=0.3]{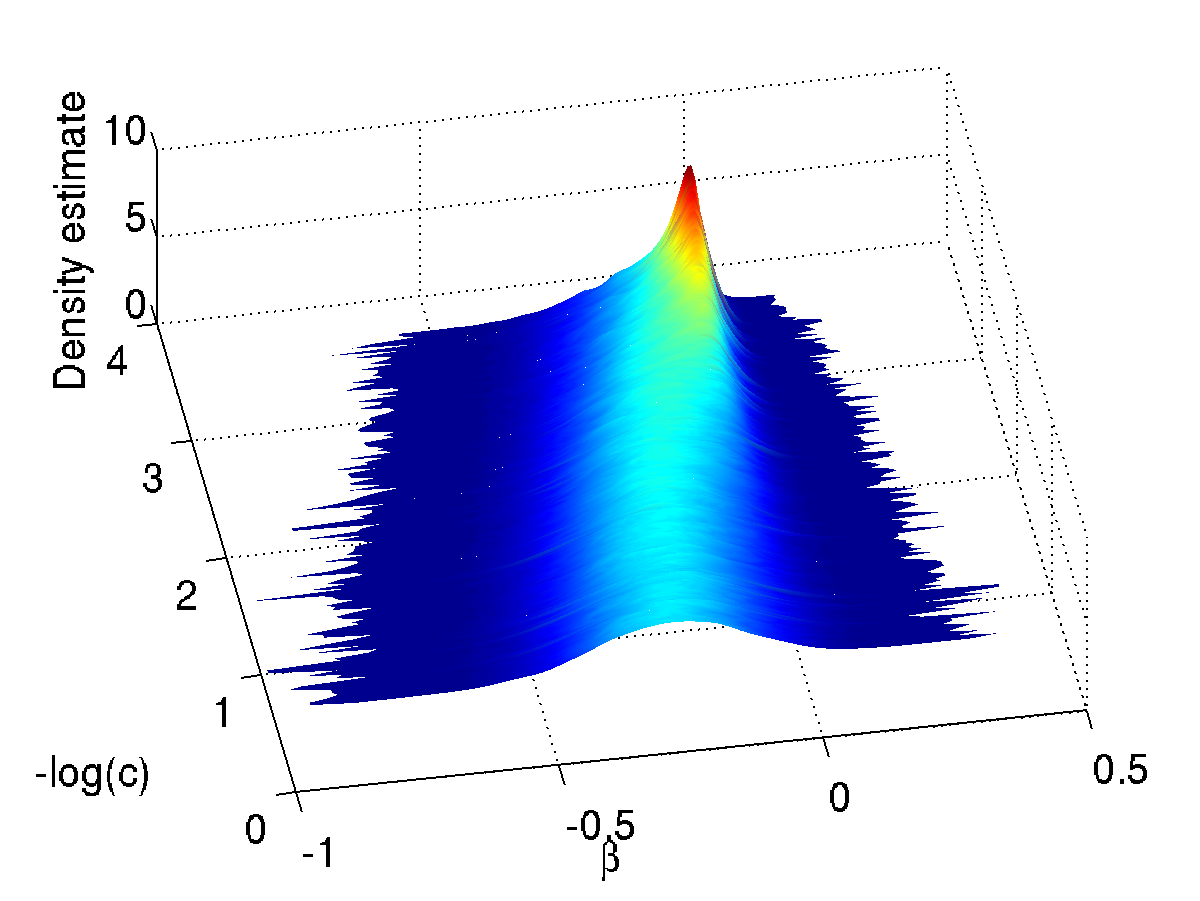}
}
\subfigure[$a=4$, $j=1$]{
	\includegraphics[scale=0.3]{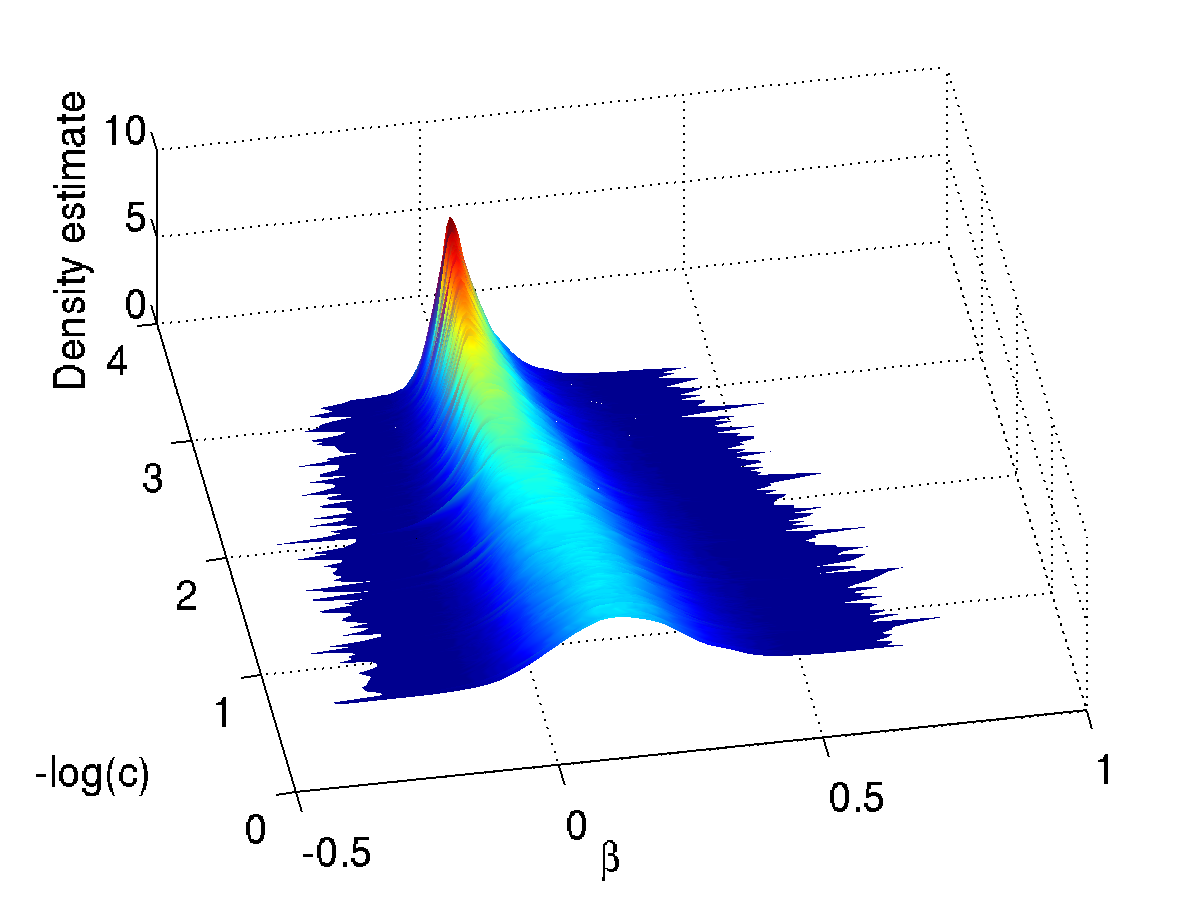}
}
\caption{Posterior density plots corresponding to Fig.~\ref{fig:a4_index_1}, $a=4$}
\label{fig:a4_index_2}
\end{figure}

Examination of the plots in Fig.\ref{fig:spa_a4}  may highlight to the statistician some interesting predictors to explore in greater detail. Individual coeffcient plots of summary statistics can be produced to provide greater information on the posterior distributions. In Fig. \ref{fig:a4_index_1} we show summary plots for four representative coefficients with their $90\%$ credible intervals (green), median (black), mean (blue) and MAP (red). These are obtained from the set of weighted SMC particles. We can see that as expected the mean and median are smooth functions of the prior scale, which the MAP can exhibit the characteristic jumping for bimodal densities. In  Fig. \ref{fig:a4_index_2} we show corresponding density estimates. These coefficients were chosen to represent markers with strong association Fig.~\ref{fig:a4_index_1}(a), weaker association Fig.~\ref{fig:a4_index_1}(b),(c) and no association Fig.~\ref{fig:a4_index_1}(d). We can see in the plots for Fig.~\ref{fig:a4_index_1}(d) and Fig.~\ref{fig:a4_index_2}(d) that for a null marker with no association signal the MAP appears to be much smother in $\log(c)$.

\begin{figure}[htp]
\center
\subfigure[$a=1$, $j=14$]{
	\includegraphics[scale=0.3]{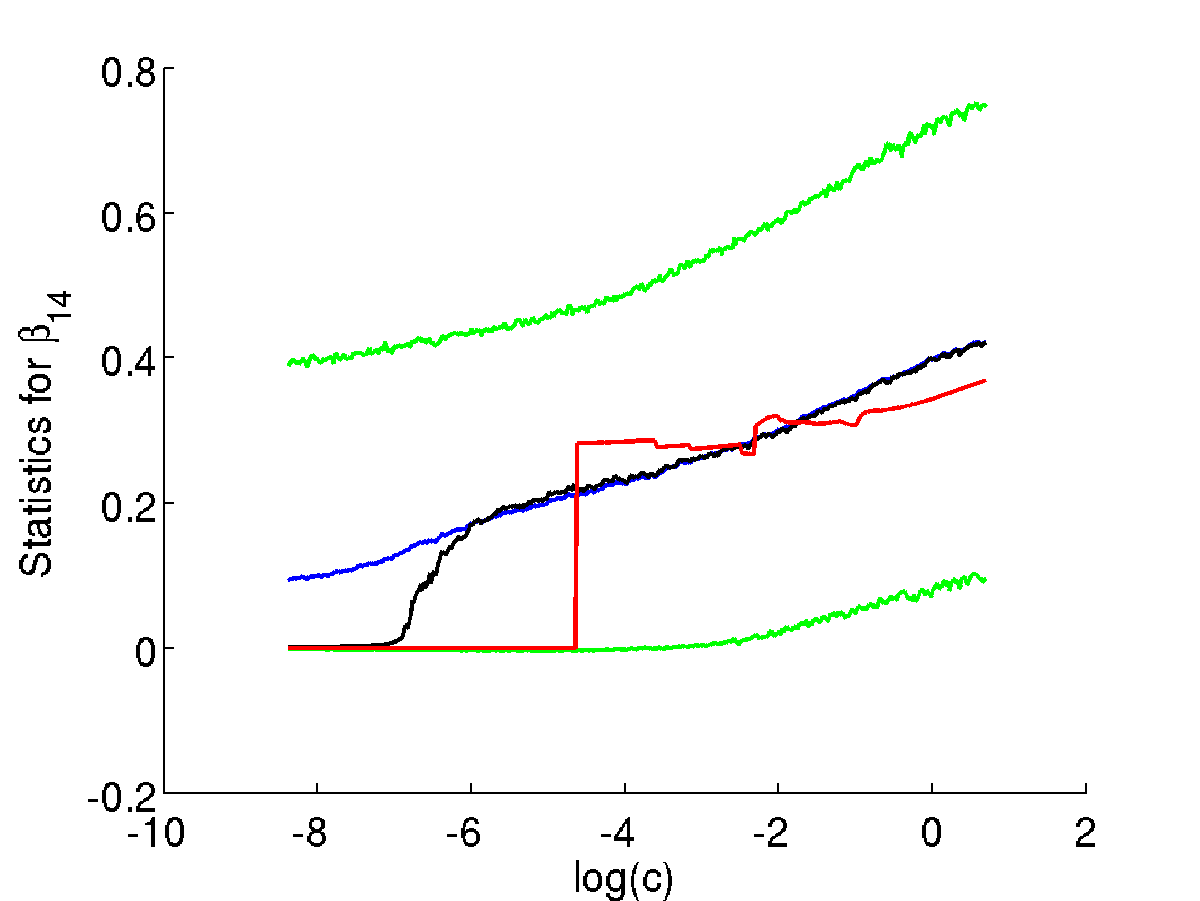}
}
\subfigure[$a=1$, $j=24$]{
	\includegraphics[scale=0.3]{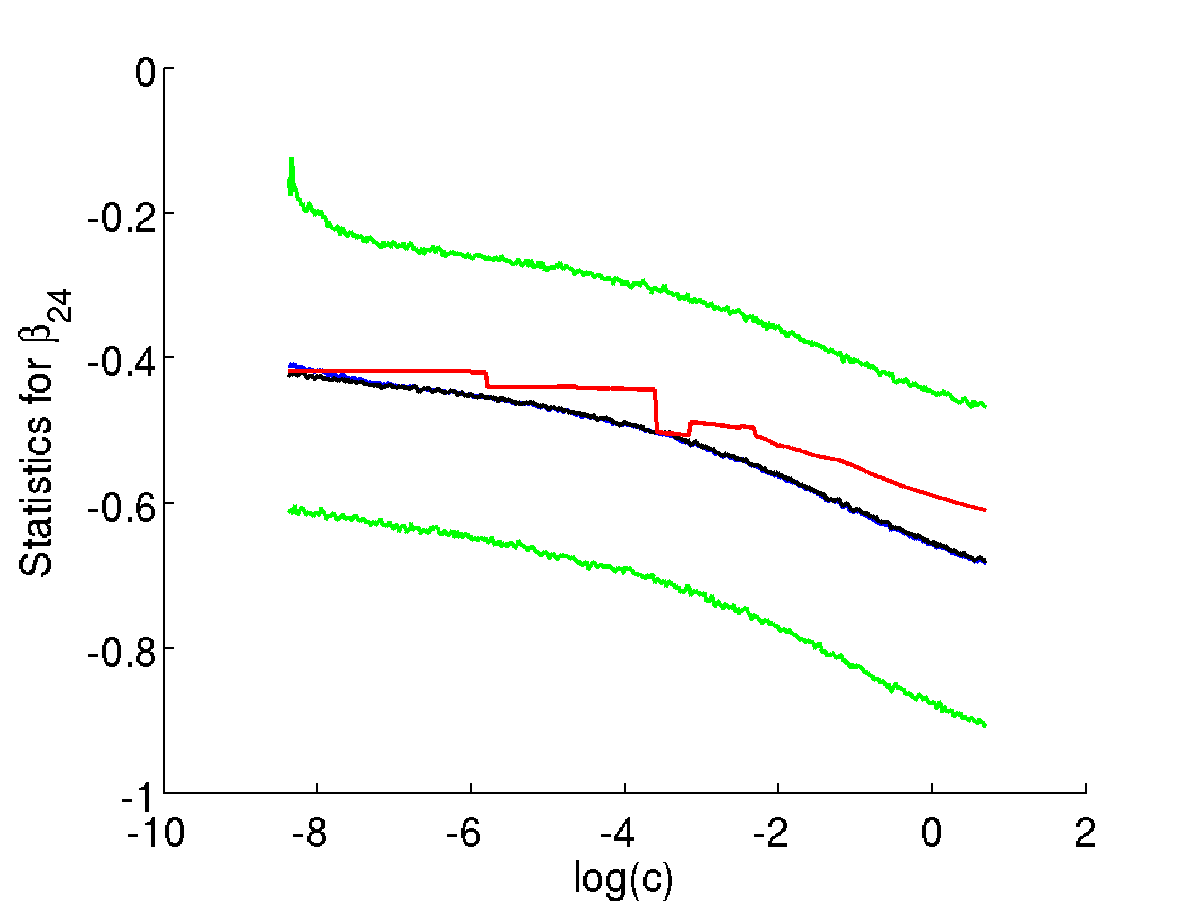}
}
\subfigure[$a=1$, $j=31$]{
	\includegraphics[scale=0.3]{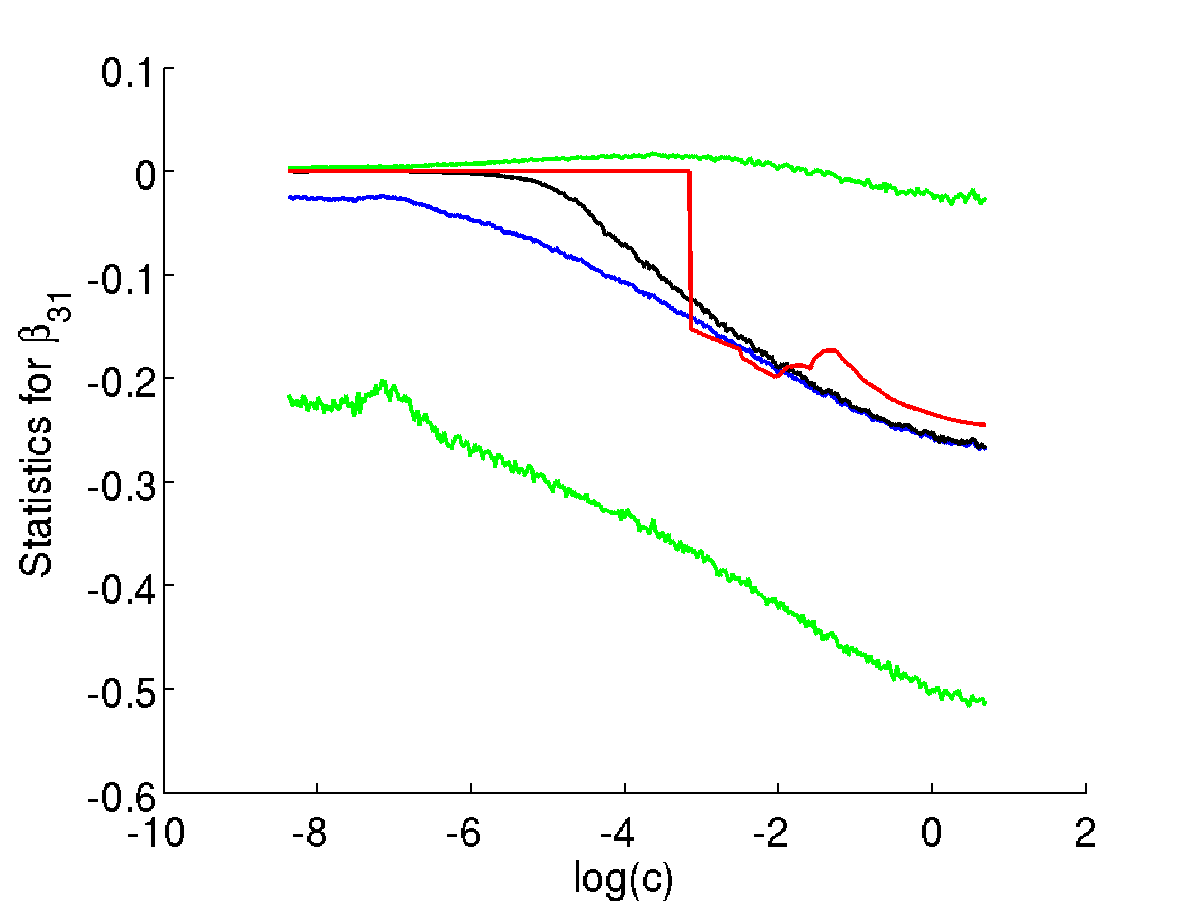}
}
\subfigure[$a=1$, $j=1$]{
	\includegraphics[scale=0.3]{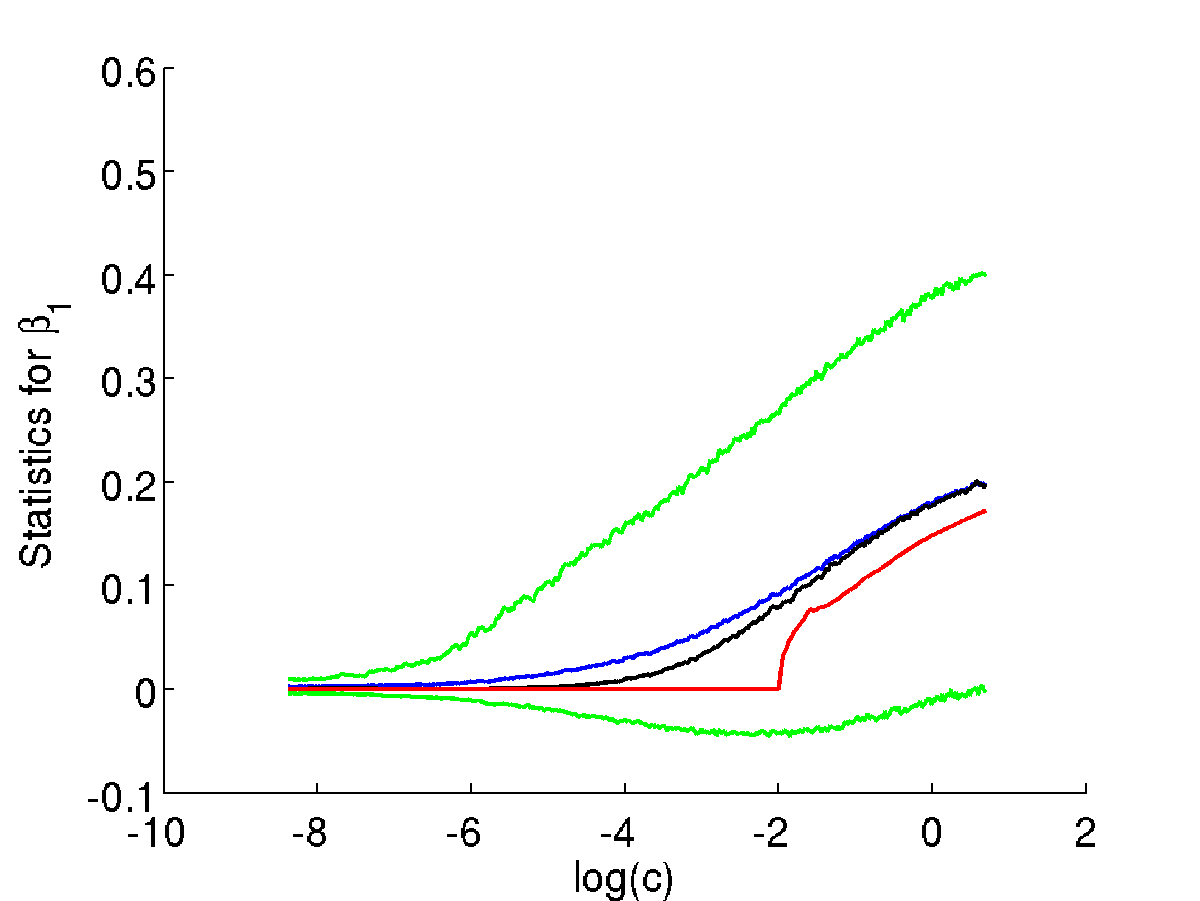}
}
\caption{Stats for individual coefficients from Fig. \ref{fig:spa_a1} with $a=1$. For each coefficient we plot $90\%$ credible intervals (green), median (black), mean (blue) and MAP (red).  Compare with Fig.~\ref{fig:a4_index_1} where $a=4$.}
\label{fig:a1_index_1}
\end{figure}

\begin{figure}[htp]
\center
\subfigure[$a=1$, $j=14$]{
	\includegraphics[scale=0.3]{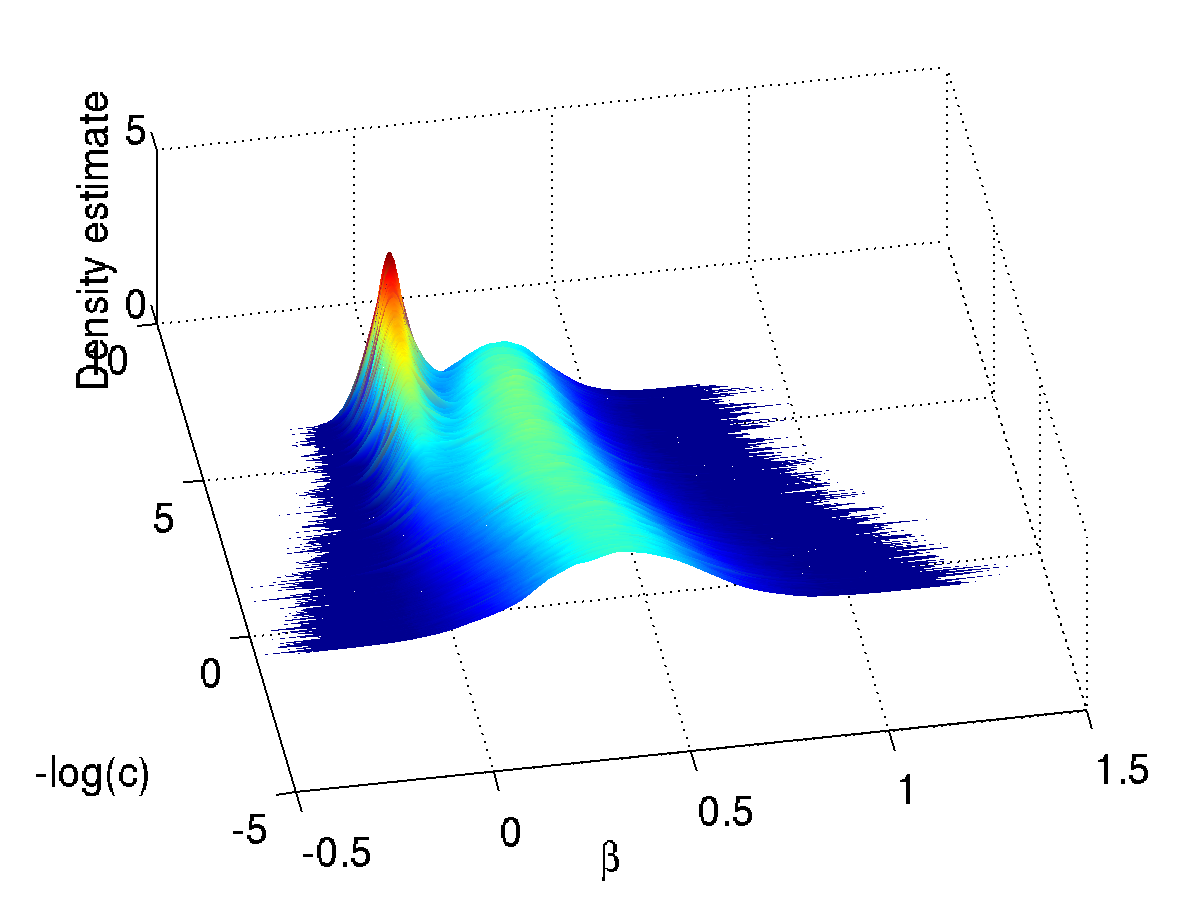}
}
\subfigure[$a=1$, $j=24$]{
	\includegraphics[scale=0.3]{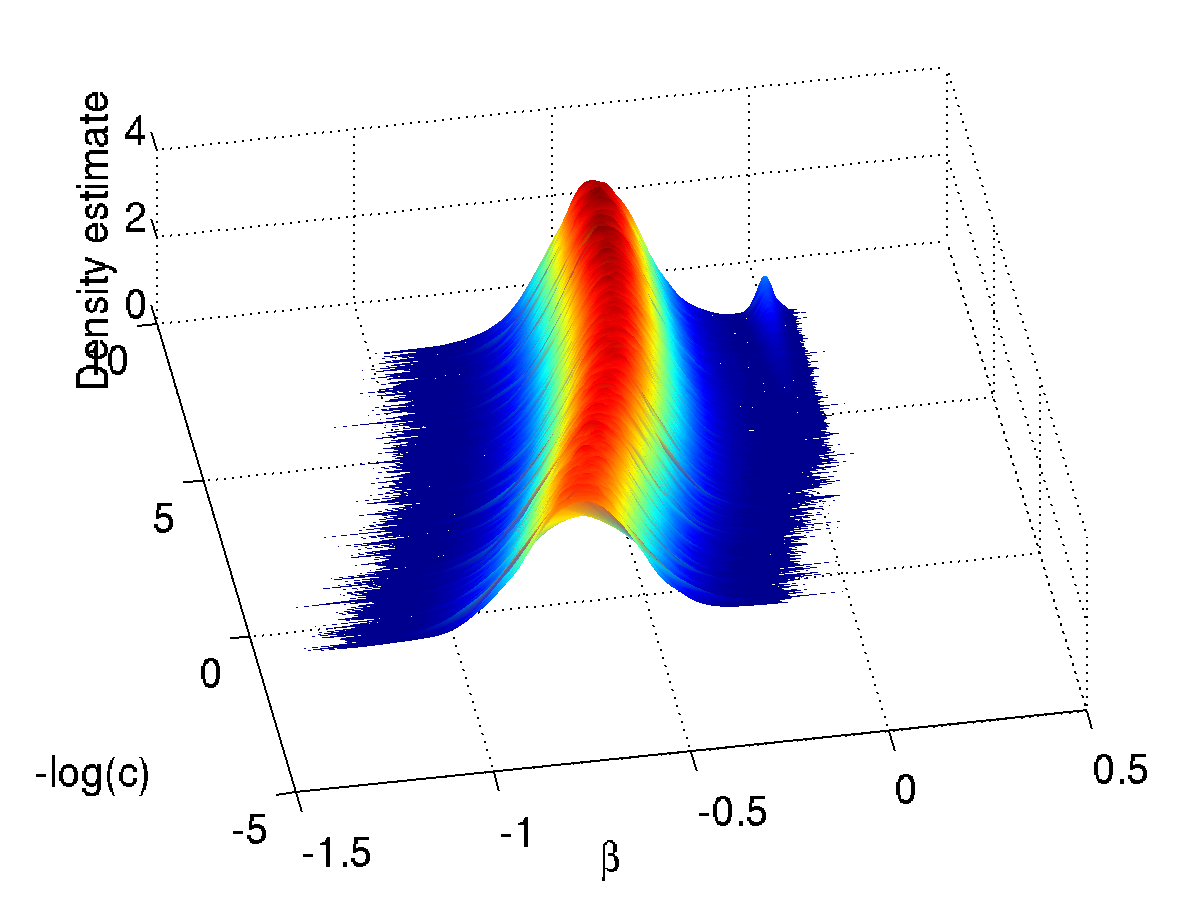}
}
\subfigure[$a=1$, $j=31$]{
	\includegraphics[scale=0.3]{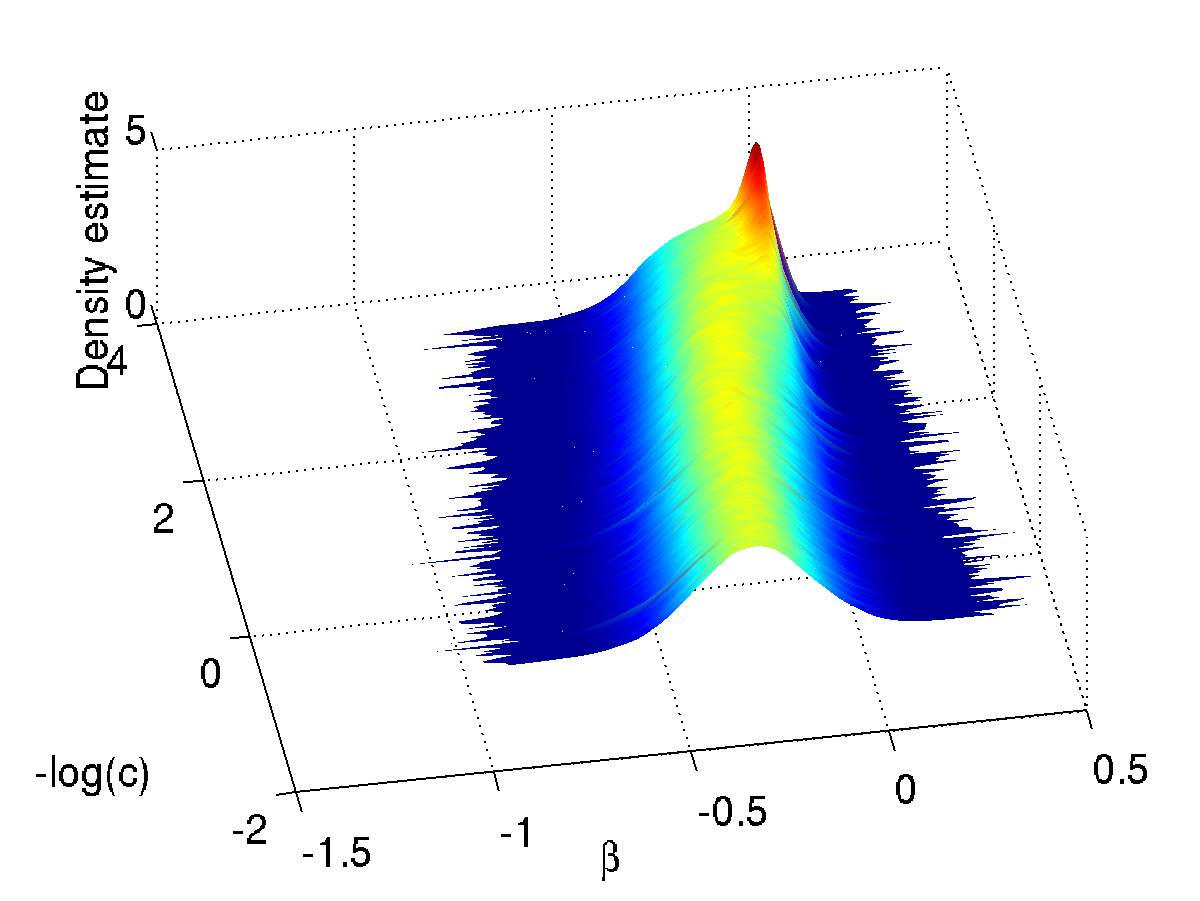}
}
\subfigure[$a=1$, $j=1$]{
	\includegraphics[scale=0.3]{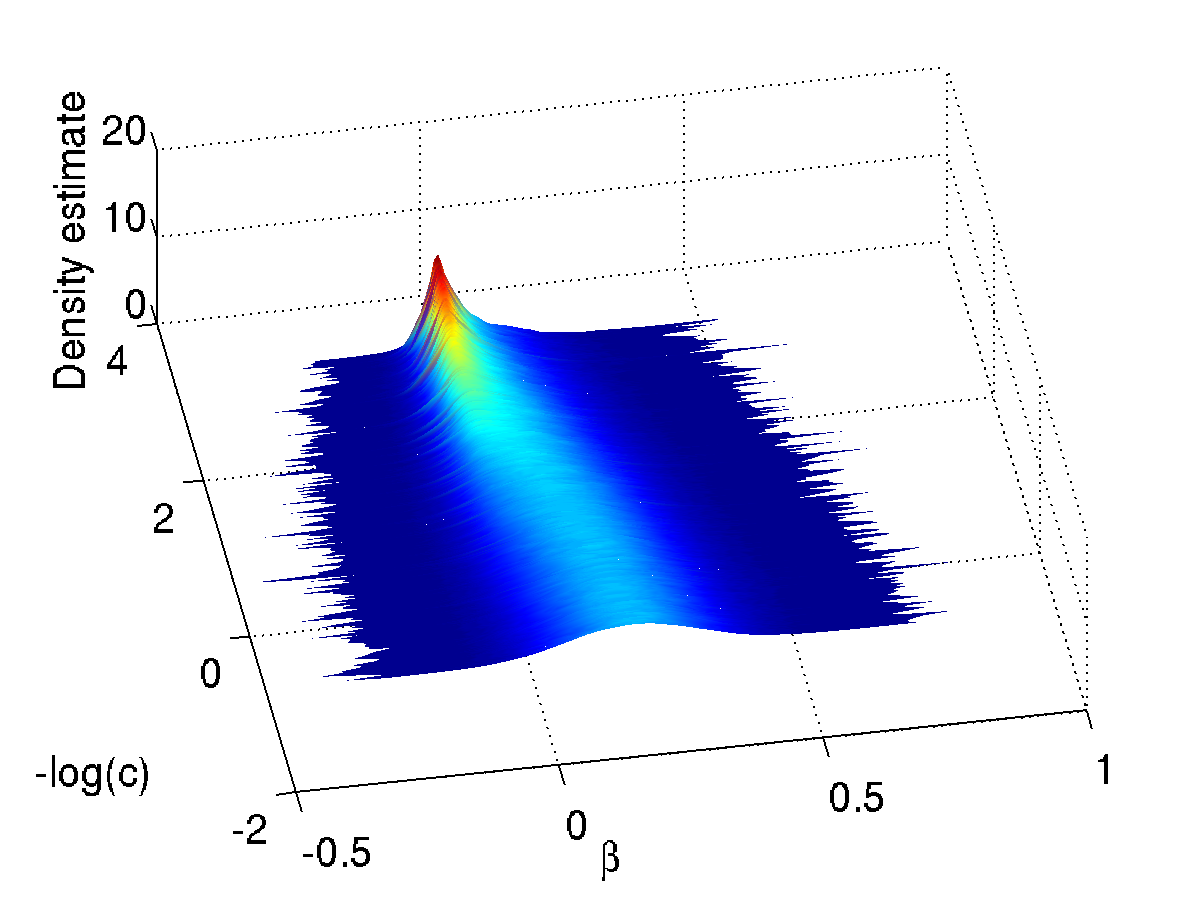}
}
\caption{Posterior density plots corresponding to Fig.~\ref{fig:a1_index_1}, $a=1$}
\label{fig:a1_index_2}
\end{figure}

Equivalent plots but for $a=1$ are shown in Fig.\ref{fig:a1_index_1} and Fig.\ref{fig:a1_index_2} where we see the greater sparsity induced by the heavier tails of the $GT(a=1,c)$ relative to $GT(a=4,c)$.

\subsection{Marginal plots}

The SMC sampler also allows us to estimate the marginal posterior probability of $\bbeta$, using (\ref{eqn:all_particles}), integrating over the uncertainty in $c$,
$$
p(\bbeta | \bX, \by, a) = \int_c p(\bbeta, c, | \bX, \by, a) d c
$$
Moreover we can also calculate the marginal posterior concentration away from zero, for given tolerance $\Delta$ as,
$$
        V = 1 - \int_c \int_{-\Delta}^{\Delta} \pi(\beta_j | \bX, \by, a, c) d\beta_j d c
$$

\begin{figure}[htp]
\center
\subfigure[]{
	\includegraphics[scale=0.3]{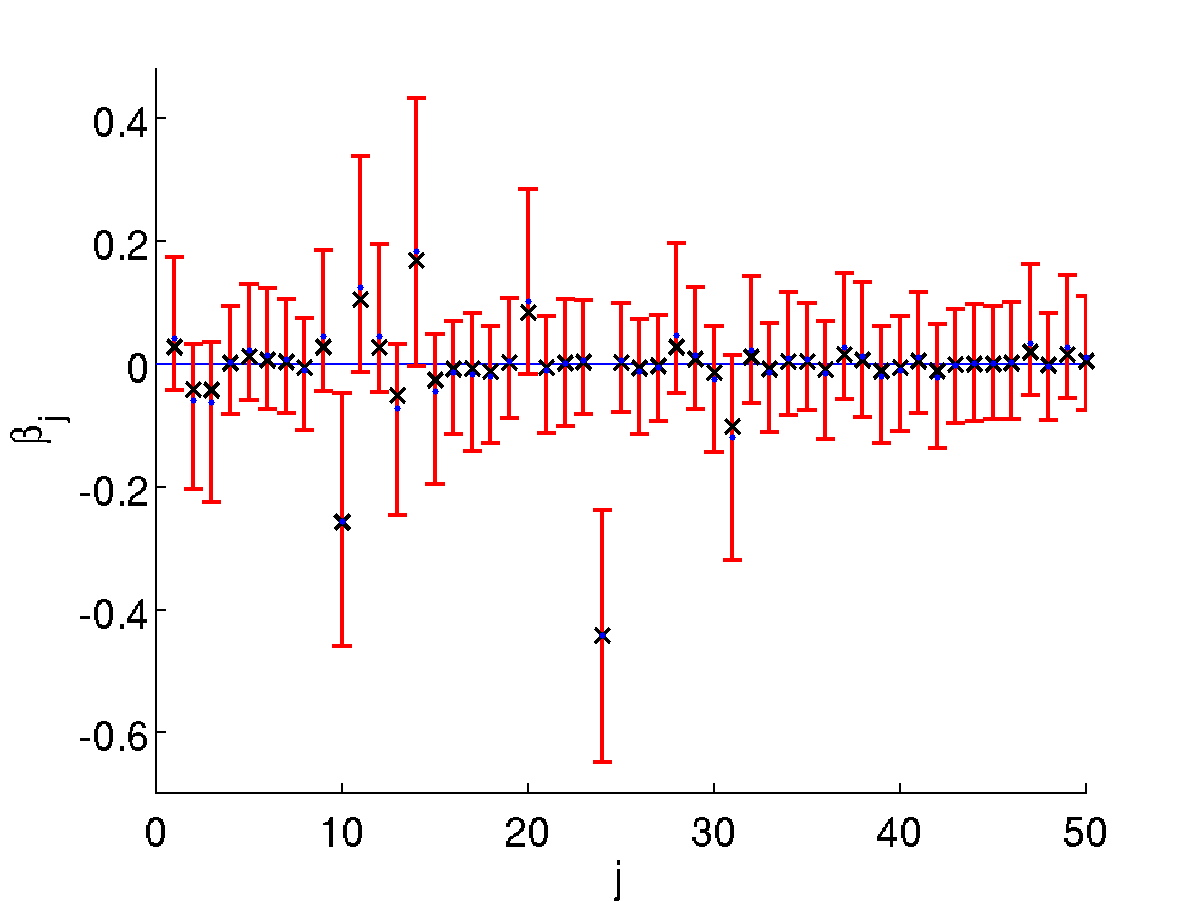}
}
\subfigure[]{
	\includegraphics[scale=0.3]{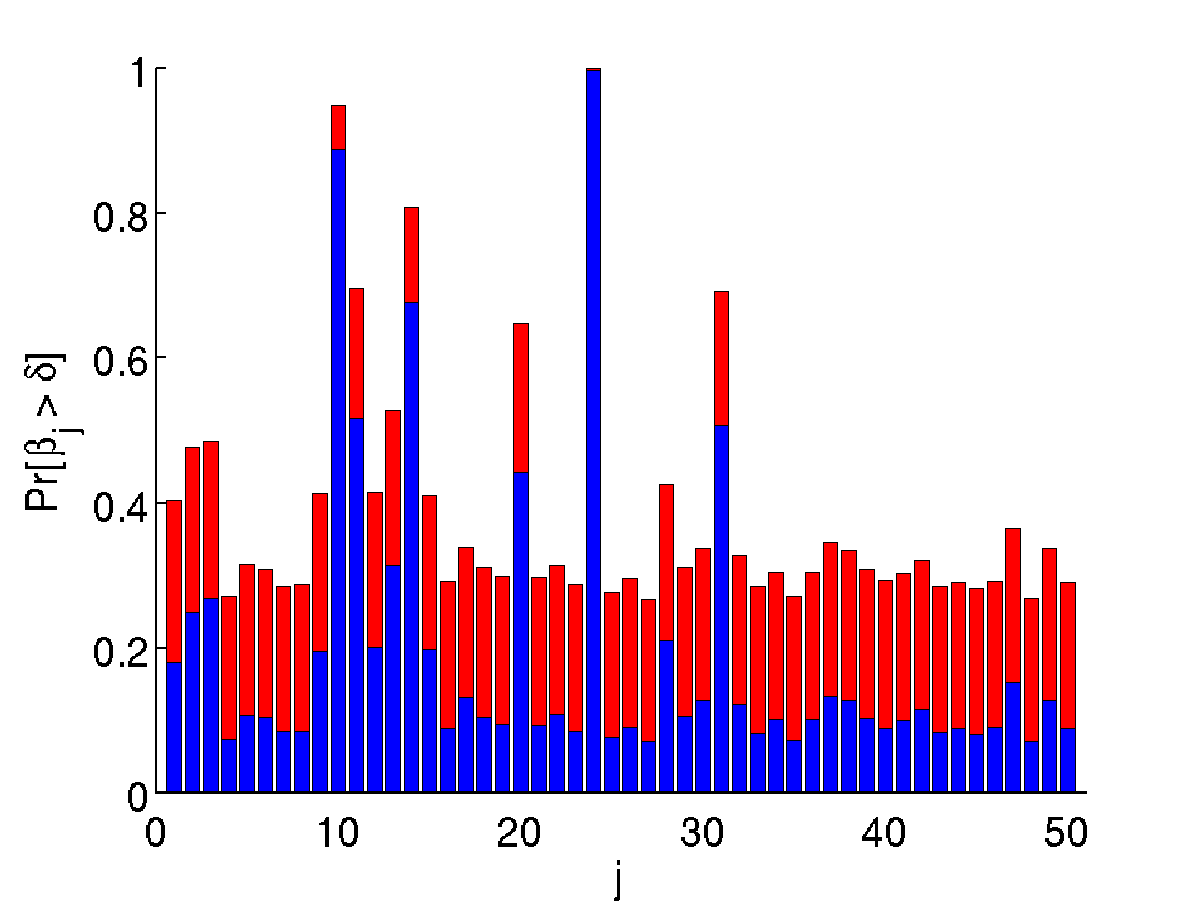}
}
\subfigure[]{
	\includegraphics[scale=0.3]{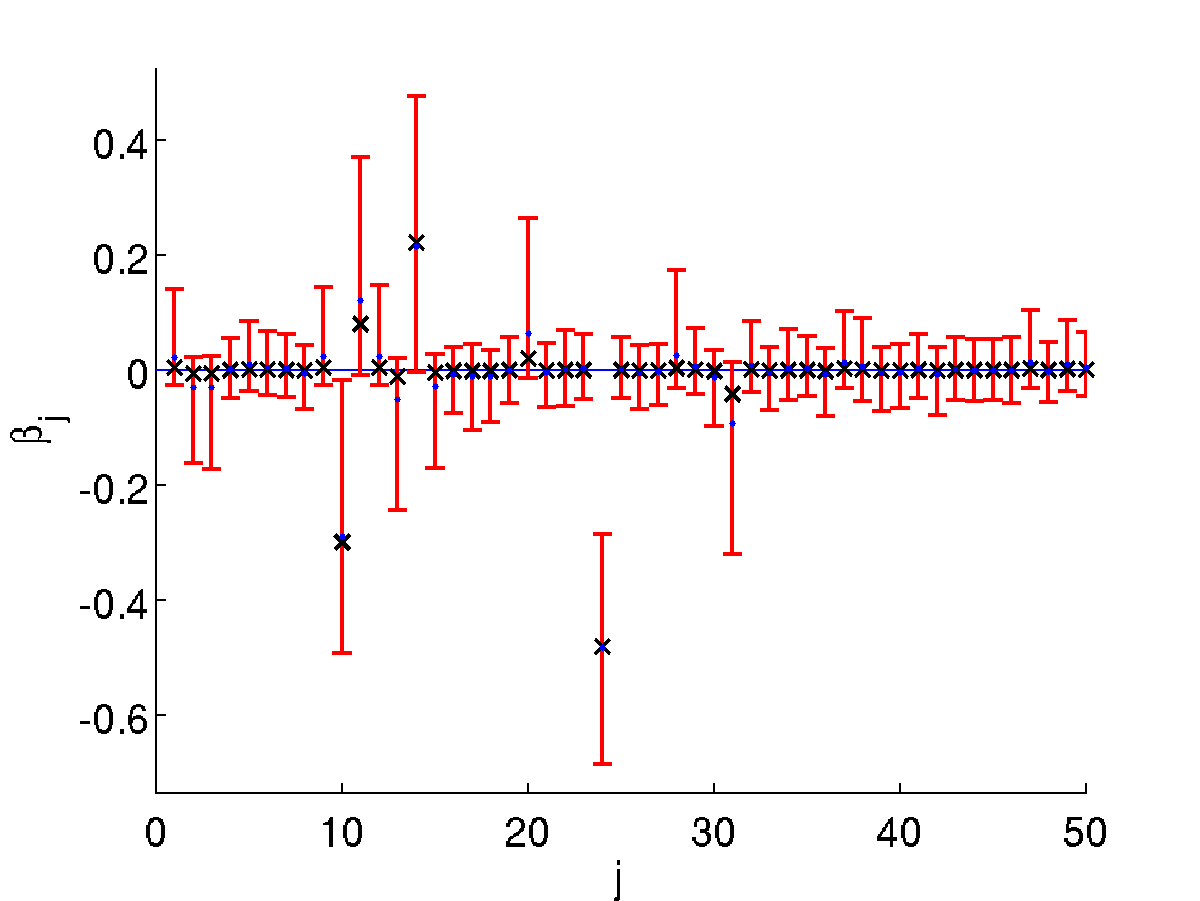}
}
\subfigure[]{
	\includegraphics[scale=0.3]{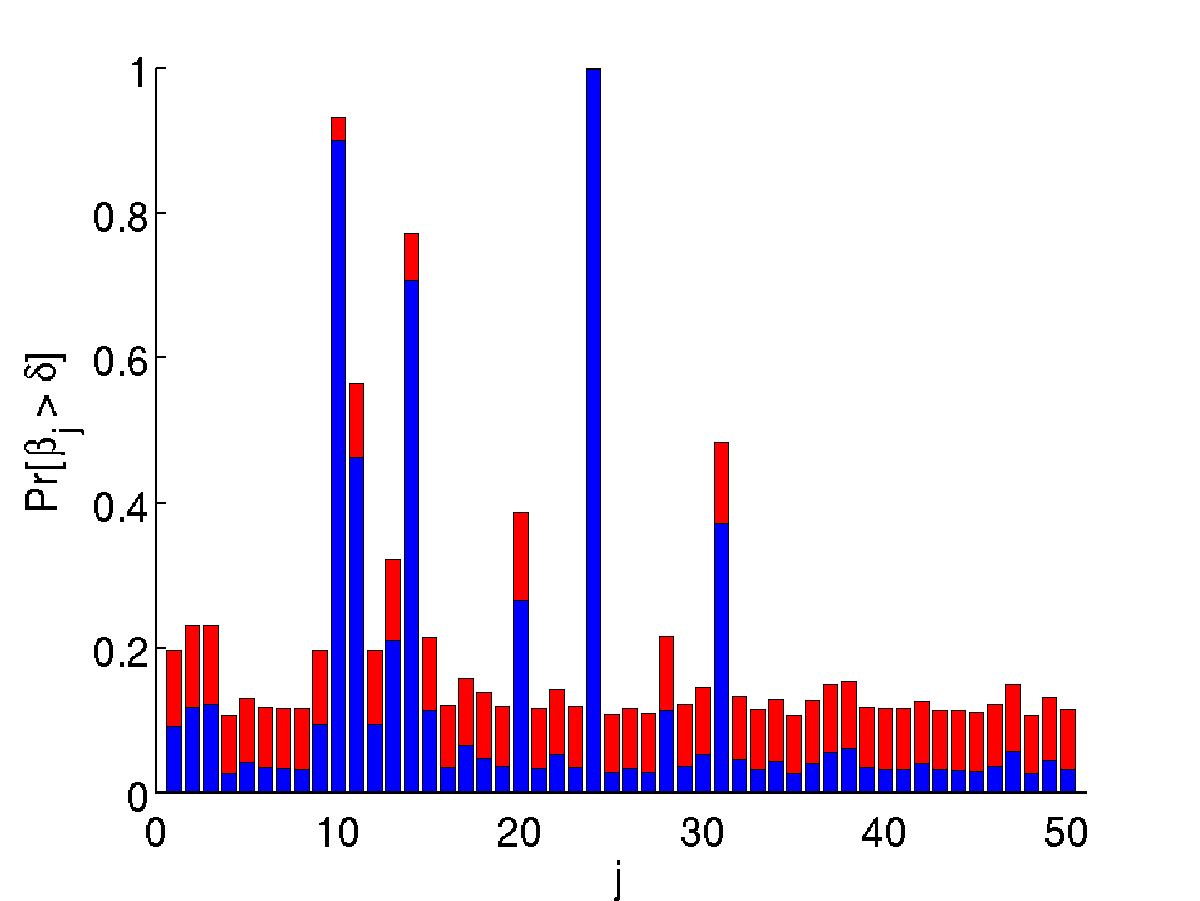}
}
\caption{Marginal plots: (a) the summary stats from marginal posterior distributions showing MAPs (crosses), Medians (stars), and $90\%$ credible intervals (bars) for $a=4$ degrees of freedom; (b) the marginal concentrations $\Delta = 0.05$ red bars, $\Delta=0.1$ (blue bars) for $a=4$; (c) marginal posteriors as in (a) but with $a=1$ prior; (d) marginal concentrations as in (b) but for $a=1$, prior.}
\label{fig:all_stats}
\end{figure}

In Fig.\ref{fig:all_stats} we plot summaries for the marginal posterior distributions of $\bbeta$ as well as the marginal concentrations, for $a=4$ and $a=1$. We can see in Fig.\ref{fig:all_stats} that the marginal plots provide a useful overview of the relative importance of each variable.

\subsection{Comparison to double-exponential prior, $a \to \infty$}

\begin{figure}[htp]
\center
\subfigure[MAPs]{
	\includegraphics[scale=0.3]{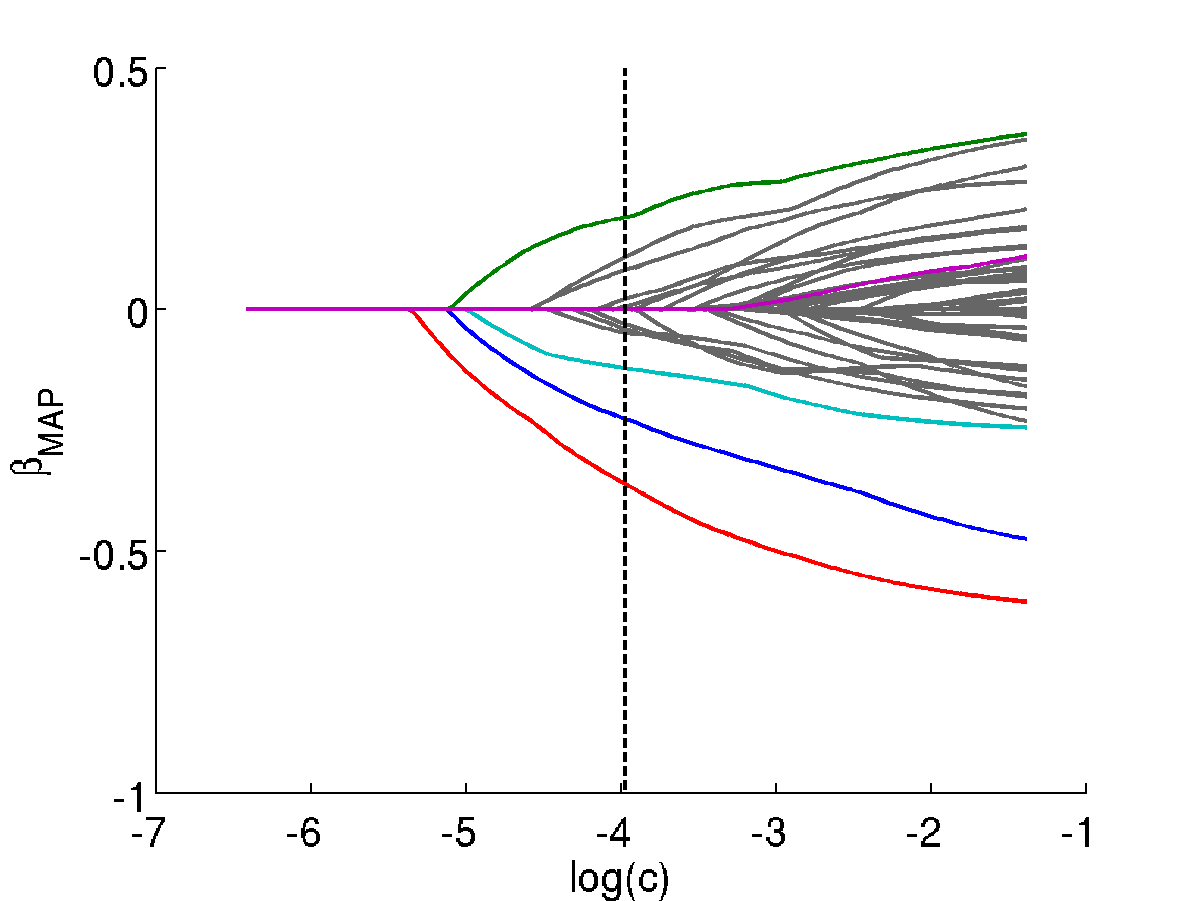}
}
\subfigure[medians]{
	\includegraphics[scale=0.3]{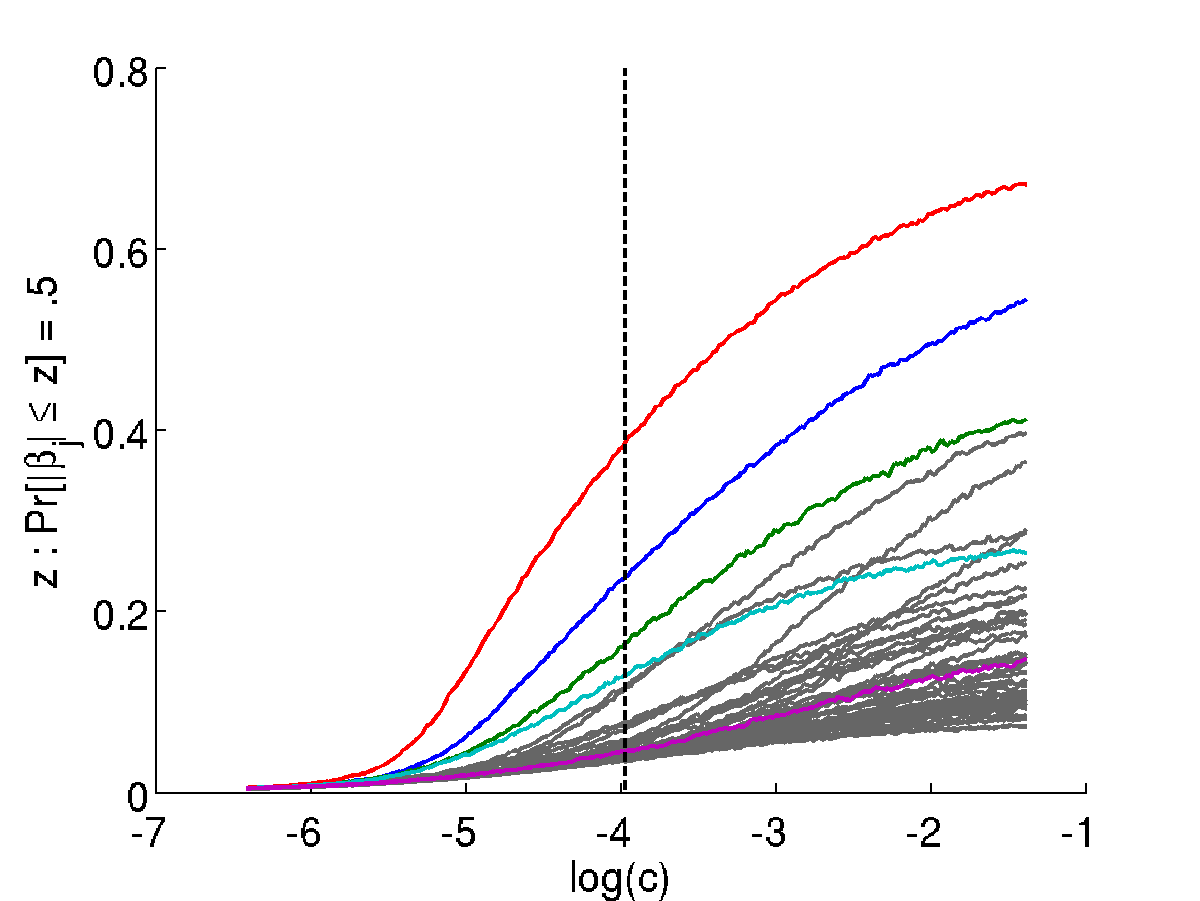}
}
\subfigure[marginal density]{
	\includegraphics[scale=0.3]{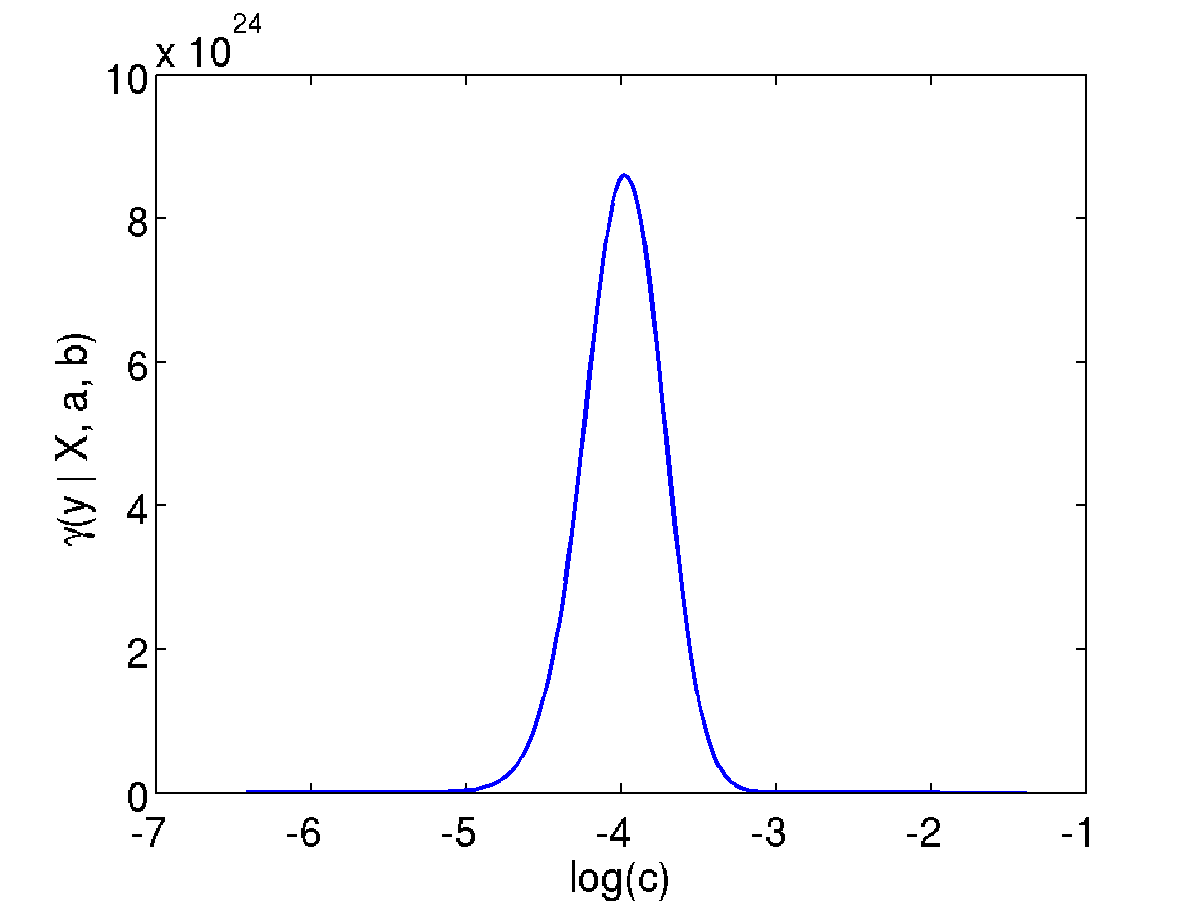}
}
\subfigure[concentrations]{
	\includegraphics[scale=0.3]{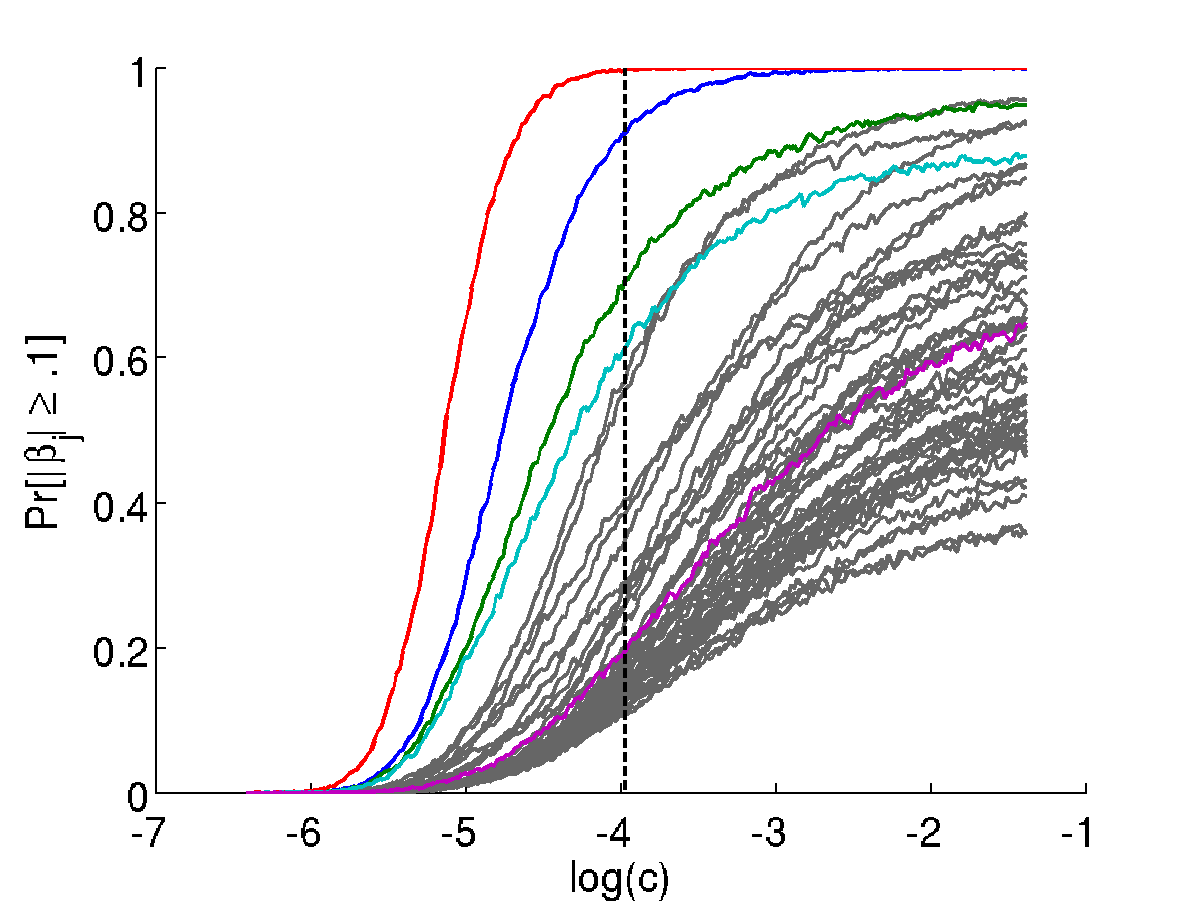}
}
\caption{SPA plots for the double-exponential prior, $Gt(a \to \infty, c)$.}
\label{fig:spa_lasso}
\end{figure}

\begin{figure}[htp]
\center
\subfigure[double-exponential $j=14$]{
	\includegraphics[scale=0.3]{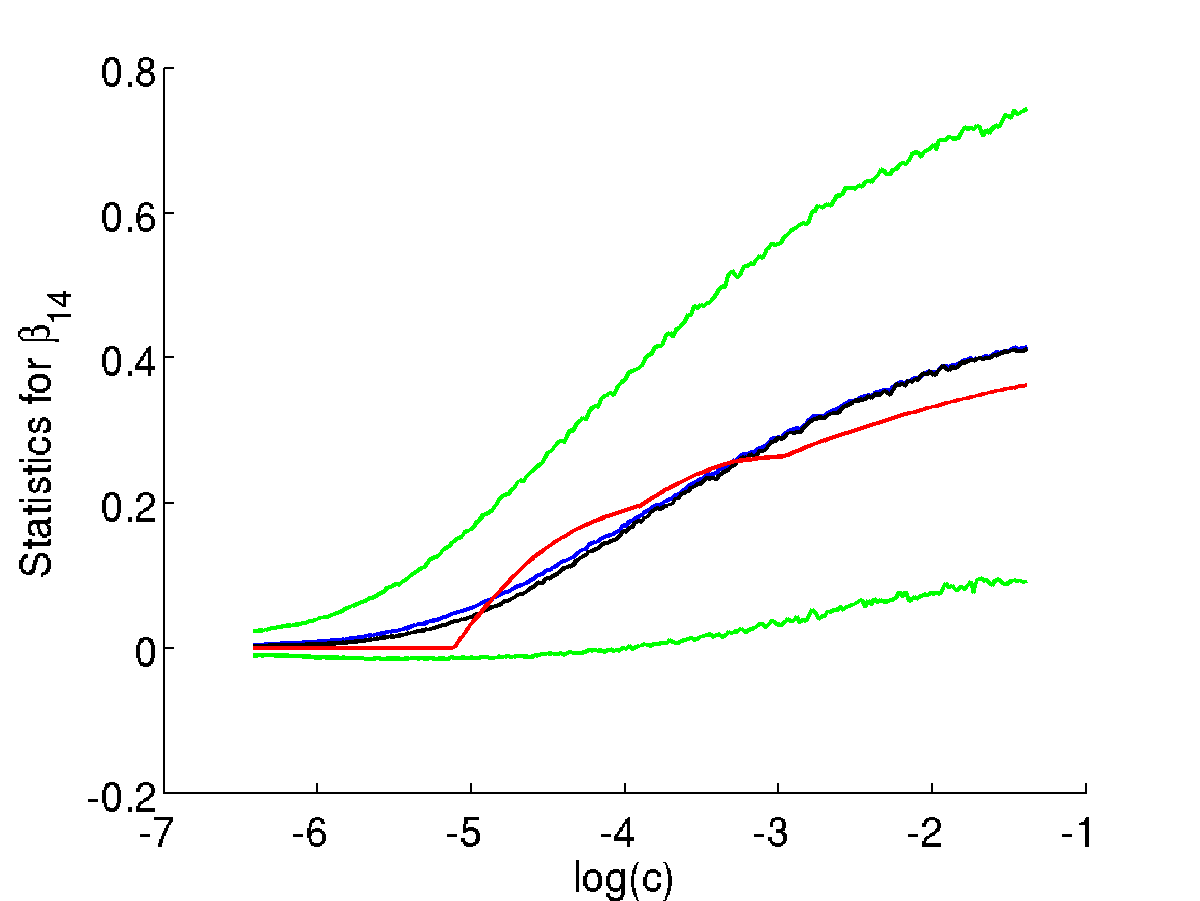}
}
\subfigure[double-exponential $j=24$]{
	\includegraphics[scale=0.3]{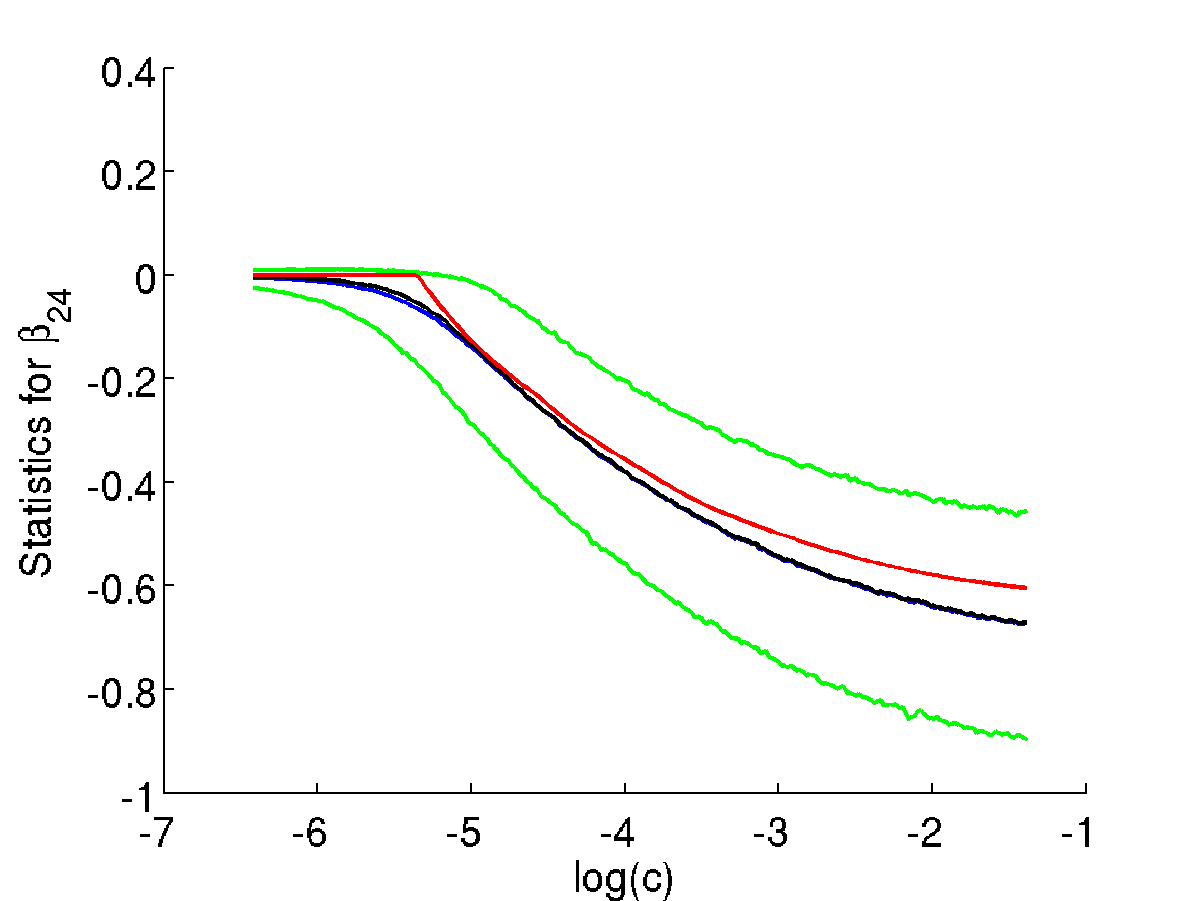}
}
\subfigure[double-exponential $j=31$]{
	\includegraphics[scale=0.3]{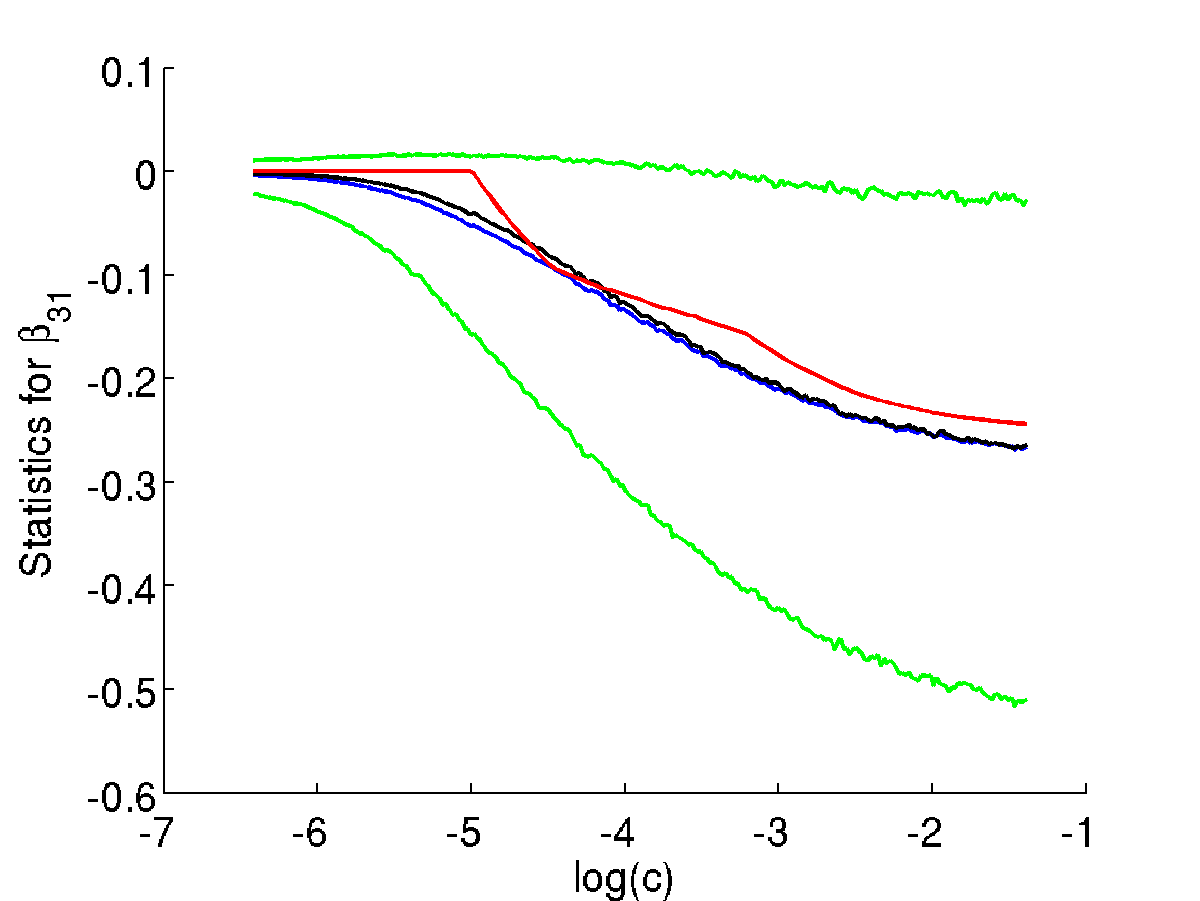}
}
\subfigure[double-exponential $j=1$]{
	\includegraphics[scale=0.3]{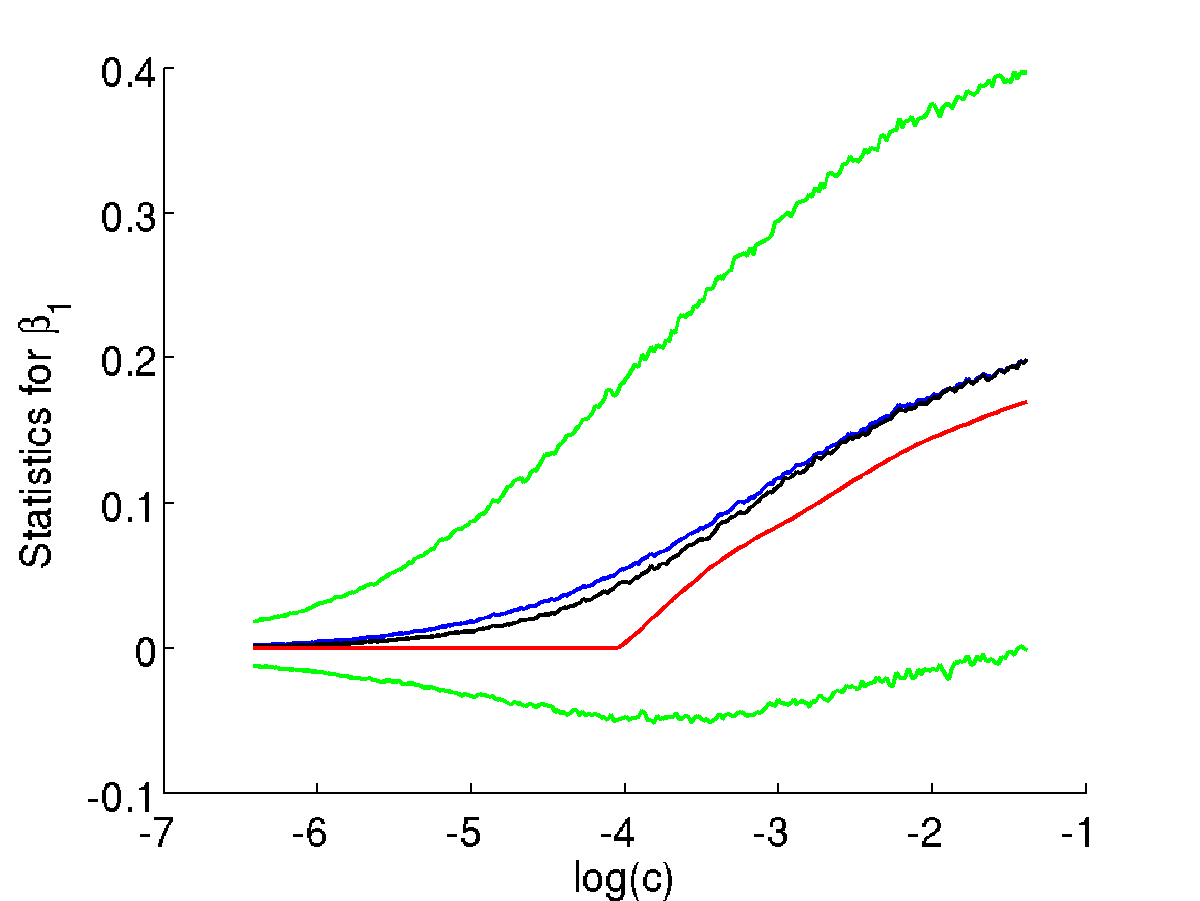}
}
\caption{Stats for individual coefficients from Fig. \ref{fig:spa_lasso} with double-exponential prior. For each coefficient we plot $90\%$ credible intervals (green), median (black), mean (blue) and MAP (red).  Compare with Fig.~\ref{fig:a4_index_1} where $a=4$.}
\label{fig:lasso_index_1}
\end{figure}

\begin{figure}[htp]
\center
\subfigure[double-exponential $j=14$]{
	\includegraphics[scale=0.3]{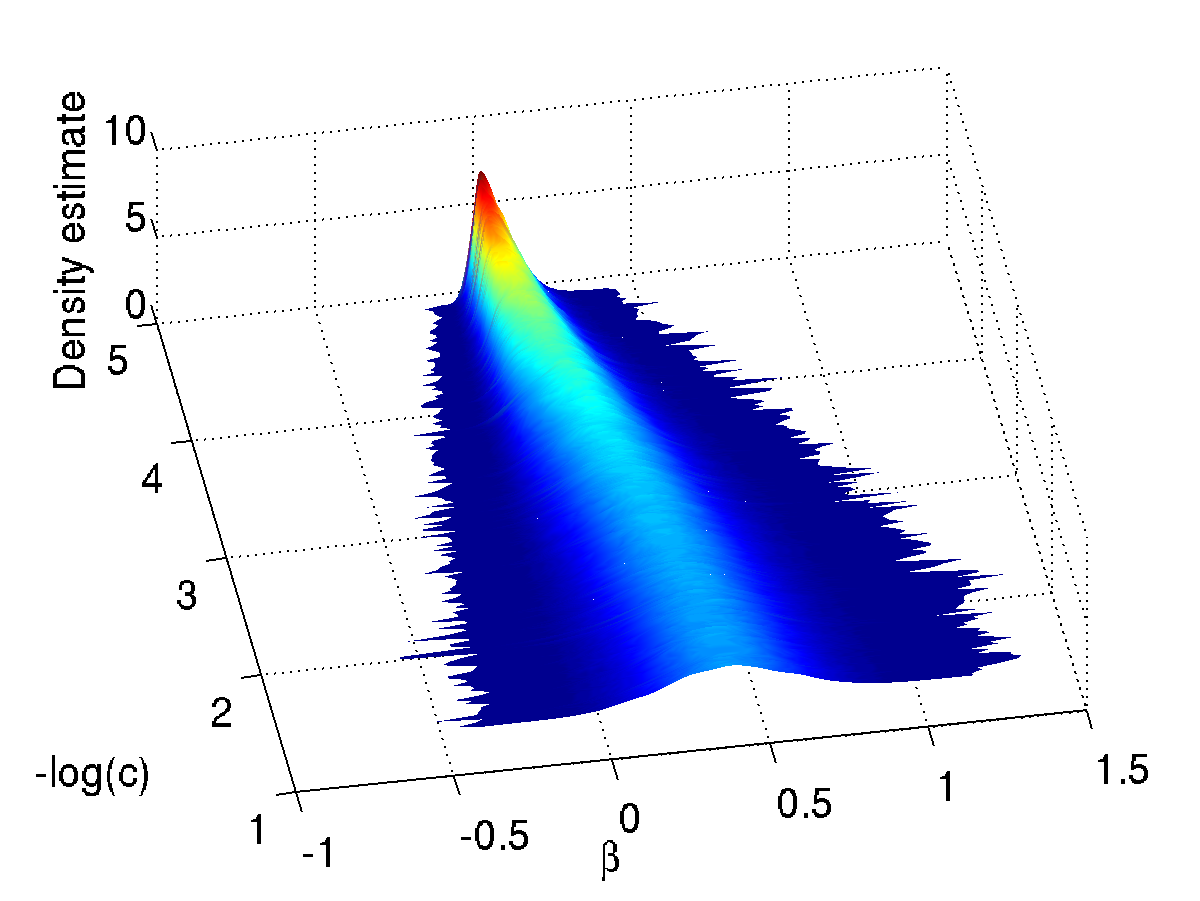}
}
\subfigure[double-exponential $j=24$]{
	\includegraphics[scale=0.3]{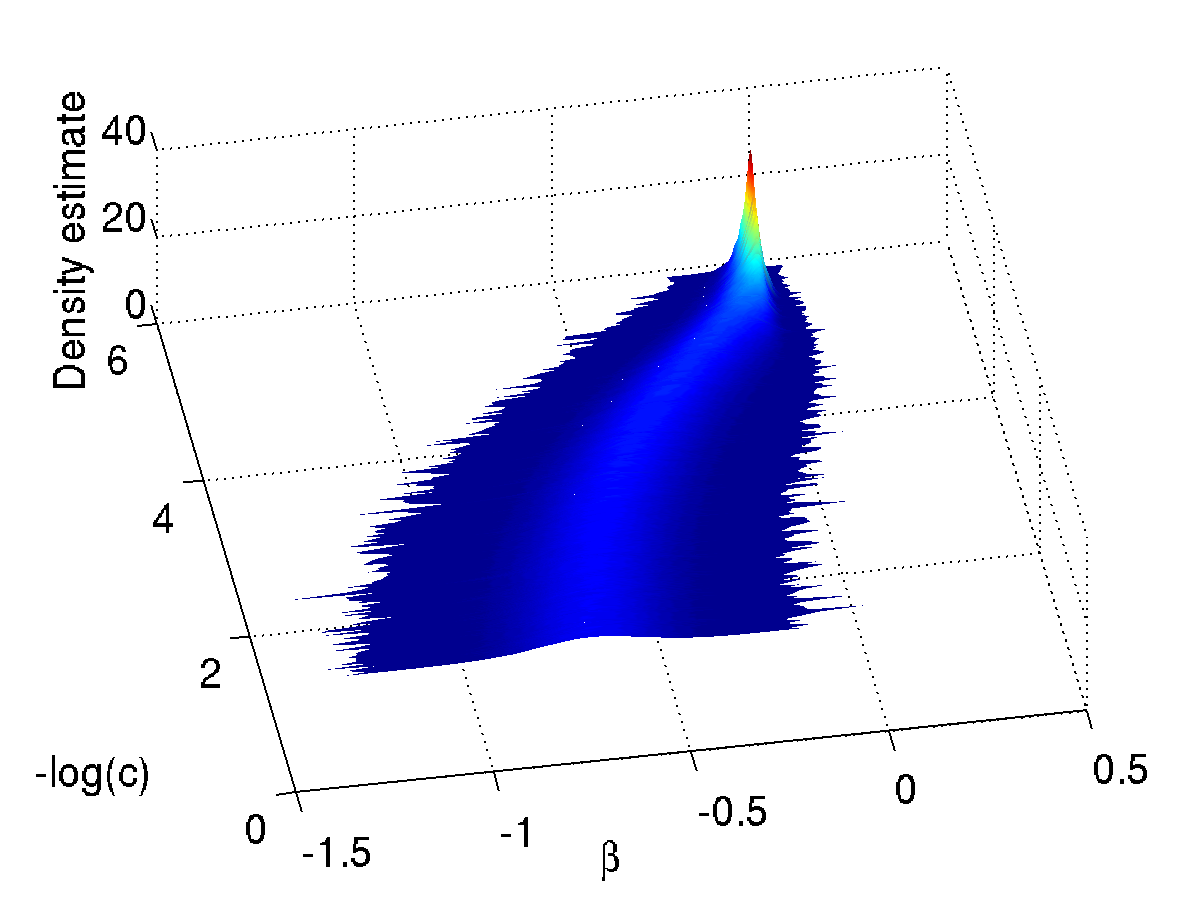}
}
\subfigure[double-exponential $j=31$]{
	\includegraphics[scale=0.3]{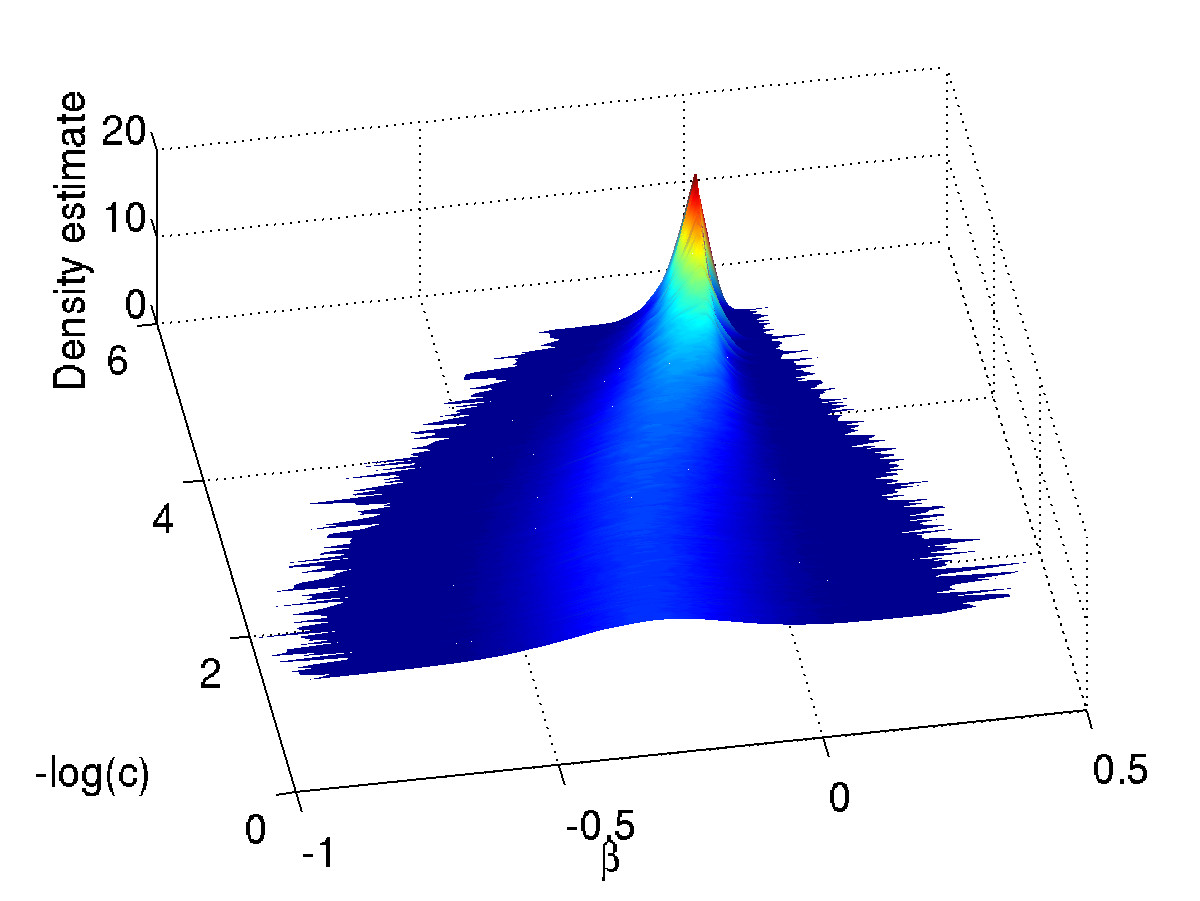}
}
\subfigure[double-exponential $j=1$]{
	\includegraphics[scale=0.3]{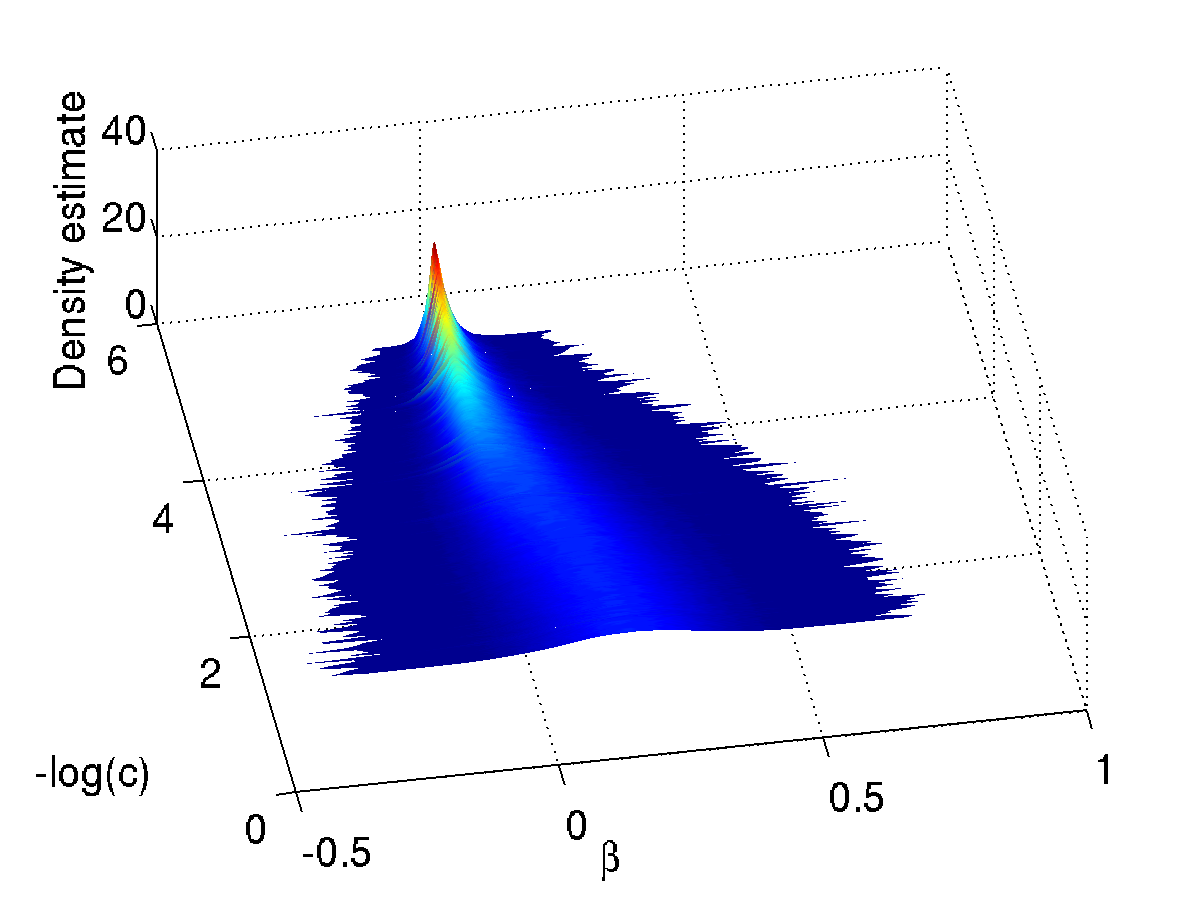}
}
\caption{Posterior density plots corresponding to Fig.~\ref{fig:lasso_index_1} with double-exponential prior.}
\label{fig:lasso_index_2}
\end{figure}

\begin{figure}[pth]
\center
\includegraphics[scale=0.6]{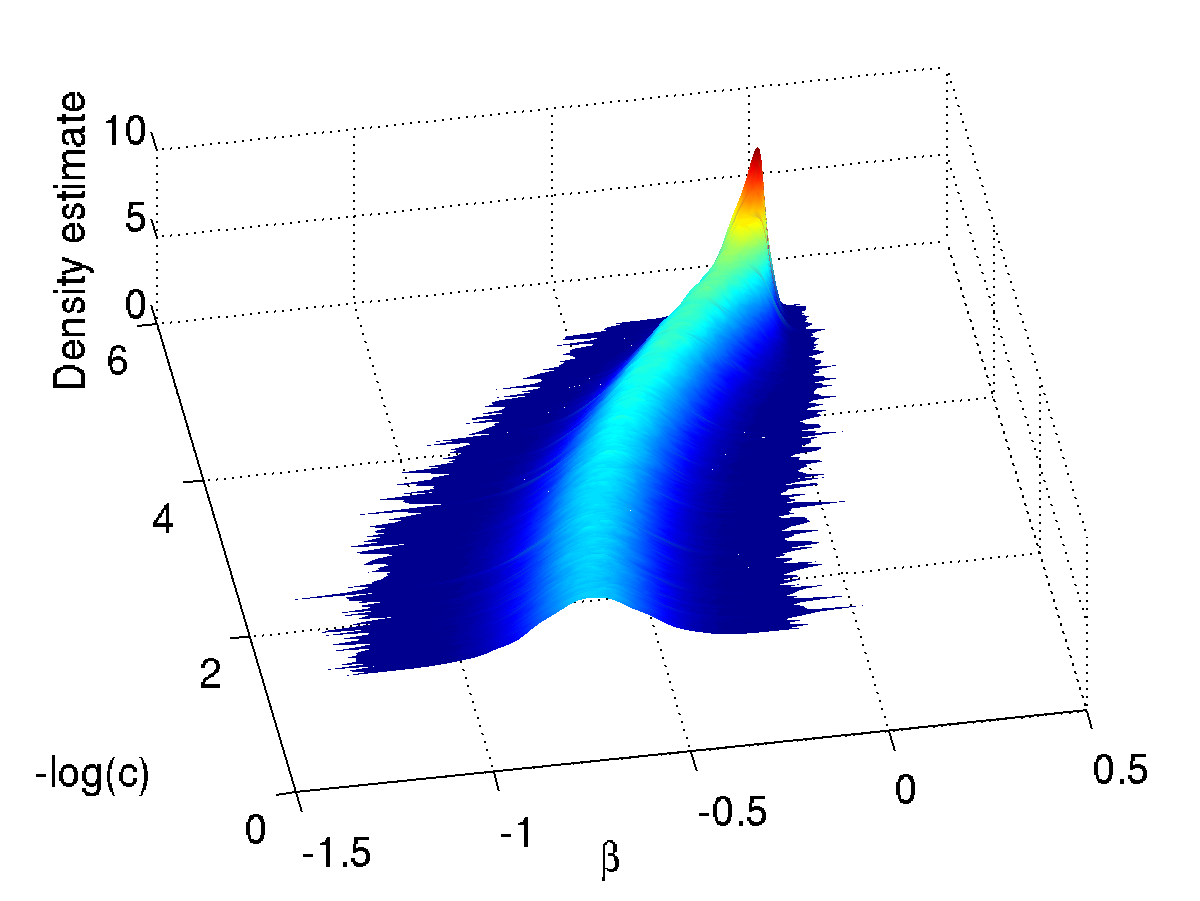}
\caption{Plot of marginal posterior of $\beta_{24}$ as a function of $c$ for double-exponential prior $a \to \infty$. Under the double-exponential prior, $a \to \infty$, we see the posterior remains log-concave and hence the MAP is a smooth function of $\log c$; as compared to Fig.~\ref{fig:mode_hop}.}
\label{fig:mode_hop_lasso}
\end{figure}

It is informative to compare the results for the $Gt(a,c)$ prior above with that of the double-exponential prior $p(\beta) \propto \exp(-\beta/c)$ which is obtained as a generalized t prior with $q=1$ and $a \to \infty$. In Fig.~\ref{fig:spa_lasso} we show SPA plots for this case. We can see the much smoother paths of the MAP compared with Fig ~\ref{fig:spa_a4}. This can also be seen in the individual coefficient plots shown in Fig. \ref{fig:lasso_index_1} and Fig. \ref{fig:lasso_index_2}. It is interesting to investigate in more detail the posterior density of $\beta_{24}$ the coefficient with strongest evidence of association. This is shown in Fig.\ref{fig:mode_hop_lasso} and should be compared to Fig.\ref{fig:mode_hop} for $a=4$. We can clearly see that the heavier tails of $a=4$ induces greater sparsity and a rapid transition from the posterior concentrated around zero to being  distributed away from zero.

%

\subsection{Large Data Set}

\begin{figure}[htp]
\center
\subfigure[MAPs]{
       \includegraphics[scale=0.3]{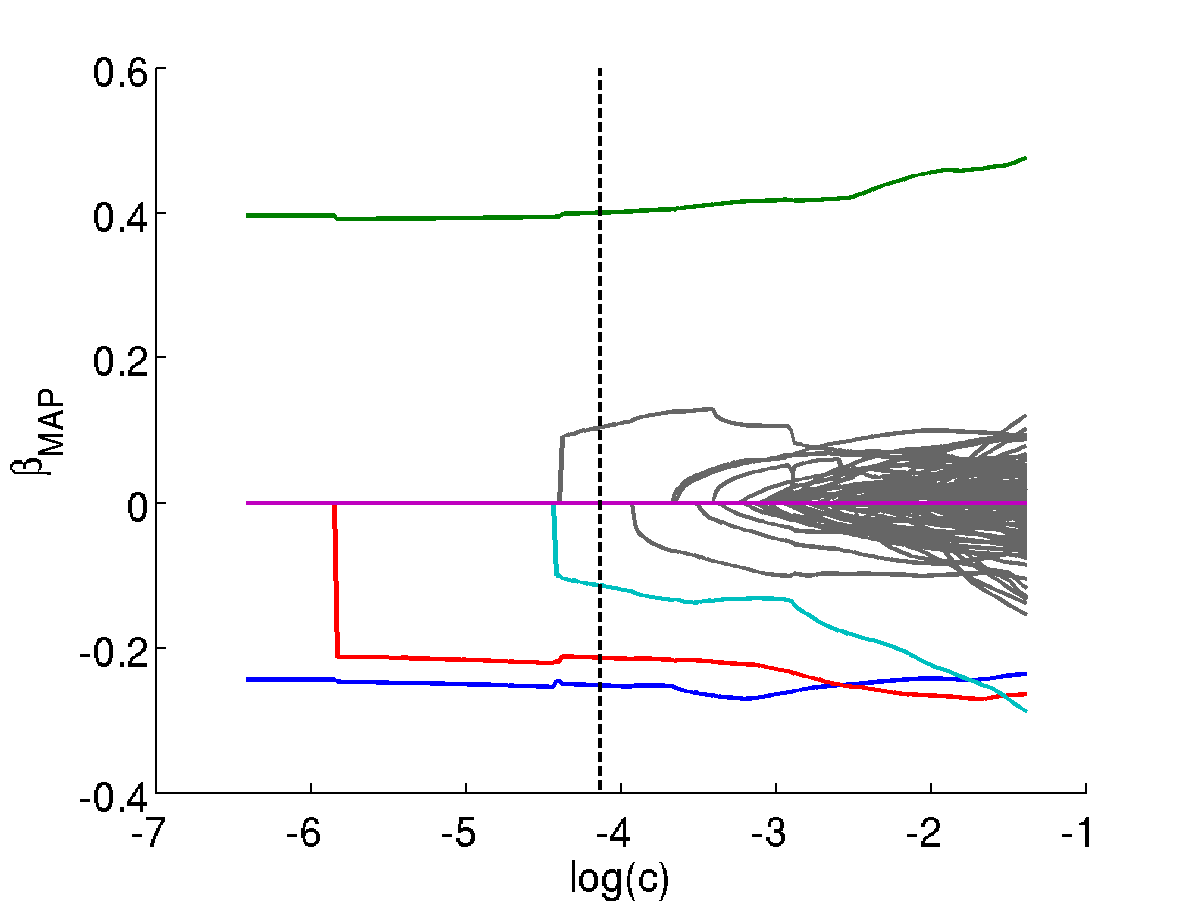}
}
\subfigure[medians]{
       \includegraphics[scale=0.3]{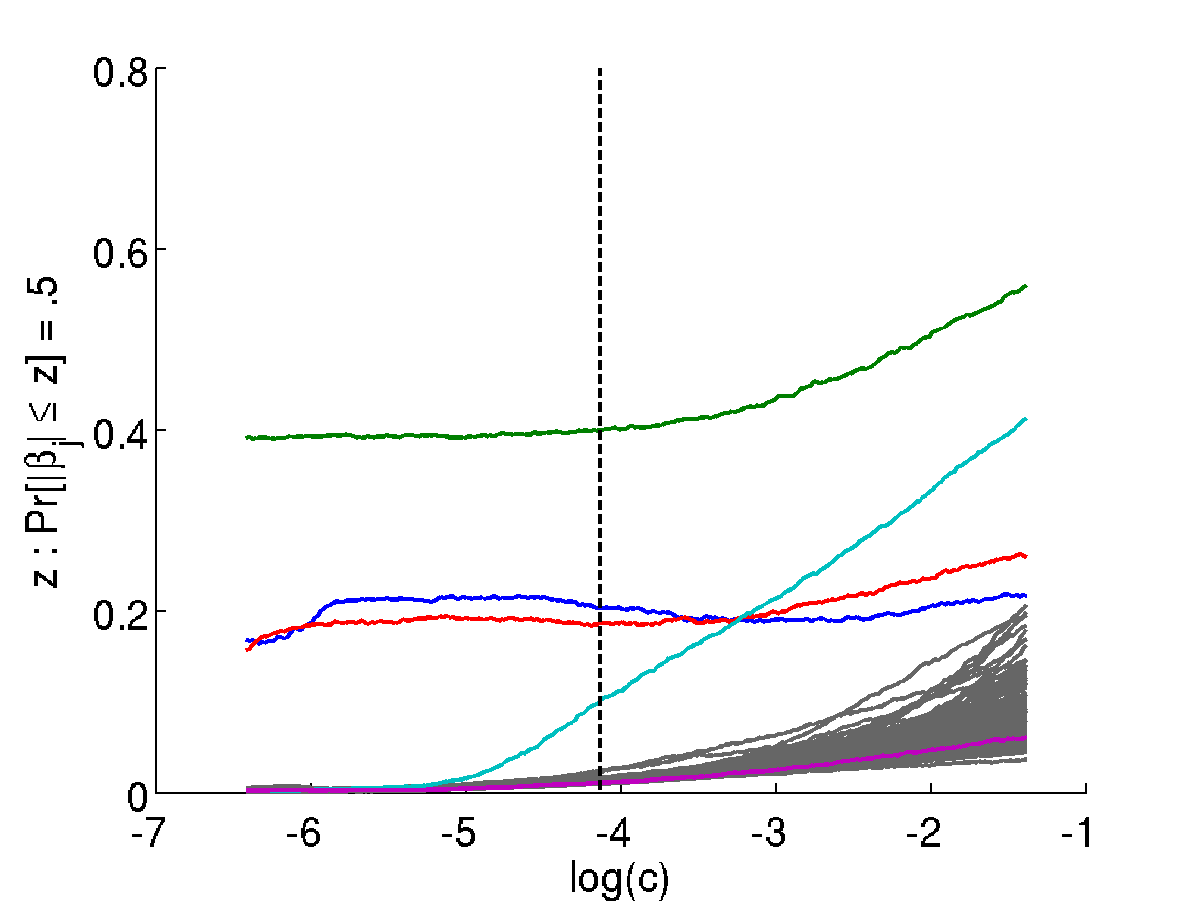}
}
\subfigure[marginal density]{
       \includegraphics[scale=0.3]{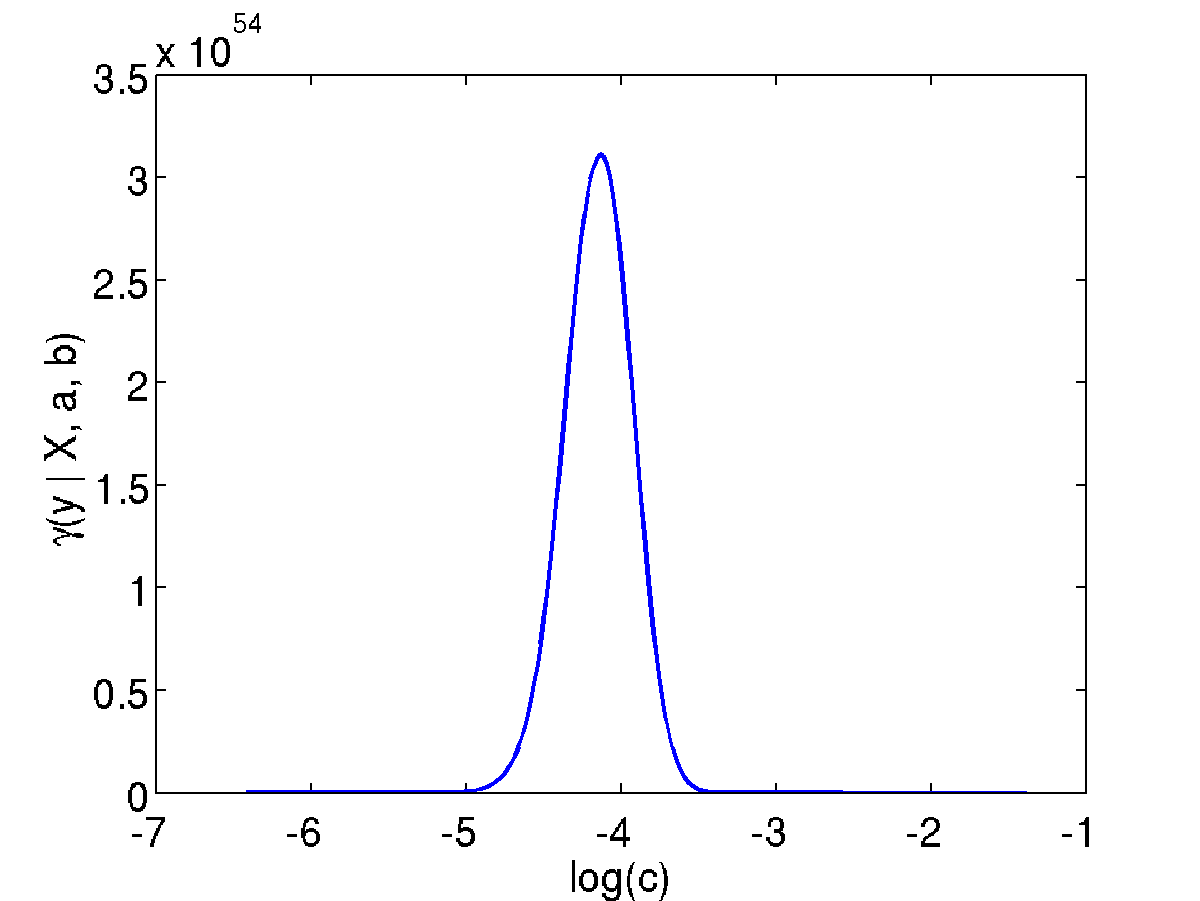}
}
\subfigure[concentrations]{
       \includegraphics[scale=0.3]{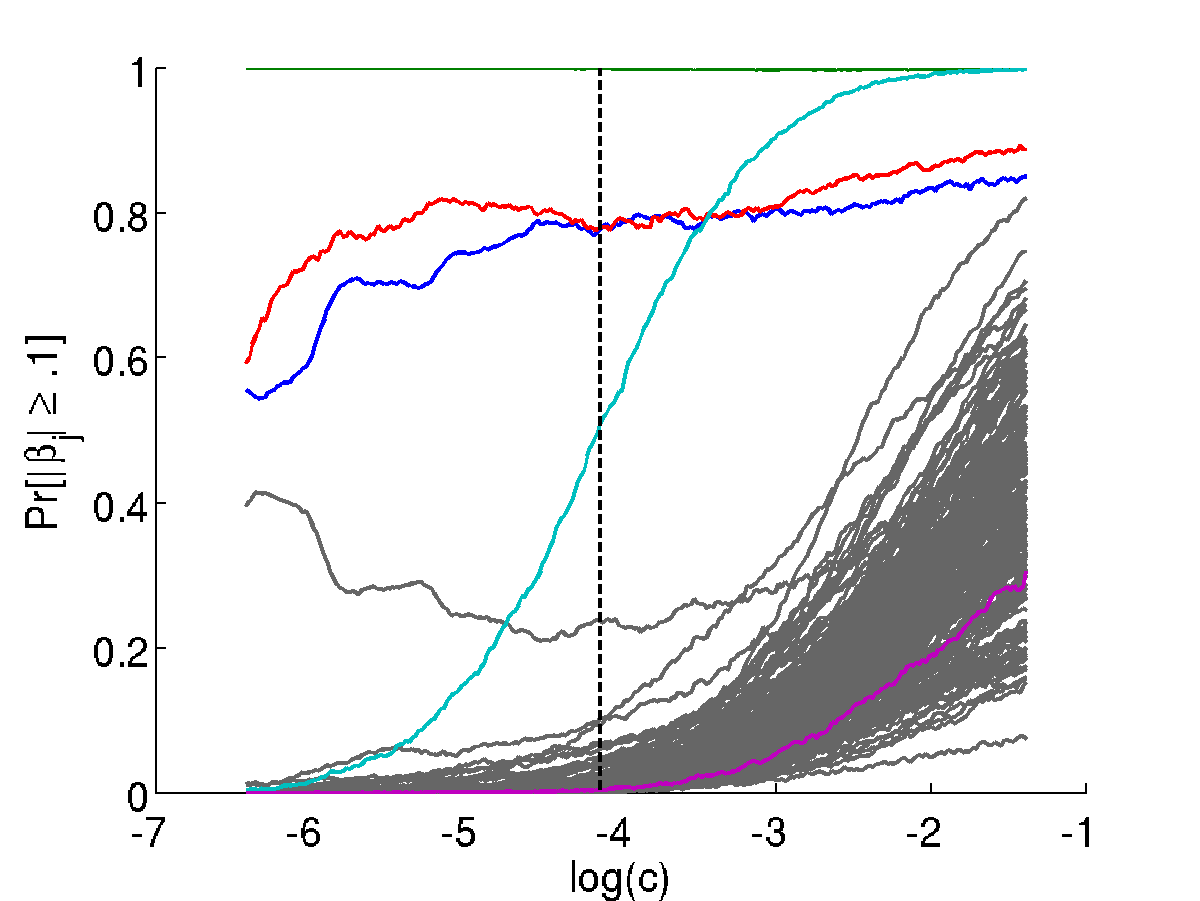}
}
\caption{SPA plots using the larger data set $n=1859$, $p=184$}
\label{fig:spa_a4n1859}
\end{figure}

\begin{figure}[htp]
\center
\subfigure[$a=4, j=22$]{
       \includegraphics[scale=0.3]{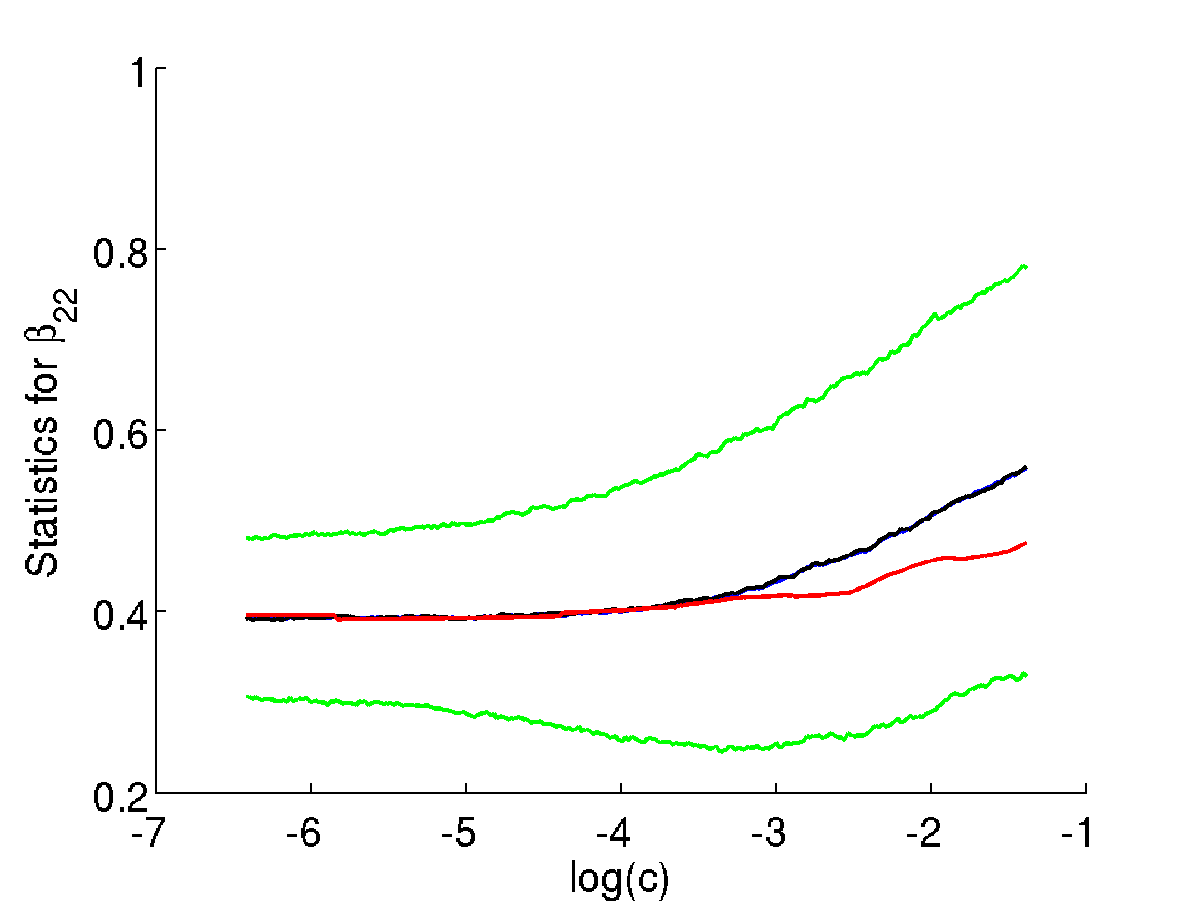}
}
\subfigure[$a=4, j=5$]{
       \includegraphics[scale=0.3]{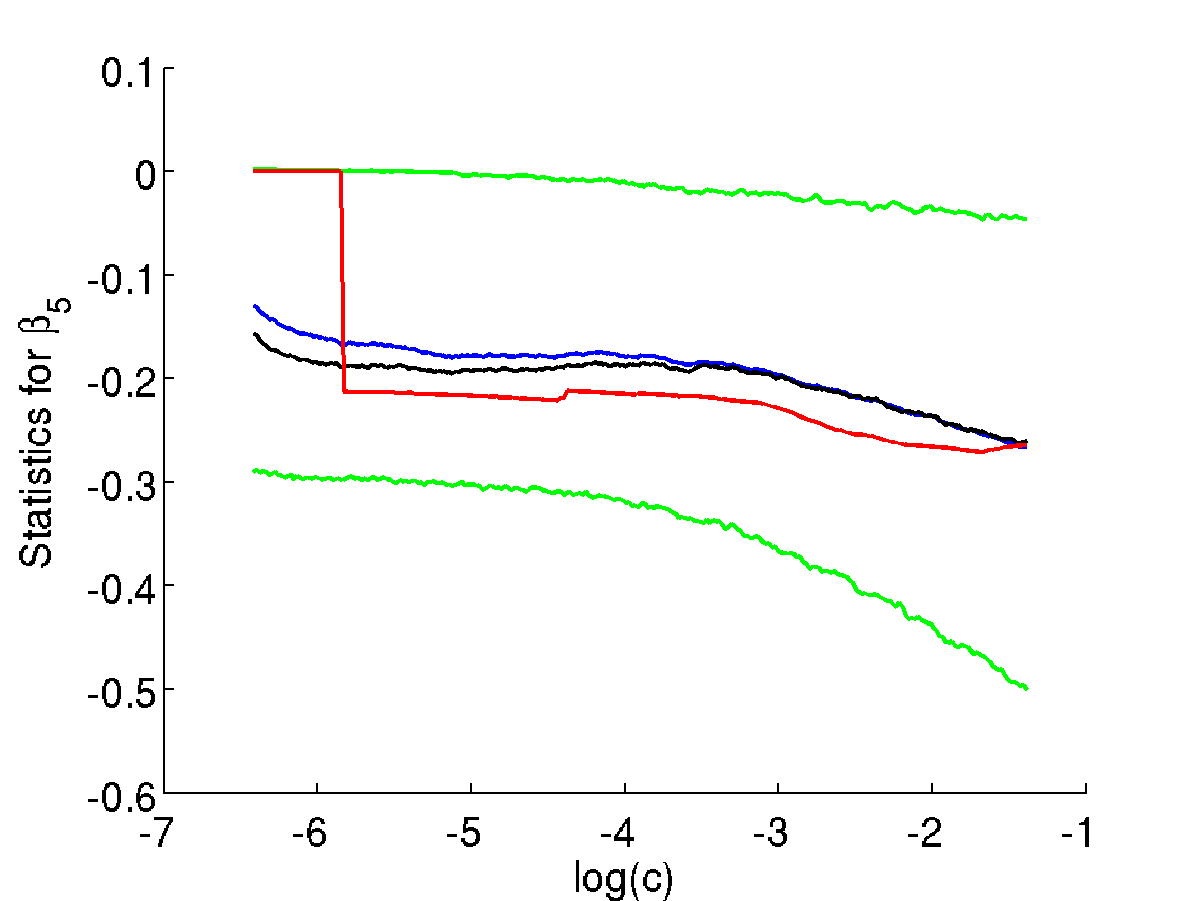}
}
\subfigure[$a=4, j=117$]{
       \includegraphics[scale=0.3]{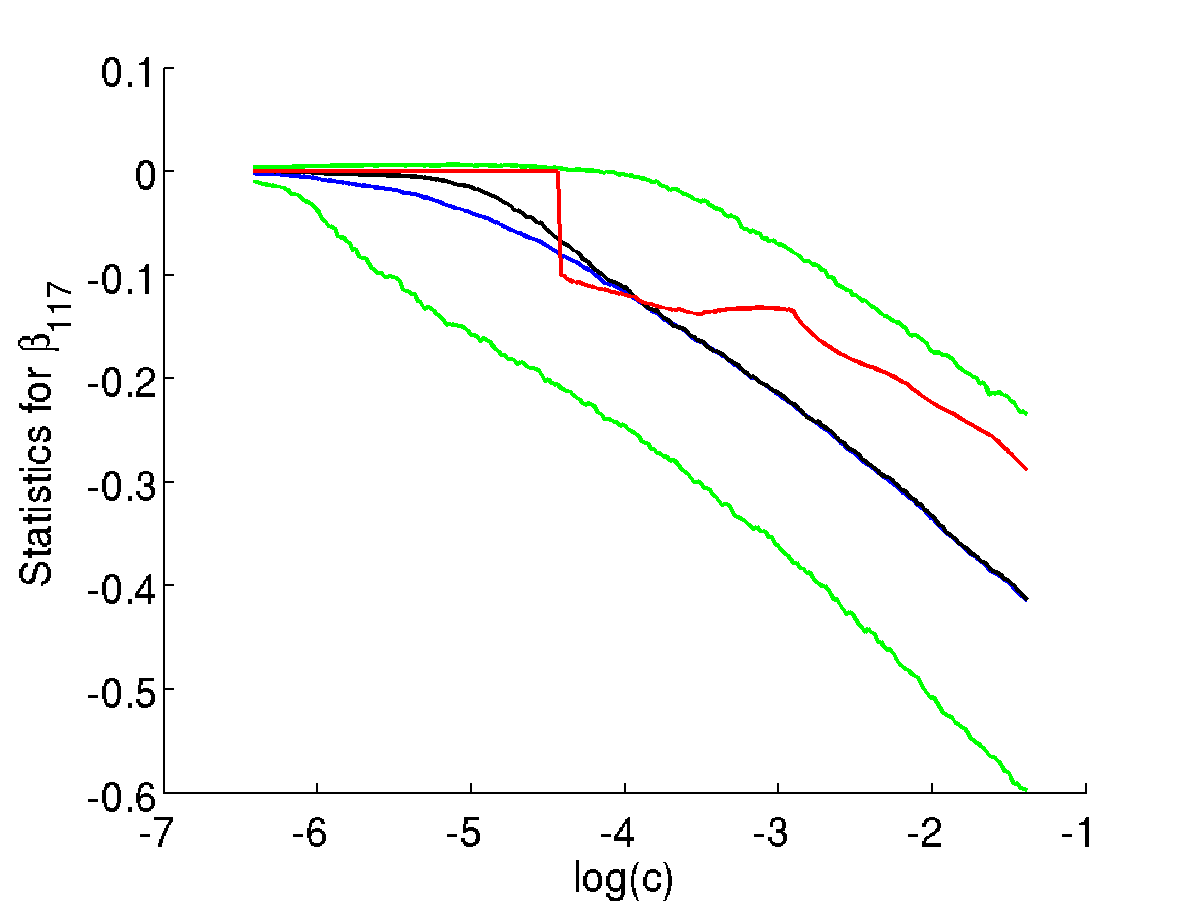}
}
\subfigure[$a=4, j=115$]{
       \includegraphics[scale=0.3]{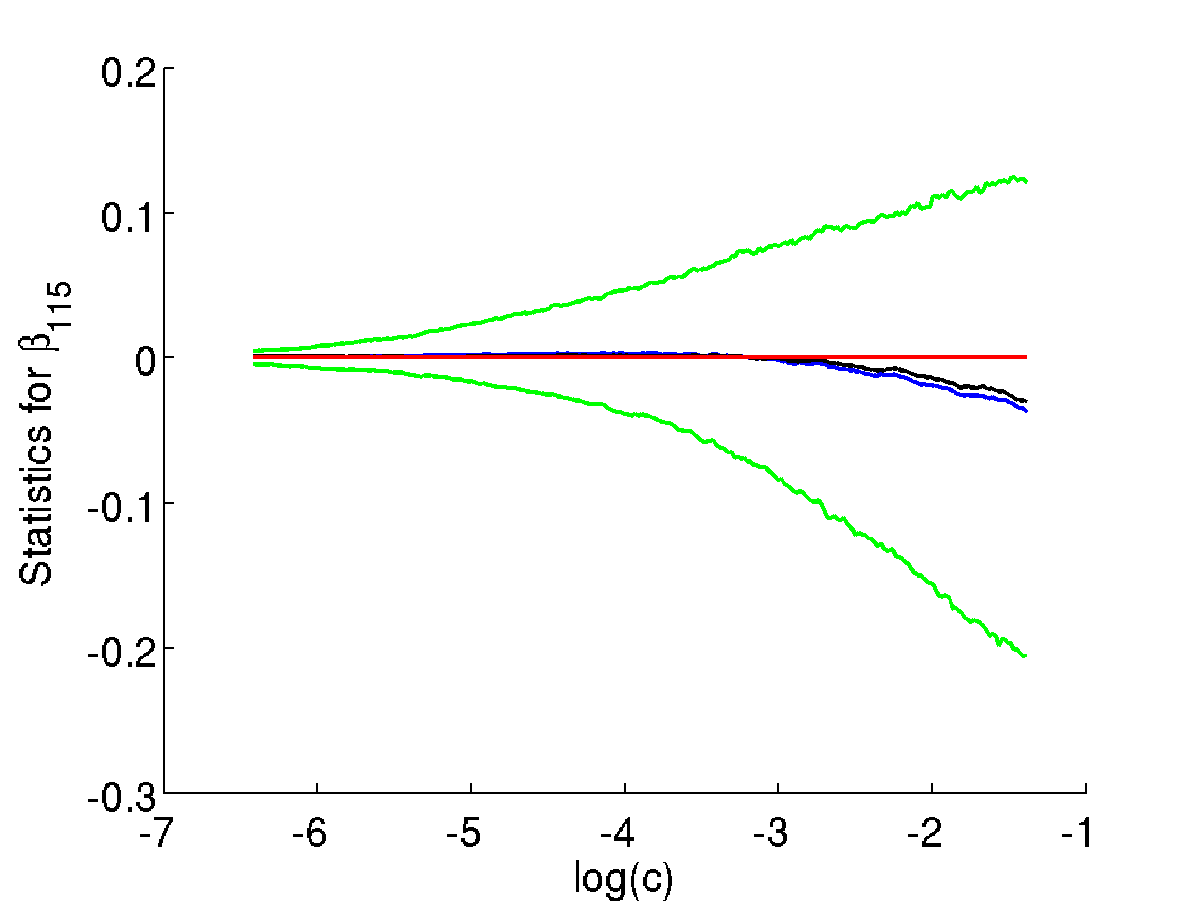}
}
\caption{Summary statistics for posterior distributions for individual coefficients, $90\%$ credible intervals (green), median (black), mean (blue) and MAP (red).}
\label{fig:a4n1859_index_1}
\end{figure}

\begin{figure}[htp]
\center
\subfigure[$a=4, j=22$]{
       \includegraphics[scale=0.3]{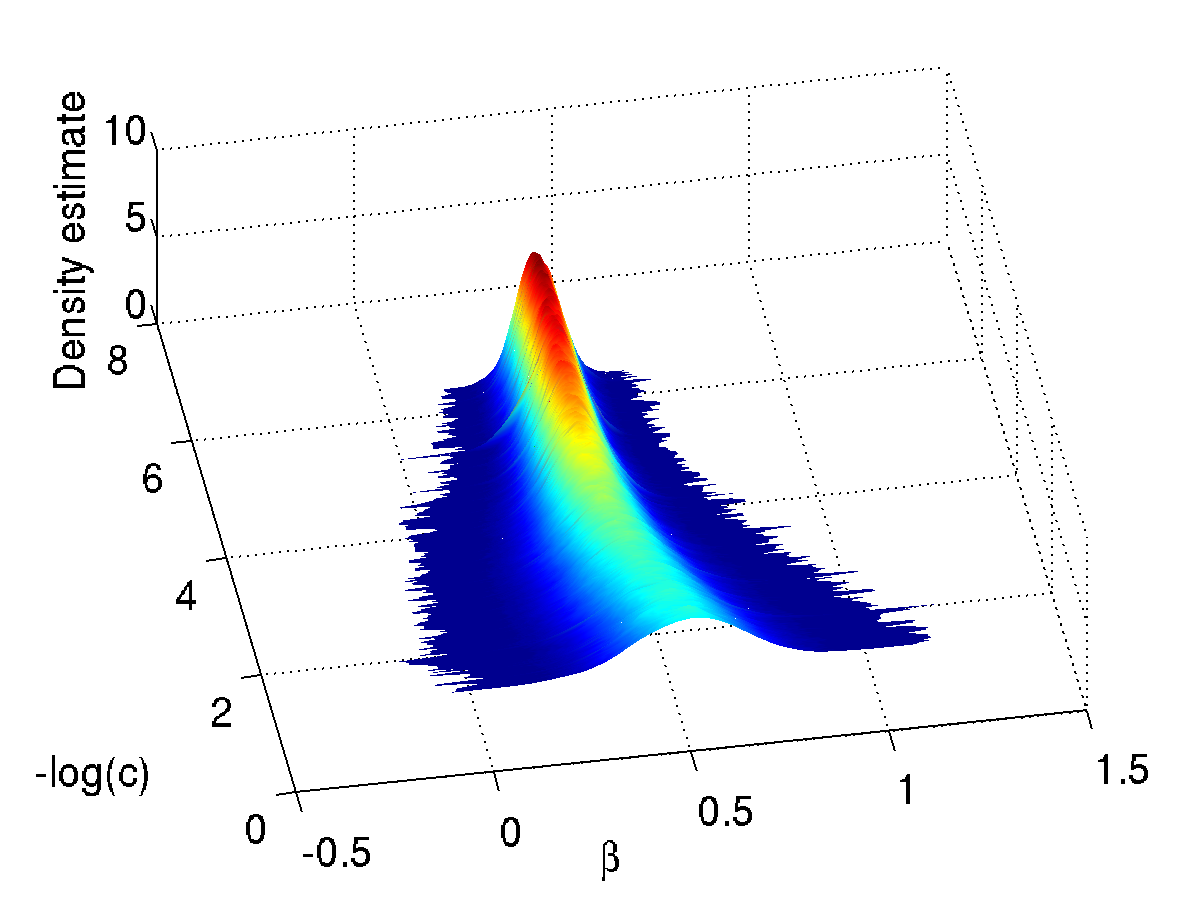}
}
\subfigure[$a=4, j=5$]{
       \includegraphics[scale=0.3]{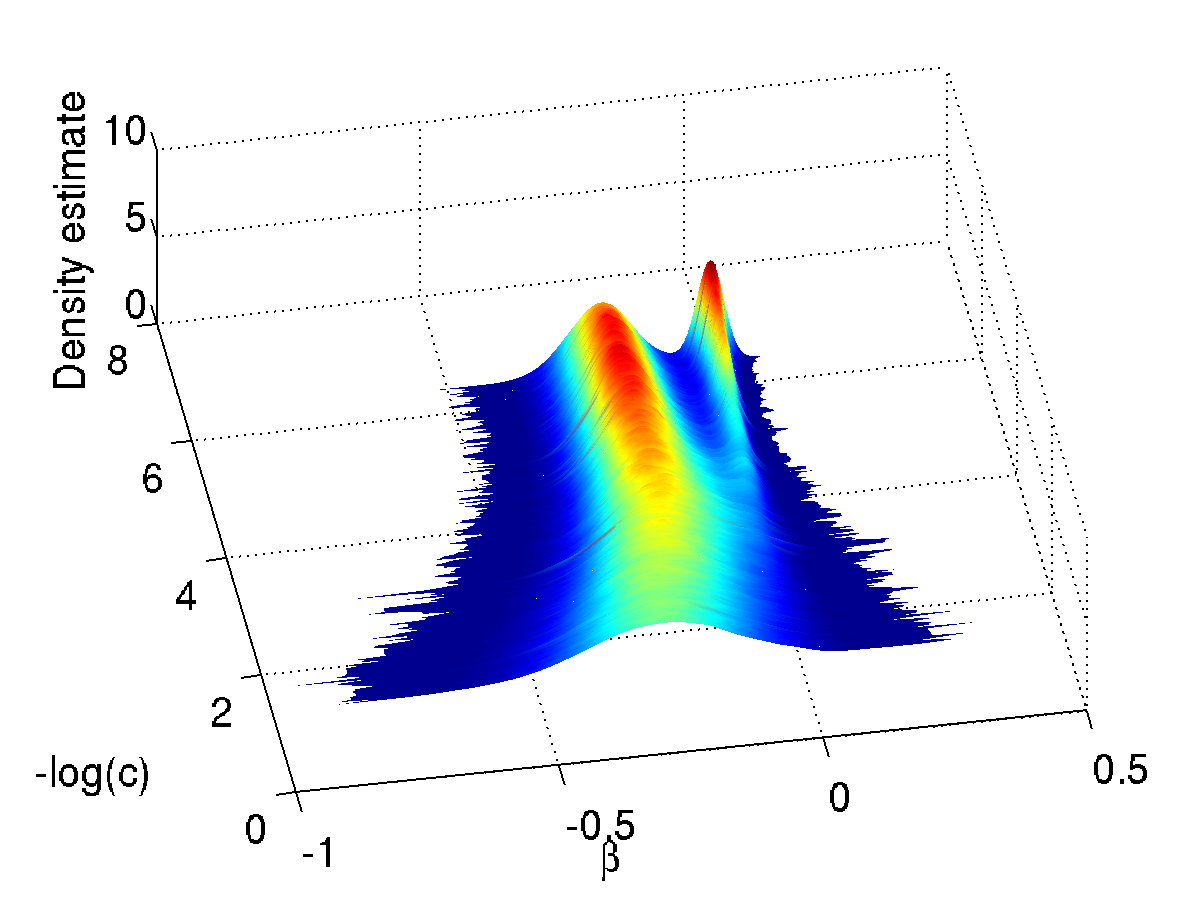}
}
\subfigure[$a=4, j=117$]{
       \includegraphics[scale=0.3]{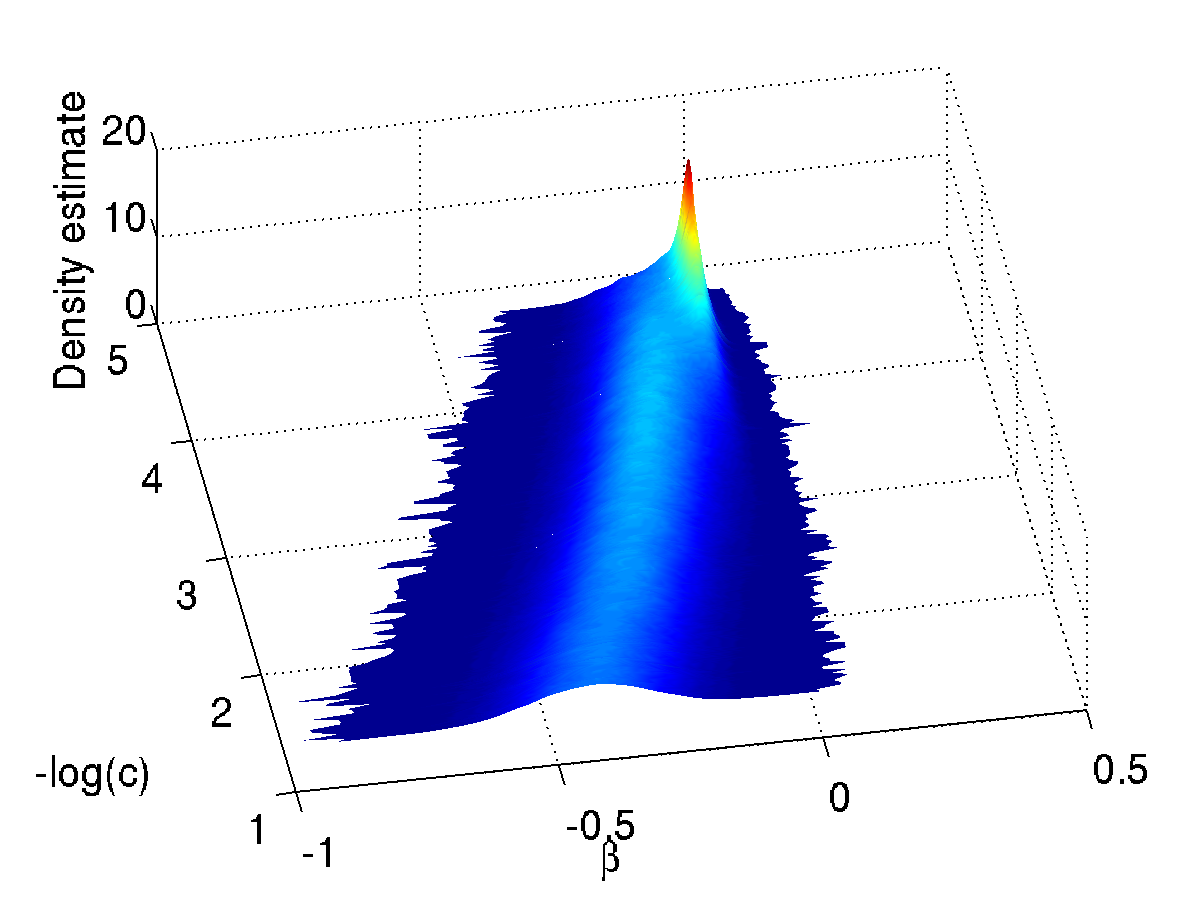}
}
\subfigure[$a=4, j=115$]{
       \includegraphics[scale=0.3]{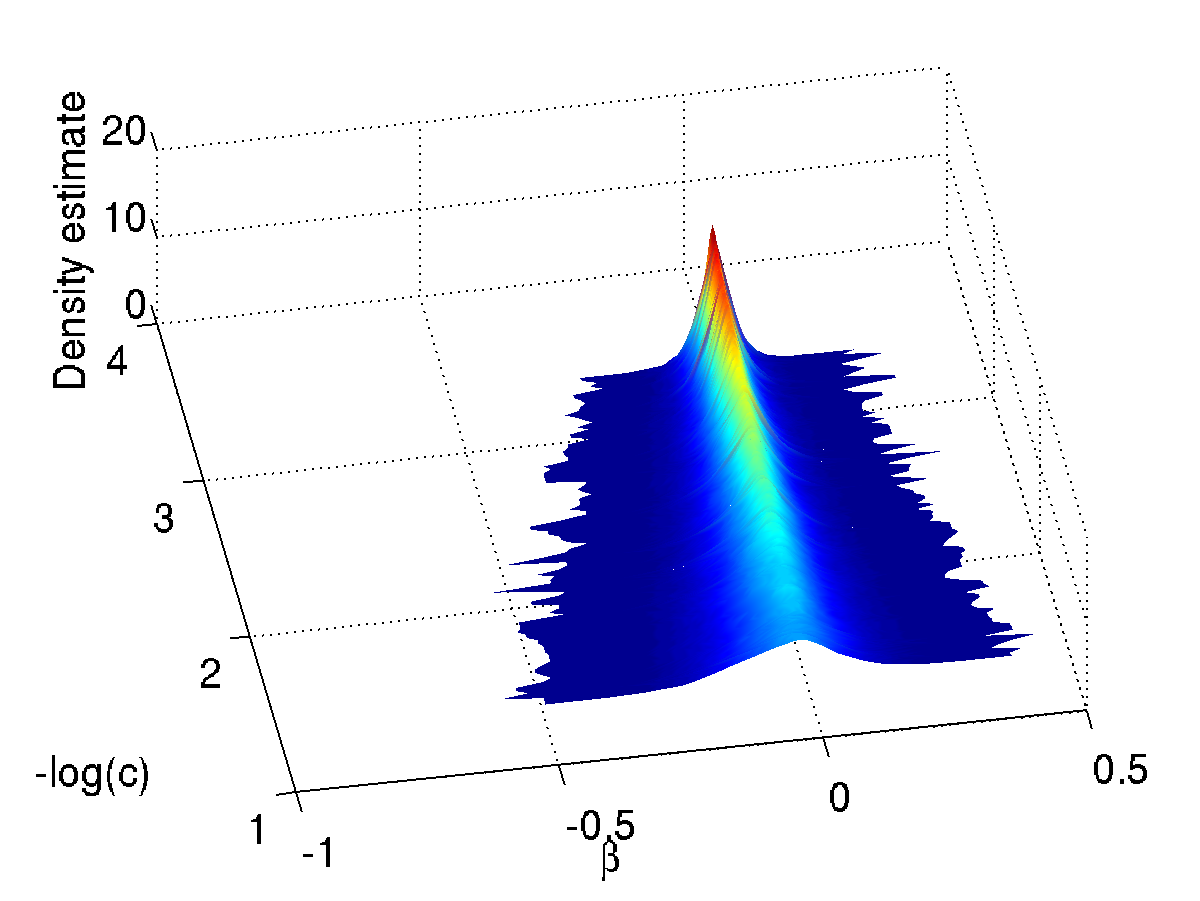}
}
\caption{Posterior density plots of coefficients in Fig.\ref{fig:a4n1859_index_1}}
\label{fig:a4n1859_index_2}
\end{figure}

We next analysed the larger data set with $n=1859$ and $p=184$. The indices of the non-zero cofficients are ${\cal{I}} = \{108, 22, 5, 117, 162\}$ with the same values retained for $\bbeta_{\cal{I}} = \{-0.2538, 0.4578, -0.1873, -0.1498, 0.0996\}$. The SPA plots are shown in Fig.\ref{fig:spa_a4n1859} with corresponding individual coefficient plots in Fig.\ref{fig:a4n1859_index_1} and Fig.\ref{fig:a4n1859_index_2}. We can see in Fig.\ref{fig:spa_a4n1859} that there is much stronger evidence that coefficients $\{108, 22, 5, 117\}$ are important. Interestingly variable $162$ is still masked by other predictors. In comparison with Fig.\ref{fig:spa_a4}(c) we observe that a smaller value of $c$ is supported by the larger data set.

\begin{figure}[htp]
\center
\subfigure[]{
       \includegraphics[scale=0.3]{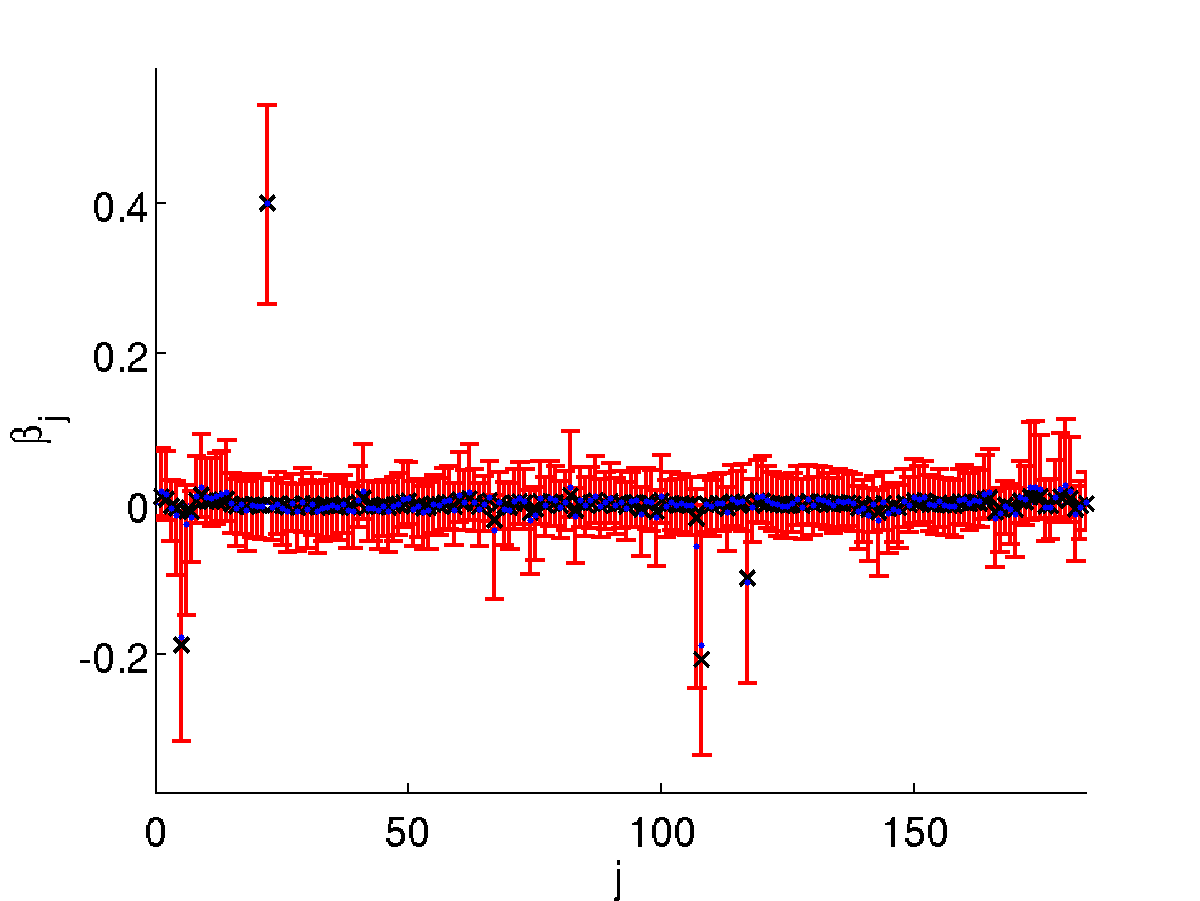}
}
\subfigure[]{
       \includegraphics[scale=0.3]{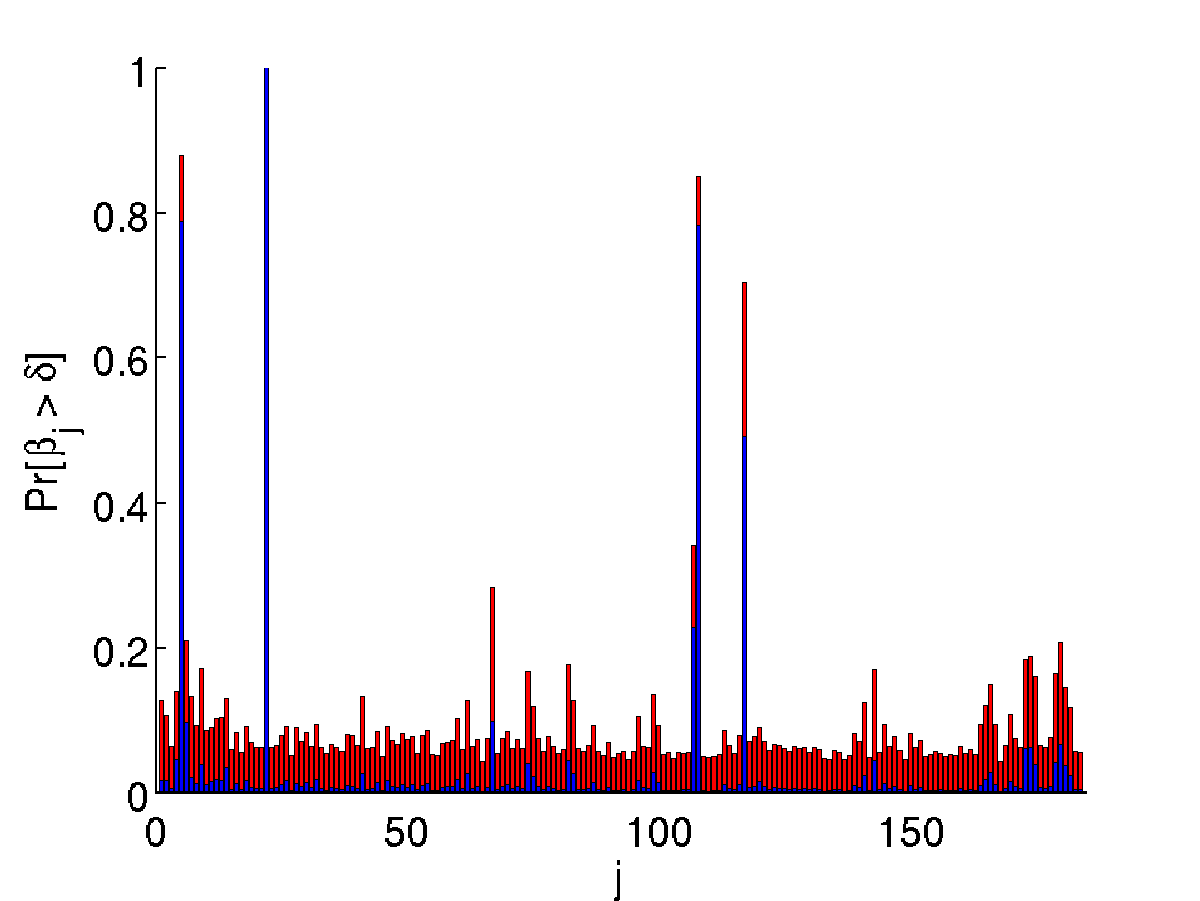}
}
\caption{Marginal plots: (a) the summary stats from marginal posterior distributions showing MAPs (crosses), Medians (stars), and $90\%$ credible intervals (bars) for $a=4$ degrees of freedom; (b) the marginal concentrations $\Delta = 0.05$ red bars, $\Delta=0.1$ (blue bars) for $a=4$}
\label{fig:all_stats_a4n1859}
\end{figure}

\begin{figure}[htp]
\center
	\includegraphics[scale=0.6]{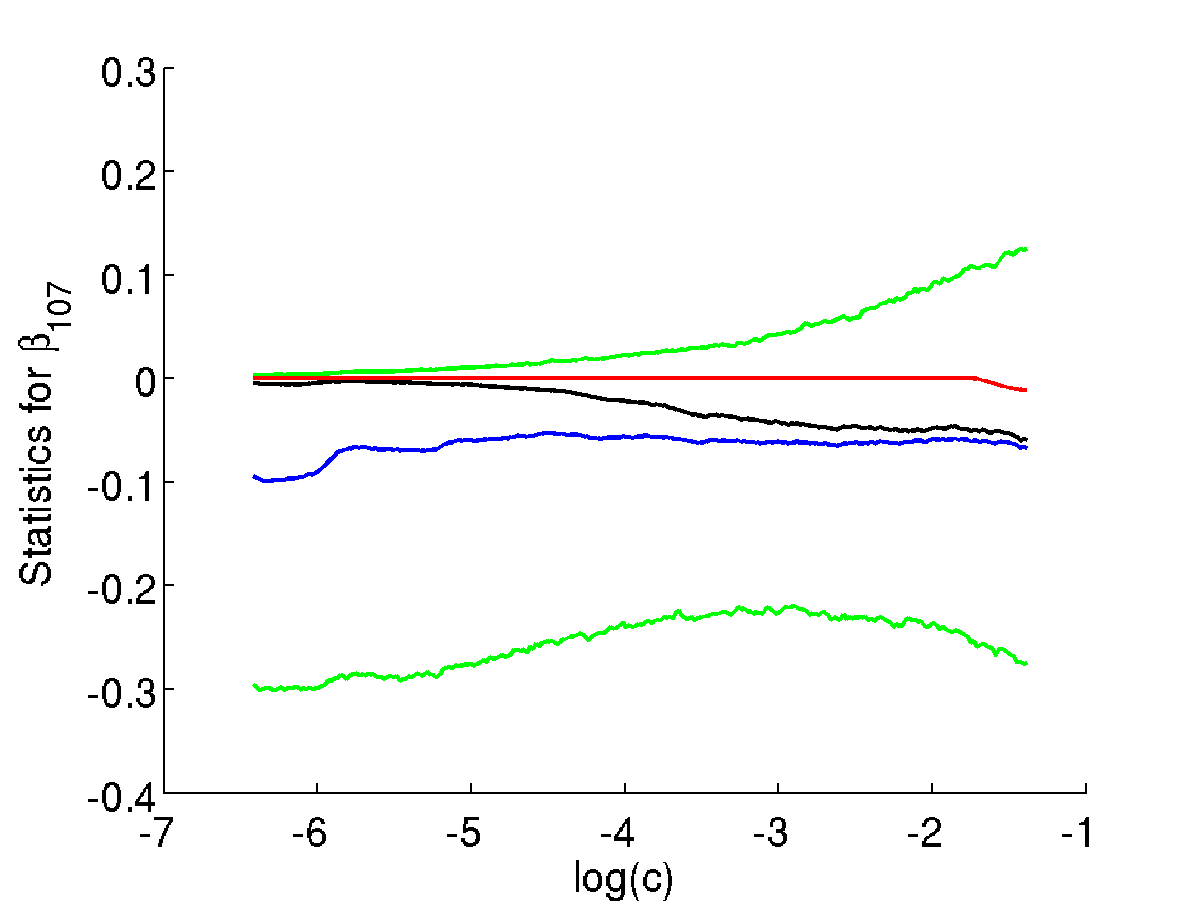}
\caption{Stats for individual coefficient $\beta_{107}$ showing $90\%$ credible intervals (green), median (black), mean (blue) and MAP (red).}
\label{fig:null_107}
\end{figure}

From Fig.\ref{fig:spa_a4n1859}(d) we can see that one null-coefficient, $\beta_{107}$, (with true $\beta_{107}=0$) appears to have some evidence of predictive importance. In Fig.\ref{fig:all_stats_a4n1859} we show summaries of the marginal distributions and concentration plots having integrated over the uncertainty in $c$. The null-coefficient SNP turns out to be situated adjacent to a predictive SNP $\beta_{108}=-0.2538$ shown as the blue line in Fig.\ref{fig:spa_a4n1859}(a), and the two markers are in high LD with correlation coefficient $C(X_{107},X_{108}) = 0.87$, leaving some ambiguity in the source of the association signal. The plot of absolute medians Fig.\ref{fig:spa_a4n1859}(b) helps partly resolve this issue pointing to only four clear association signals around the mode $\tilde{c}$ of the marginal probability. We can drill down a little further and plot summary statistics for the posterior distribution for $\beta_{107}$ shown in Fig.\ref{fig:null_107}. In Fig.\ref{fig:null_107} we can see that the MAP (red line) holds at zero for almost all values of $c$, adding support that this is a ``null'' marker, although the credible intervals (green lines) show there is considerable uncertainty in the actual value. In such a way we can see how complex association signal can be explored, gaining a better understanding of the source of association signal.

\section{Discussion}

We have presented an exploratory approach using generalized t priors that we call Bayesian sparsity-path-analysis to aid in the understanding of genetic association data. The approach involves graphical summaries of posterior densities of coefficients obtained by sweeping over a range of prior precisions, including values with low posterior probability, in order to characterize the space of models from  the sparse to the most complex.

This use of SMC methodology is ideally suited to the inference task, by indexing the SMC on the scale of the prior. The resulting collection of weighted particles provides us with approximations for the coefficient posterior distributions for each scale in addition to estimates of the marginal likelihood and allows for improved robustness in MAP computation. The simulations are computationally demanding and would take days worth of run-time using conventional single-threaded CPU processing. To alleviate this we make use of many-core GPU parallel processing producing around a 20-40 fold improvement in run-times. This has real benefit in allowing for data analysis within a working day for the larger data set we considered.

\section*{Acknowledgements}
The authors wish to acknowledge the organisers of the CRM, Montreal, workshop on  "Computational statistical methods for genomics and system biology". CH wishes to acknowledge support for this research by the MRC, OMI and Wellcome Trust Centre for Human Genetics, Oxford. The code is available from AL or CH.

\bibliographystyle{unsrt}
\bibliography{sip,sip_smc}

\newpage
\section*{Figures}
\center
\includegraphics[scale=0.55]{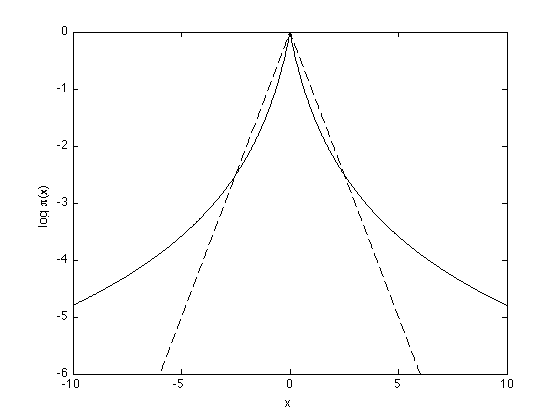}
\captionof{figure}{Plot of log probability under the $Gt(1,1)$ (solid line) and the Lasso prior $DE(0,1)$ (dashed line).}
\label{fig:gent_dexp}

\center
\includegraphics[scale=0.55]{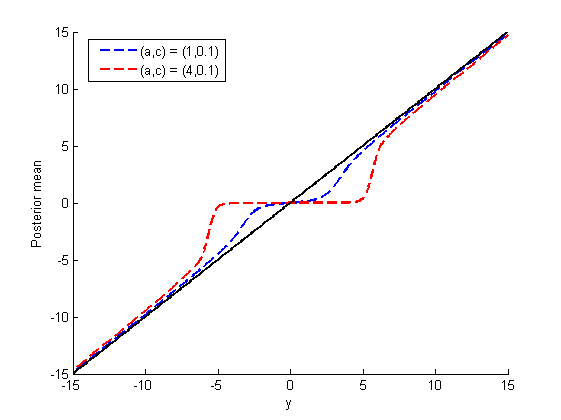}
\captionof{figure}{Estimate of the posterior mean under a single normal observation, $y$, for the $Gt(a=1,b-0.1)$ (blue dashed) and $Gt(a=4,0.1)$ (red dashed). We can see the sparsity by the setting to zero around $y=0$ and the approximate unbiasedness with reduction in shrinkage as $|y|>>0$.}
\label{fig:shrink}

\end{document}